\documentclass[reqno,centertags,12pt]{amsart}

\usepackage[utf8]{inputenc}
\usepackage[T1]{fontenc}
\usepackage[american]{babel}
\usepackage{times}
\usepackage{dsfont}
\usepackage{mathtools}
\usepackage{bbm}
\usepackage[hidelinks]{hyperref}
\usepackage{tikz}
\usetikzlibrary{angles, quotes,decorations.pathreplacing}
\usepackage{pgfplots}
\usetikzlibrary{pgfplots.dateplot}
\usepackage{pgfplots}
\pgfplotsset{compat=1.18}

\usepackage{xcolor}

\hypersetup{pdftitle={}, pdfauthor={D. Hundertmark}, pdfsubject={35P15; 35J10, 81Q10}, pdfkeywords={}}

\usepackage[utf8]{inputenc}

\usepackage[varqu,varl]{inconsolata}

\usepackage[T1]{fontenc}
\usepackage{textcomp}

\usepackage{amsthm}
\usepackage{enumitem,multirow}

\usepackage[colorinlistoftodos]{todonotes}
\usepackage{amsmath}
\usepackage{amssymb}
\usepackage{amsthm}
\usepackage{physics}

\usepackage[utf8]{inputenc}
\usepackage[sb]{libertine}
\usepackage[varqu,varl]{inconsolata}
\usepackage[libertine]{newtxmath}
\useosf

\usepackage[margin={3cm},marginparwidth={2.5cm}]{geometry}

\newcommand{\N}{{\mathbb{N}}}
\newcommand{\R}{{\mathbb{R}}}

\newcommand{\C}{{\mathbb{C}}}

\newcommand{\calS}{{\mathcal{S}}}

\newcommand{\wti}{\widetilde  }

\newcommand{\re}{\mathrm{Re}}

\renewcommand{\abs}[1]{\left| #1 \right|}
\renewcommand{\norm}[1]{\left\Vert #1 \right\Vert}
\allowdisplaybreaks
\DeclareMathOperator{\supp}{supp}

\newcommand{\hatt}{\widehat}

\newtheorem{theorem}{Theorem}[section]
\newtheorem{proposition}[theorem]{Proposition}
\newtheorem{lemma}[theorem]{Lemma}

\theoremstyle{definition}
\newtheorem{definition}[theorem]{Definition}

\newtheorem{remark}[theorem]{Remark}

\numberwithin{theorem}{section}
\numberwithin{equation}{section}

\DeclareMathOperator {\ess}{ess}

\newcounter{smalllist}

\usepackage{color}
\thanks{\copyright 2025 by the authors. Faithful reproduction of this article,
       in its entirety, by any means is permitted for non--commercial purposes}

\title[On the Excess Charge Problem of Atoms]{On the Excess Charge Problem of Atoms}

\author[D.~Hundertmark]{Dirk Hundertmark}
\address[D.~Hundertmark]{Department of Mathematics, Institute for Analysis, Karlsruhe Institute of Technology, 76128 Karlsruhe, Germany\\
	Department of Mathematics, Altgeld Hall, University of Illinois at Urbana-Champaign, 1409 W. Green Street, Urbana, IL 61801}
\email{dirk.hundertmark@kit.edu}

\author[N.~Pattakos]{Nikolaos Pattakos}
\address[N.~Pattakos]{Department of Mathematics, Institute for Analysis, Karlsruhe Institute of Technology, 76128 Karlsruhe, Germany}
\email{nikolaos.pattakos@gmail.com}

\author[M.~R.~Schulz]{Marvin Raimund Schulz}
\address[M.~R.~Schulz]{Department of Mathematics, Institute for Analysis, Karlsruhe Institute of Technology, 76128 Karlsruhe, Germany}
\email{marvin.schulz@kit.edu}

\begin{document}

\begin{abstract}
This paper establishes new bounds on the maximum number of electrons 
$ N_c(Z) $ that an atom with nuclear charge $Z$ can bind. Specifically, 
we show that 
\begin{equation*}
	N_c(Z) < 1.1185Z + O(Z^{1/3}) 
\end{equation*}
with an explicit bound on the lower order term $O(Z^{1/3})$. 
This result improves long--standing bounds by Lieb and Nam obtained 
in 1984, respectively 2012.  
Our bounds show the fundamental difference between fermionic and 
bosonic atoms for finite $Z$ since for bosonic atoms it is known that 
$\lim N_c(Z)/Z = t_c \approx 1.21$ in the limit of large nuclear 
charges $Z$. 
\end{abstract}

\maketitle
\setcounter{tocdepth}{1}
{\hypersetup{linkcolor=black} 
\tableofcontents }


\section{Introduction} \label{ch5:intro}
It is widely accepted that a single atom in a vacuum can bind at most one or two additional electrons. From a heuristic perspective, this is evident: atoms are electrically neutral, and while an additional electron may coexist with the atom, the addition of more electrons becomes challenging due to Coulomb repulsion caused by the net-negative charge of the resulting configuration. Deriving this behavior from the many--body Schr\"odinger equation remains a challenging open question for many decades, \cite{BS:1984, BS:2000}. 

An atom of nuclear charge $Z$ should be able to bind at least $N<Z+1$ electrons, since the farthest out electron will experience a net Coulomb attraction to the nucleus, the remaining $N-1$ electrons cannot fully screen the charge of the nucleus when $N-1<Z$. This was made rigorous in the early work of Zhislin \cite{Z:1969}, 
so $\liminf_{Z\to\infty} \frac{N_c(Z)}{Z}\geq1$ was known for a long time, already. That there is a critical number $N_c(Z)<\infty$ that an atom of charge $Z$ can bind was first shown independently by Ruskai \cite{R:1982} and Sigal \cite{S:1982}. Later, Lieb, Sigal, Simon, and Thirring \cite{LSST:1988} proved 
\begin{align*}
    \lim_{Z\to\infty} \frac{N_c(Z)}{Z}=1\, .
\end{align*}
In fact, they proved the result with $\lim$ replaced by $\limsup$. 
That $N_c(Z)\ge Z$ was shown by Zhislin \cite{Z:1960} much earlier. 
The result by  Lieb, Sigal, Simon and Thirring uses a compactness argument and does not provide any quantitative bounds on how big $N_c(Z)$ is for finite nuclear charge $Z$. 
Fefferman and Seco \cite{FS:1990} and Seco, Sigal and Solovej \cite{SSS:1990} proved in 1990 
\begin{align*}
    N_c(Z)\le Z +O(Z^{5/7}) 
    \quad \text{as } Z\to \infty\, .
\end{align*}
Non--asymptotic bounds are rare, for a long time the only non-asymptotic bound was due to Lieb \cite{L:1984:b} who proved the famous bound 
\begin{align*}
    N_c(Z)<2Z+1 
    \quad \text{for all } Z\ge 1\, .
\end{align*}
Lieb's result is independent of the statistic of the particles, i.e., independent of whether they are 
fermions of bosons, and it also holds, suitably modified, for systems of atoms, i.e., molecules. 
While Lieb's bound certainly overcounts $N_c(Z)$, it shows that Hydrogen ($Z=1$) can bind only two electrons, which is observed in nature.  

It took $28$ years until Nam's breakthrough result  \cite{N:2012} could significantly improve Lieb's longstanding bound. Nam  showed  for a single atom with fermionic statistics that 
\begin{align*}
    N_c(Z)<1.22Z + 3 Z^{1/3}\, 
    \quad \text{for all } Z\ge 1\, .
\end{align*}
In this work, we provide even tighter bounds on $N_c$. In particular, we prove  
    \begin{equation*}
        \begin{split}
          N_c(Z) < 1.1185 Z + 4 Z^{1/3}
          \quad\text{for all } Z\ge 4\, . 
        \end{split}
    \end{equation*}
We would like to stress that this shows that for large nuclear charges $Z$ fermionic atoms do indeed behave much differently from bosonic atoms. Bosonic atoms are known to allow for a surcharge of $21 \, \%$, i.e., large bosonic atoms can bind roughly 
$N \sim 1.21 Z$ bosonic particles for large $Z$. The leading order coefficient in Nam's bound is just above $1.21$ whereas in our bound it is lower, with a good safety margin. 

To achieve our improvements, we significantly extend Nam's approach and, in addition, prove some of the conjectures made in 
\cite{N:2012}. 

\smallskip\noindent
\textbf{Acknowledgements:} 
The authors wish to thank Ioannis Anapolitanos, Leonid Chaichenets, and Peer Kunstmann for insightful discussions, which greatly contributed 
to the development and refinement of this manuscript. 
Funded by the Deutsche Forschungsgemeinschaft (DFG, German Research Foundation) – Project-ID 258734477 – SFB 1173.

\section{Basic notation and main result}
We consider a nucleus of charge $Z$ placed at the origin and $N$ particles located at positions \mbox{$x_{1}, \dots x_{N} \in \R^3$}. In the limit of infinite nuclear mass and after appropriate rescaling the Schr\"odinger operator of a single atom is given by  
\begin{equation} \label{ch5:2:hamilton}
    H_{N,Z} \coloneqq  \frac{1}{2} \sum_{k=1}^N P^2_k -  \sum_{i=1}^N \frac{Z\alpha}{\abs{x_i}} + \sum_{1\leq i<j \leq N}  \frac{\alpha}{\abs{x_i-x_j}} \, .
\end{equation}
Here $\alpha = e^2/(\hbar c)$ is the fine structure constant and $P_j=-i \hbar \nabla_j$ denotes the usual three-dimensional momentum operator with respect to the variable $x_j$. 
\begin{remark}
     We choose atomic units, that is $\hbar = m_e = c = e =1$ and consequently $\alpha=1$. Another usual choice of units is setting the electron mass $m_e =1/2$ as for example in \cite{L:1976}, such that the factor in the kinetic energy vanishes. The question on the maximal excess charge is independent of the choice of these units. 
\end{remark}
The Hilbert space of $N$ particles of spin $ S\in \N_0/2$ is the $N$--fold tensor product
\begin{equation} \label{ch5:2:many_particle_hilbert}
    \mathcal{H}_N \coloneqq \bigotimes_{i=1}^N  L^2(\R^3;\C^{2S+1}) \, .
\end{equation}
Let $z_1,\dots z_N$ be the combined position--spin coordinate of a single particle, $z_j=(x_j,s_j)$. Considering identical particles we only consider stats $\psi \in \mathcal{H}_N$ such that
\begin{equation} \label{ch5:2:pauli_princ}
    \abs{\psi(  \dots z_i , \dots, z_j, \dots ) }^2 = \abs{\psi(  \dots z_j , \dots, z_i, \dots ) }^2 .
\end{equation}
We distinguish between fermions or bosons.  
For fermions, the state space is the subspace of totally antisymmetric functions $\mathcal{H}^f_N$. We call  $\psi \in \mathcal{H}_N  $ totally antisymmetric if
\begin{equation*}
    \psi(  \dots z_i , \dots, z_j, \dots ) = -  \psi(  \dots z_j , \dots, z_i, \dots ) 
    \quad \text{for all } i\neq j\in \{1,\ldots,N\}\,.  
\end{equation*}
For bosons, we consider the subspace of totally symmetric functions denoted by $\mathcal{H}^b_N$. We call  $\psi \in \mathcal{H}_N  $ totally symmetric if
\begin{equation*}
    \psi(  \dots z_i , \dots, z_j, \dots ) =  \psi(  \dots z_j , \dots, z_i, \dots ) 
    \quad \text{for all } i\neq j\in \{1,\ldots,N\}\,.  
\end{equation*}
With regards to real atoms, the particles of interest are electrons and thus the fermionic case with spin $S=1/2$ is the natural one. Thus we discuss $H_{N,Z}$ on $\mathcal{H}^f_N$ first and discuss the case of bosons later. The bottom of the spectrum of $H_{N,Z}$ is called ground-state energy
\begin{equation} \label{ch5:2:def_ground state}
E_{N,Z} \coloneqq \inf \sigma(H_{N,Z}) \, .
\end{equation}
At first, $E_{N,Z}$ does not need to be an eigenvalue of $H_{N,Z}$. The HVZ theorem,  proved by Zhislin \cite{Z:1960}, van Winter \cite{vW:1964} and Hunziker \cite{H:1966} shows that the essential spectrum of $H_{N,Z}$ is given by    $\sigma_{\ess}\left(H_{N,Z}\right)= [E_{N-1,Z},\infty)$.  
Thus the \emph{binding condition} 
\begin{equation}\label{ch5:2:binding_inequality}
    E_{N,Z}<E_{N-1,Z}
\end{equation}
ensures that there is some discrete spectrum below the essential spectrum, hence the atom can bind $N$ electrons. 
In particular, \eqref{ch5:2:binding_inequality} ensures that 
$E_{N,Z}$ is an eigenvalue of $H_{N,Z}$, i.e., there exists a function $\psi_{N,Z}$ such that
\begin{equation} \label{ch5:2:grnd_exists}
    H_{N,Z} \psi_{N,Z} = E_{N,Z} \psi_{N,Z} \, .
\end{equation}
Thus it is natural to call 
 \begin{equation*}
     N_c(Z) \coloneqq \max \{N\in \N: E_{N,Z}<E_{N-1,Z}\} \, 
 \end{equation*}
 the critical number of particles that an atom of charge $Z$ can bind.
 
 In this paper, we derive new bounds on $N_c(Z)$. The strategy of our proof is inspired by \cite{L:1976} and, in particular, \cite{N:2012}.  
 One assumes \eqref{ch5:2:grnd_exists} with $E_{N,Z}<E_{N-1,Z}$ for fixed nuclear charge $Z>0$ (not necessarily an integer) which will then lead 
 to a contradiction if $N$ is too large.
We sketch the original  Bunguria--Lieb argument together with Nam's improvement in Section \ref{ch5:3:sec:bln_argument}. 
 
Our main result is 
\begin{theorem} \label{ch5:2:main_thm} Let $s\in (1,3]$ then there exists $c(s)>0$ such that
    \begin{equation*}
        N_c(Z) < b(s)\, Z + c(s)Z^{1/3},
    \end{equation*}
    where 
    \begin{equation} \label{ch5:2:def_of_b}
        b(s) \coloneqq \max_{0\le t\le 1} \frac{1+t^{s-1}}{1+t^s}.  
    \end{equation}
\end{theorem}
\begin{remark} We conjecture that Theorem \ref{ch5:2:main_thm} holds for any $s\geq 1$ with the same $b(s)$ and some $c(s)<\infty$ with $c(s)\to \infty$ as $s\to \infty$.
\end{remark}
We prove Theorem \ref{ch5:2:main_thm} in Section \ref{ch5:7:ch5:7:proof_main_thrm}. Finding good bounds on $c(s)$ is technical, we study the cases $s=2$ and $s=3$ in greater detail.  In particular, for $s=2$ we show
\begin{proposition} \label{ch5:2:case_s=2}  For any $Z\geq 2$
    \begin{equation}\label{ch5:2:new_bnd_for_s=2}
        N_c(Z)<  \frac{1}{2}(\sqrt{2}+1) \, Z + 2.96 Z^{1/3} \, .
    \end{equation}
    where  $1.2071<\frac{1}{2}(\sqrt{2}+1) <1.2072$.
\end{proposition}
For $s=3$ we show
\begin{proposition} \label{ch5:2:corol_s=3} For any $Z\geq 4$
    \begin{equation} \label{ch5:2:s=3_inequality}
        \begin{split}
        N< b(3) Z +  3.90 Z^{1/3} + 0.0134 +  0.184 Z^{-1/3} +  0.0196 Z^{-2/3},
        \end{split}
    \end{equation}
    with 
    \begin{equation*}
        1.1184 < b(3) = \frac{2}{3}\frac{\sqrt[3]{1+\sqrt{2}}}{(1+\sqrt{2})^{2/3}-1} < 1.1185 \,.
    \end{equation*}
\end{proposition}
\begin{remark}
    Proposition \ref{ch5:2:case_s=2} improves the bound $N_c(Z) < 2Z+1$ 
    in \cite{L:1984:b} for $Z>5.3$. Since the constant  
$(\sqrt{2}+1)/2$ is slightly better than the one in \cite{N:2012} it improves the result of \cite{N:2012} for all $Z\geq 2$. 
For $Z\geq 35.8$ the bound in \eqref{ch5:2:s=3_inequality} is better than \eqref{ch5:2:new_bnd_for_s=2}. In particular Proposition \ref{ch5:2:case_s=2} shows 
\begin{equation*}
    N< 1.12 Z + 4 Z^{1/3}, \quad Z\geq 4 \, .
\end{equation*}
\end{remark} 
We prove Propositions \ref{ch5:2:case_s=2} and \ref{ch5:2:corol_s=3} in Section \ref{ch5:7:ch5:7:proof_main_thrm}.\\
Proposition \ref{ch5:2:corol_s=3} significantly improves upon the results of \cite{N:2012} for large $Z$. More importantly, while it falls short of proving the asymptotic neutrality of fermionic atoms, it provides the first quantitative result showing that, for atoms, the distinction between fermions and bosons is crucial.
In \cite{BL:1983} Benguria and Lieb showed that in contrary to fermions for bosonic atoms
\begin{equation*}
    \lim\limits_{Z\to \infty} \frac{N_c(Z)}{Z} = t_c > 1\, .
\end{equation*}
In fact, Benguria and Lieb proved the lower bound 
$\liminf_{Z\to \infty} \frac{N_c(Z)}{Z} \ge  t_c$. 
Solovej showed the complementary bound 
$\limsup_{Z\to \infty} \frac{N_c(Z)}{Z} = t_c$ 
in \cite{S:1990:b}. 

Numerically one finds $t_c \approx 1.21$ (see \cite{B:1984}). 
Thus large bosonic atoms can have an excess charge of $21\, \% $. 
On the other hand, as proven in \cite{LSST:1988},  fermionic atoms are asymptotically neutral. Our bound \eqref{ch5:2:s=3_inequality} shows that fermionic atoms are very much different from bosonic atoms not only in the limit of infinite nuclear charge $Z$, but also, quantitatively, for large but finite nuclear charge. 
The constant $\frac{1}{2}(\sqrt{2}+1) < 1.2072$  
in our bound \eqref{ch5:2:new_bnd_for_s=2} is just a bit smaller 
than $t_c\approx 1.21$, 
so it does not allow for this conclusion if one also allows for possible 
numerical uncertainties in the calculation of the precise value of $t_c$. 
However, the constant $b(3)<1.1185$ in the leading order term of 
\eqref{ch5:2:s=3_inequality} is certainly much smaller than $t_c$, 
including possible numerical errors in the calculation of $t_c$. 

Thus our results, in particular Proposition \ref{ch5:2:corol_s=3}, show that 
fermionic atoms are quantitatively quite different from bosonic atoms, not only in the limit of large nuclear charges but already for medium values of $Z$. We apply our method to bosonic atoms in a forthcoming paper.
\section{The Benguria--Lieb--Nam Argument: Playing with weights} \label{ch5:3:sec:bln_argument}
Based on an idea by Benguria, Lieb showed $N_c<2Z+1$ for any $Z\geq 1$ in \cite{L:1984:b}. One starts by taking the scalar product of  the 
solution of the  Schr\"odinger equation
\begin{equation} \label{ch5:3:schrodinger_eq}
    H_{N,Z} \psi_{N,Z} = E_{N,Z} \psi_{N,Z}
\end{equation}
 with $\abs{x_k} \psi_{N,Z}$ where $x_k$ is the position of the 
 $k^{\text{th}}$ electron. Using an idea of Benguria allows to control the terms involving the electron--electron repulsion which together with a crucial positivity of the weighted kinetic energy term  
$\Re \langle |x_k|\psi, P^2\psi\rangle$ leads to the bound $N<2Z+1$  
under the binding condition \eqref{ch5:2:binding_inequality}. 
This led to the first non--asymptotic quantitative bound for the number of particles an atom can bind for arbitrary $Z>0$  
We sketch more of the main ideas shortly. 

In \cite{N:2012} Nam used a similar approach but changed the weight from 
$|x_k|$ to $|x_k|^2$. 
This change in weight complicated the analysis considerably. Mainly because $\Re \langle |x_k|\psi, P^2\psi\rangle$ is no longer positive anymore but also because with Nam's choice of weight the analysis of the terms including the electron--electron repulsion gets considerably more involved. 
Nevertheless, Nam was able to prove $N_c(Z)< 1.22 Z + 3Z^{1/3}$ 
with the help of his new choice of weight, a breakthrough compared to the bound of Benguria--Lieb. 

We refer to \cite{N:2022} for a recent and comprehensive review, which includes a discussion of what we would call the 
\emph{Benguria-Lieb-Nam Argument}. 
In this work, we follow a similar strategy but modify the power in the ansatz to a general power $\abs{x}^s$ with $s\in (1,3]$. The case 
$s=1$ is the case treated by Lieb in \cite{L:1984:b}. 
The idea is the following. Assume we have a solution $\psi_{N,Z}$ 
of  \eqref{ch5:3:schrodinger_eq}. 
Let $k\in \{1, 2, \dots, N\}$ and multiply the Schr\"odinger equation 
from the left by $\abs{x_k}^s \overline{\psi_{N,Z}}$, then, in the quadratic form sense, 
\begin{equation} \label{ch5:3:schrodinger}
    0=\langle \abs{x_k}^s \psi_{N,Z},( H_{N,Z} - E_{N,Z} )\psi_{N,Z} \rangle
    =\Re \langle \abs{x_k}^s \psi_{N,Z},( H_{N,Z} - E_{N,Z} )\psi_{N,Z} \rangle\, .
\end{equation}
Note that $\psi_{N,Z}$ is a many--particle function and the inner product above is the scalar product in the many--particle Hilbert space  
$\mathcal{H}_N$ given in \eqref{ch5:2:many_particle_hilbert}. 
For an introductory explanation, see, for example, 
\cite[Chapter 3]{book:LS:2009}. We ignore the spin, for now, since it is 
irrelevant to the argument. It is only relevant for bounds of the kinetic energy. 

We want to single out the $k^{\text{th}}$ particle. 
Let $N(k) \coloneqq \{1,2,3,\dots N\} \setminus \{k\}$. Then 
the atomic operator $H_{N,Z}$ for $N$ particles, defined \eqref{ch5:2:hamilton},  can be written as 
\begin{equation} \label{ch5:3:H_N_toH_N-1}
    H_{N,Z} 
      = 
        \frac{1}{2}P^2_{k} - \frac{Z}{\abs{x_k}} 
        + \sum_{\substack{i\in N(k)}}  \frac{1}{\abs{x_i-x_k}} 
        + H^{(k)}_{N-1,Z} \, .
\end{equation}
with 
\begin{equation*}
    H^{(k)}_{N-1,Z} 
      = 
        \sum_{\substack{i=1\\i \neq k}}^{N}\left(\frac12 P^2_{i} 
        -\frac{Z}{|x_{i}|}\right)+  
        \sum_{\substack{i,j \in N(k)\\ i < j}} \frac1{|x_{i}-x_{j}|} \, .
\end{equation*}
the operator of an $N-1$ particle system where the $k^\text{th}$ 
particle is removed. Combining \eqref{ch5:3:schrodinger} and \eqref{ch5:3:H_N_toH_N-1} we find
\begin{equation} \label{ch5:3:get_rid_of_many_particle}
    \begin{split}
    0=&\Re \left\langle \abs{x_k}^s \psi_{N,Z}, \left[ \frac{1}{2}P^2_{k} - \frac{Z}{\abs{x_k}} + \sum_{\substack{i\in N(k)}} \frac{1}{\abs{x_i-x_k}} \right] \psi_{N,Z} \right\rangle  \\
    &+ \Re\langle \abs{x_k}^s \psi_{N,Z}, ( H^{(k)}_{N-1,Z} - E_{N,Z} )\psi_{N,Z} \rangle \,.
    \end{split}
\end{equation}
From \eqref{ch5:2:binding_inequality} we have $H^{(k)}_{N-1,Z} \geq E_{N-1}$. Since $H^{(k)}_{N-1,Z}$ commutes with $\abs{x_k}^s$ and  for fixed $x_k$ the function $\abs{x_k}^s\psi$ has the same symmetry as $\psi$ in the other $N-1$ variables, we have 
\begin{equation} \label{ch5:3:use_binding_inequality}
  \begin{split}
    \Re \langle \abs{x_k}^s & \psi_{N,Z}, ( H^{(k)}_{N-1,Z} - E_{N,Z} )\psi_{N,Z} \rangle \\
      &=  
        \langle \abs{x_k}^{s/2} \psi_{N,Z}, ( H^{(k)}_{N-1,Z} - E_{N,Z} )\abs{x_k}^{s/2}\psi_{N,Z} \rangle 
    \geq 0 \, .      
  \end{split}
\end{equation}
Combining \eqref{ch5:3:get_rid_of_many_particle} and \eqref{ch5:3:use_binding_inequality}  we arrive at
\begin{equation} \label{ch5:3:to_be_summed_over_k}
    \begin{split}
        0
          &\geq 
            \frac{1}{2}
            \Re\left\langle \abs{x_k}^s \psi_{N,Z}, 
             P^2_{k}\psi_{N,Z} \right\rangle 
            -Z \left\langle 
                 \psi_{N,Z}, \abs{x_k}^{s-1} \psi_{N,Z} 
                \right\rangle \\
        &\phantom{\ge ~} 
          +\sum_{\substack{i\in N(k)}} \left\langle \psi_{N,Z}, \frac{ \abs{x_k}^s}{\abs{x_i-x_k}}  \psi_{N,Z} \right\rangle   \, .  
    \end{split}
\end{equation}
Of course, 
$\left\langle \abs{x_k}^s \psi_{N,Z}, 
  P^2_{k}\psi_{N,Z} \right\rangle 
  = \left\langle \nabla_{k}(\abs{x_k}^s \psi_{N,Z}), 
               \nabla_{k}\psi_{N,Z} \right\rangle$ 
in the quadratic form sense, where $\nabla_k$ is the gradient with respect to the position of the $k^\text{th}$ particle. 

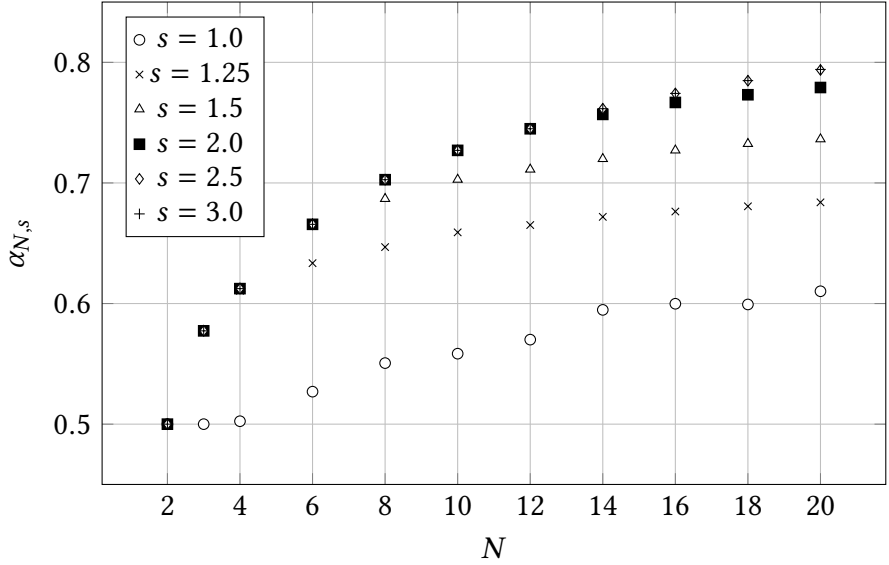
\begin{figure}[!ht]
    \centering
    \begin{tikzpicture}
\begin{axis}[
    xlabel={$N$},
    ylabel={$\alpha_{N,s}$},
    legend pos=north west,
    grid=both,
    width=12cm,
    height=8cm,
    ymin=0.45,
    ymax=0.85,
]

\addplot[only marks, mark=*, mark options={scale=1, fill=white}] table [x index=0, y index=1, col sep=space] {
	2.000000 0.500000
	3.000000 0.500000
	4.000000 0.502421
	6.000000 0.526938
	8.000000 0.550623
	10.000000 0.558434
	12.000000 0.570078
	14.000000 0.594753
	16.000000 0.599823
	18.000000 0.599243
	20.000000 0.610181
};
\addlegendentry{$s = 1.0$}

\addplot[only marks, mark=x, mark options={scale=1}] table [x index=0, y index=1, col sep=space] {
2.000000 0.500000
3.000000 0.577350
4.000000 0.610569
6.000000 0.633486
8.000000 0.646812
10.000000 0.659028
12.000000 0.665092
14.000000 0.671887
16.000000 0.676301
18.000000 0.680615
20.000000 0.683910
};
\addlegendentry{$s = 1.25$}

\addplot[only marks, mark=triangle*, mark options={scale=1, fill=white}] table [x index=0, y index=1, col sep=space] {
2.000000 0.500000
3.000000 0.577350
4.000000 0.612372
6.000000 0.665685
8.000000 0.686859
10.000000 0.702812
12.000000 0.711203
14.000000 0.719955
16.000000 0.727052
18.000000 0.732468
20.000000 0.736316	
};
\addlegendentry{$s = 1.5$}

\addplot[only marks, mark=square*, mark options={scale=1}] table [x index=0, y index=1, col sep=space] {
2.000000 0.500000
3.000000 0.577350
4.000000 0.612372
6.000000 0.665685
8.000000 0.702689
10.000000 0.727011
12.000000 0.744928
14.000000 0.756828
16.000000 0.766747
18.000000 0.773127
20.000000 0.779129
};
\addlegendentry{$s = 2.0$}

\addplot[only marks, mark=diamond*, mark options={scale=1, fill=white}] table [x index=0, y index=1, col sep=space] {
2.000000 0.500000
3.000000 0.577350
4.000000 0.612372
6.000000 0.665685
8.000000 0.702689
10.000000 0.727027
12.000000 0.744928
14.000000 0.761565
16.000000 0.774217
18.000000 0.784777
20.000000 0.793688
};
\addlegendentry{$s = 2.5$}

\addplot[only marks, mark=+, mark options={scale=1}] table [x index=0, y index=1, col sep=space] {
2.000000 0.500000
3.000000 0.577350
4.000000 0.612372
6.000000 0.665685
8.000000 0.702689
10.000000 0.727026
12.000000 0.744928
14.000000 0.761566
16.000000 0.774229
18.000000 0.784858
20.000000 0.794109
};
\addlegendentry{$s = 3.0$}

\end{axis}
\end{tikzpicture}
    \caption{Numerical approximation of the values of $\alpha_{N,s}$ for various $s$ and $N$. 
    Starting from an initial sample set of vectors the values of $\alpha_{N,s}$ have been obtained using a Broyden–Fletcher–Goldfarb–Shanno (BFGS) algorithm implemented in Python. For $s=2.5$ and $s=3.0$ the values of $\alpha_{N,s}$ seem to be almost identical for $N\leq 20$. In the plot $\alpha_{16,1}>\alpha_{18,1}$ which is contrary to Lemma \ref{ch5:3:lem:monotonicity in N} and due to the fat that numerical approximation is difficult for small $s\geq 1$.
    }
    \label{ch5:3:fig:alpha_N_S}
\end{figure}

Due to the symmetry of the ground state, see \eqref{ch5:2:pauli_princ}, 
the first two terms of \eqref{ch5:3:to_be_summed_over_k} do not 
depend on $k$. Consequently by summing over 
$k\in \{1,2,\dots,N\}$ we arrive at
\begin{equation} \label{ch5:3:summed_over_k}
    \begin{split}
        0
        & \geq 
          \frac{1}{2} 
           \Re \left\langle 
             \nabla_1(\abs{x_1}^s \psi_{N,Z}),  
             \nabla_{1}\psi_{N,Z} 
           \right\rangle -Z \left\langle \abs{x_1}^{s-1} \psi_{N,Z}, \psi_{N,Z} \right\rangle \\
        &\phantom{\ge ~}
          +\frac{1}{N}
          \sum_{k=1}^N\sum_{\substack{j\in N(k)}} 
          \left\langle 
            \psi_{N,Z}, 
            \frac{ \abs{x_k}^s}{\abs{x_j-x_k}}  
            \psi_{N,Z} 
          \right\rangle   \, .  
    \end{split}
\end{equation}
Symmetrizing the double sum above yields
\begin{equation} \label{ch5:3:symmetrize}
    \begin{split}
        \frac{1}{N}\sum_{k=1}^N\sum_{\substack{i\in N(k)}} \left\langle \psi_{N,Z}, \frac{ \abs{x_k}^s}{\abs{x_i-x_k}}  \psi_{N,Z} \right\rangle&=\frac{1}{2N}\sum_{k=1}^N\sum_{\substack{j\in N(k)}} \left\langle \psi_{N,Z}, \frac{  \abs{x_j}^s+ \abs{x_k}^s}{\abs{x_j-x_k}}  \psi_{N,Z} \right\rangle \\
        &= \frac{1}{N} \sum_{\substack{j,k=1\\j < k}}^{N} \left\langle \psi_{N,Z}, \frac{  \abs{x_j}^s+ \abs{x_k}^s}{\abs{x_j - x_k}}  \psi_{N,Z} \right\rangle \, .
    \end{split}
\end{equation}
For $N\in \N$ with $N\geq 2$ we define
\begin{equation} \label{ch5:3:def_alpha}
    \alpha_{N,s}
      \coloneqq 
        \inf \left\{ 
          \frac{\sum_{1\le j<k\le N} 
            \frac{\abs{x_j}^s + \abs{x_k}^s}{|x_{j}-x_{k}|}}
            {(N-1)\sum_{k=1}^N \abs{x_k}^{s-1} } : x_k \in \R^{3} 
            \text{ for } k=1,\ldots,N
        \right\}
\end{equation}
which for $s=2$ was introduced by Nam in \cite[Equation (1)]{N:2012}. Combining \eqref{ch5:3:summed_over_k} and \eqref{ch5:3:symmetrize} and using the definition of $\alpha_{N,s}$ yields
\begin{equation} \label{ch5:3:main_inequality}
    \alpha_{N,s}(N-1) 
      < 
        Z 
        -  \frac{1}{2}
           \frac{\Re\left\langle \nabla_1(\abs{x_1}^s \psi_{N,Z}),  \nabla_{1}\psi_{N,Z} \right\rangle}{\left\langle \abs{x_1}^{s-1} \psi_{N,Z}, \psi_{N,Z} \right\rangle} \, .
\end{equation}
\begin{remark}
    The bound \eqref{ch5:3:main_inequality} is the starting point for our analysis. The problem 
    is now reduced to find good lower bounds for $\alpha_{N,s}$ and the 
    second term in the right--hand side of \eqref{ch5:3:main_inequality}.
    Numerical approximation of the values of $\alpha_{N,s}$ for various 
    are given in Figure \ref{ch5:3:fig:alpha_N_S}. 
\end{remark}

When $s=1$ we have $\alpha_{N,s} \ge  1/2$ since 
$\abs{x}+\abs{y}\ge \abs{x-y}$ by the triangle inequality. 
Also 
$\Re\left\langle \nabla_1(\abs{x_1} \varphi),  \nabla_{1}\varphi \right\rangle 
= \left\langle \nabla_1(\abs{x_1}^{1/2} \varphi),  \nabla_{1}(\abs{x_1}^{1/2}\varphi) \right\rangle 
- \tfrac{1}{4}\left\langle \varphi, \abs{x_1}^{-1}\varphi) \right\rangle > 0$, using the 
IMS localization formula and Hardy's inequality in dimension three. 
Thus the second term on the right--hand--side of \eqref{ch5:3:main_inequality} can be dropped when $s=1$ 
and one recovers Lieb's result in the case of a single atom directly.  

It is not too hard to see that for all $s\ge 1$ one has  
$\alpha_{2,s}=1/2$ as well. 
Besides these two cases, $s=1$ and general $N\in\N$, respectively, $N=2$ 
and  general $s\ge 1$,  it is nontrivial to find good lower bounds for 
$\alpha_{N,s}$. Such a bound was derived for 
$\alpha_{N,2}$ in \cite{N:2012}. For Nam's argument it was essential 
to have $s=2$.  
He also showed that $\alpha_{N,2}$ is monotone increasing in $N$. 
This holds for general $s\ge 1$. 
\begin{lemma}\label{ch5:3:lem:monotonicity in N}
    $\alpha_{N,s}$ is increasing in $N\in\N$ for all $s\ge 1$. 
\end{lemma}
\begin{proof}
  In fact, Nam's original proof carries over with minor changes 
  in notation. For the convenience of the reader, we give the 
  short argument. 
  Singling out the particle $m$, we have  
  \begin{align*}
    \sum_{1\le j<k\le N} 
      \frac{\abs{x_j}^s + \abs{x_k}^s}{|x_{j}-x_{k}|}
      &= \sum_{m=1}^N\left( 
            \frac{1}{N-2} \sum_{\substack{1\le j<k\le N\\ j\neq m, k\neq m}}\frac{\abs{x_j}^s + \abs{x_k}^s}{|x_{j}-x_{k}|}
          \right) \\
      &\ge \sum_{m=1}^N\left( 
            \alpha_{N-1,s} \sum_{\substack{k=1\\k\neq m}}^N \abs{x_k}^s 
          \right) 
        = \alpha_{N-1,s} (N-1)  \sum_{k=1}^N \abs{x_k}^s
  \end{align*}
  using the definition of $\alpha_{N-1,s}$. This shows that 
  $\alpha_{N,s}\le \alpha_{N+1,s}$ for all $N\in\N$. 
\end{proof}
Lemma \ref{ch5:3:lem:monotonicity in N} shows that $\alpha_{N,s}$ is increasing in $N$, so once it is bounded, it has a limit. 
Dividing the denominator and numerator in the definition of $\alpha_{N,s}$ 
by $N^2$ suggests the following mean-field type approximation    
\begin{align}\label{ch5:3:def:preliminary def beta_s}
  \beta_s 
    &\coloneqq
      \inf  \left\{ 
        \frac{\iint_{\R^3 \times \R^3} \frac{\abs{x}^s + \abs{y}^s }{2\abs{x-y}}
        d\mu(x) d\mu(y)}{\int_{\R^3} \abs{x}^{s-1} d\mu(x)} : \mu \in P\left(\R^3\right)  
      \right\}\, ,
\end{align}
where $P(\R^3)$ is the set of probability measures on $\R^3$. 
In fact, for technical reasons, we will additionally assume further regularity of the probability measures $\mu$ in the definition of $\beta_s$, see \eqref{ch5:4:uniform_expression} below. 
\begin{lemma} \label{ch5:3:lem:alpha bounded by beta}
    For all $N\in\N$ and $s\ge 1$ we have $\alpha_{N,s}\le \beta_s$. 
\end{lemma}
\begin{proof}
Again, the proof follows the argument in \cite{N:2012}. By symmetry, 
\begin{align*}
  \iint_{\R^3 \times \R^3} & \frac{\abs{x}^s + \abs{y}^s }{2\abs{x-y}}
        d\mu(x) d\mu(y)
   = \frac{1}{N}\int_{\R^{3N}} \frac{1}{N-1}\sum_{1\le j<k\le N} 
       \frac{\abs{x_j}^s + \abs{x_k}^s }{2\abs{x_j-x_k}}
        d\mu(x_1)\ldots d\mu(x_N) \\
  &\ge \frac{\alpha_{N,s}}{N} 
          \int_{\R^{3N}} 
             \sum_{k=1}^N |x_k|^{s-1} 
          d\mu(x_1)\ldots d\mu(x_N) 
    =  \alpha_{N,s}
          \int_{\R^3}  |x|^{s-1} d\mu(x) \, . 
\end{align*}
Thus $\alpha_{N,s}\le \beta_s$ for all $N\in\N$ and $s\ge 1$. 
\end{proof}
\begin{remark}
Given the monotonicity of $\alpha_{N,s}$ in $N$ and the fact that mean-field expressions such as \eqref{ch5:3:def:preliminary def beta_s} often are an excellent approximation for many particle expression such as the one for $\alpha_{N,s}$, one expects that $\alpha_{N,s}$ converges to $\beta_s$ in the limit of large $N$. This is indeed the case, Lemma \ref{ch5:5:lem:bound on alpha for large s} shows that 
$\lim_{N\to\infty}\alpha_{N,s} = \beta_s$ for all $s\ge 2$, 
see Remark \ref{ch5:5:remark_alpha_beta}. 
\end{remark}
That leaves us with the task of finding the value of $\beta_s$, or, at least, good lower bounds for it. 
Nam conjectured in \cite{N:2012} that the infimum in the definition of $\beta_2$ is achieved by radially symmetric probability measures $\mu$. 
If this is the case, then one can easily find 
\emph{excellent lower bounds} for $\beta_2$, and also $\beta_s$. 
In the next section, we show that indeed the infimum is achieved 
with radial measures, not only for $s=2$ but even 
in the range $2\le s\le 3$. 
This allows us to tighten the lower bound 
for $\beta_2$ and it also yields excellent lower bounds for $\beta_s$ in the range $2\le s\le 3$. 

In Section \ref{ch5:5:ch5:5:sec:lwr_bnds} we then show how to derive lower 
bounds for $\alpha_{N,s}$ in terms of $\beta_s$. This
works for all $s\ge 2$, but using the lower bounds for $\beta_s$ 
derived in Section \ref{ch5:4:sec:variational problem} requires 
to restrict our studies to $2\le s\le 3$.  
In the limit $s\to 1$ our estimate reduces to the estimate 
found by Lieb and in the case $s=2$ we find a new 
improved estimate sharpening the results in \cite{N:2012}. 
We can further improve the results by choosing $s\in (1,3]$ 
optimally.

In Section \ref{ch5:6:sec_upper}, we derive upper bounds for the 
right--hand side of \eqref{ch5:3:main_inequality}. 
In Section \ref{ch5:7:ch5:7:proof_main_thrm}, we prove the main 
Theorem \ref{ch5:2:main_thm} along with 
Propositions \ref{ch5:2:case_s=2} and \ref{ch5:2:corol_s=3}. 
\section{Symmetry of minimizing sequences} 
\label{ch5:4:sec:variational problem}
First, we give a more careful definition of $\beta_s$. From now on we set, for $s\ge 1$,  
\begin{equation} \label{ch5:4:uniform_expression} 
\beta_{s}
  \coloneqq
    \inf  \left\{ 
      \frac{\iint_{\R^3 \times \R^3} \frac{\abs{x}^s + \abs{y}^s }{2\abs{x-y}}
      d\mu(x) d\mu(y)}{\int_{\R^3} \abs{x}^{s-1} d\mu(x)} : \mu \in D_s(\R^3)  
    \right\}.
\end{equation}
where $D_s(\R^3)= P(\R^3) \cap H^{-1}(\R^3)\cap L_{s-1}(\R^3)$. 
Here $P(\R^3)$ is the set of all probability measures in $\R^3$, $H^{-1}$ is the usual  Sobolev space of negative order  
\begin{align*}
    H^{-1}(\R^d) 
      = \left\{
            f\in \calS^*(\R^d):\, 
            \|f\|_{H^{-1}}= \left(\int_{\R^d} \frac{|\hatt{f}(k)|^2}{1+|k|^2}\, d k \right)^{1/2}<\infty
        \right\}
\end{align*}
where $\calS^*(\R^d)$ is the space of tempered distributions, 
$\hatt{f}$ the Fourier transform of a tempered distribution $f$, 
and $L_t(\R^3)$ the set of all finite signed or complex--valued 
measures for which $\int_{\R^3}|x|^{t} d|\mu|(x)<\infty$, where 
$|\mu|$ is the total variation of $\mu$. 

First, we show that for $1\le s\le 3$ the infimum in 
\eqref{ch5:4:uniform_expression} can be computed using only radially symmetric probability measures. 
\begin{definition}
  Let $\rho\in P(\R^3)$. The radial part of $\rho$ is the measure 
  $\overline{\rho}$ given by 
  \begin{equation}\label{ch5:4:radialization}
    \int_{\R^3} f(x) d\overline{\rho}(x) \coloneqq \int_{\R^3} \int_{\text{SO}(3) }f(U^{-1}x) \ dU d\rho(x),
  \end{equation}
  for any bounded measurable function $f$  where $dU$ is the normalized Haar measure on $SO(3)$.
  We say that a probability measure $\rho$ is radial if $\rho=\overline{\rho}$. 
    The set of radial probability measures on $\R^3$ is given by 
  \begin{align*}
      P_\text{rad}(\R^3) 
        \coloneqq
          \left\{
            \rho\in P(\R^3): \rho=\overline{\rho}
          \right\}\, .
  \end{align*}
\end{definition}
With this definition, we have 
\begin{theorem}\label{ch5:4:thm:radial lower bound}
  For all $1\le s\le 3$ we have 
  \begin{align}\label{ch5:4:eq:radial lower bound}
      \beta_s=\beta^\text{rad}_s 
        \coloneqq 
          \inf\left\{ 
            \frac{\iint_{\R^3 \times \R^3} \frac{\abs{x}^s + \abs{y}^s }{2\abs{x-y}}
            d\mu(x) d\mu(y)}{\int_{\R^3} \abs{x}^{s-1} d\mu(x)} : \mu \in D_{s,\text{rad}}(\R^3) 
    \right\}. 
  \end{align}
  where $D_{s,\text{rad}}(\R^3)= P_\text{rad}(\R^3)\cap H^{-1}(\R^3)\cap L_{s-1}(\R^3)$. 
\end{theorem}
The proof of this theorem is based on 
\begin{lemma} \label{ch5:4:lemma_lax_milgram}
    Let $\rho \in P(\R^3) \cap H^{-1}(\R^3)$ and 
    \begin{equation}\label{ch5:4:interaction}
        I_s(\rho) 
          \coloneqq 
            \iint_{\R^3 \times \R^3} 
              \frac{\abs{x}^s + \abs{y}^s }{2\abs{x-y}}
              d\rho(x)d\rho(y) \, .
    \end{equation}
    Then  for any $s\in(1,3]$ 
    \begin{align*}
      I_s(\rho) \geq I_s(\overline{\rho})
    \end{align*} 
    where $\overline{\rho}$ is the radial part of $\rho$. 
\end{lemma}
\begin{remark}\label{ch5:4:rem:positive definiteness I_0}
$I_0$ is the Coulomb energy, which is known to be positive definite, 
that is, $I_0(\rho)\ge 0$ for all signed and even complex--valued 
measures (see \cite{G:2003}, or \cite[Theorem 5.1]{book:LS:2009}). 
From positive definiteness the bound $I_0(\rho)\ge I_0(\overline{\rho})$ 
follows easily. 
Indeed let $\nu=\rho-\overline{\rho}$ be the non--radial part of $\rho$ and also define the bilinear version $I_0(\rho_1,\rho_2)= \iint \frac{1}{|x-y|}d\rho_1(x)d\rho_2(y)$. Then 
\begin{align*}
  I_0(\rho)= I_0(\overline{\rho}+\nu,\overline{\rho}+\nu) 
   = I_0(\overline{\rho},\overline{\rho}) +2I_0(\overline{\rho},\nu) +I_0(\nu,\nu) 
\end{align*}
Since $\overline{\rho}$ is a radial measure, Newtons theorem shows 
that its potential, given by 
$V_{\overline{\rho}}(y)= \int \frac{1}{|x-y|}d \overline{\rho}(x)= 
\int  \frac{1}{\max(|x|,|y|)}d\overline{\rho}(x)= V_{\overline{\rho}}(|y|)$ is also radial, hence 
\begin{align*}
  I_0(\overline{\rho},\nu) = \frac{1}{2}\int V_{\overline{\rho}}(|y|)d\nu(y) =0
\end{align*}
since the non--radial part $\nu$ is orthogonal to radial functions. 
Hence 
\begin{align*}
    I_0(\rho) = I_0(\overline{\rho}) + I_0(\nu) \ge I_0(\overline{\rho})\,.
\end{align*}
Unfortunately, we don't know of any such simple argument based on positive 
definiteness, or variations thereof, for $I_s$ when $s>0$. 

Since the proof of Lemma \ref{ch5:4:lemma_lax_milgram} is a bit lengthy, we postpone it to the end of this section and will first show how it implies Theorem \ref{ch5:4:thm:radial lower bound} 
and discuss its consequences. 
We will always assume $s\ge 1$ in the following.
\end{remark}

\begin{proof}[Proof of Theorem \ref{ch5:4:thm:radial lower bound}]
In the definition of $\beta_s$, the infimum is taken over quotients of the form $I_s(\rho)/(\int|x|^{s-1}d\rho(x))$. The denominator does not change under radial symmetrization, i.e., we have 
\begin{align*}
    \int |x|^{s-1} d\rho(x) = \int |x|^{s-1} d\overline{\rho}(x)
\end{align*}
where $\overline{\rho}$ is the radial part of $\rho$. Note that 
$\overline{\rho}$ is again a probability measure and it is also in $H^{-1}$. By Lemma \ref{ch5:4:lemma_lax_milgram} we have 
$I_s(\rho)\ge I_s(\overline{\rho})$ when $1\le s\le 3$. So for this range of parameters $s$ we have 
\begin{align*}
   \frac{I_s(\rho)}{ \int|x|^{s-1}d\rho(x) } 
   \ge \frac{I_s(\overline{\rho})}{ \int|x|^{s-1}d\overline{\rho}(x) },  
\end{align*}
which implies \eqref{ch5:4:eq:radial lower bound}. 
\end{proof}
Equipped with Theorem \ref{ch5:4:thm:radial lower bound} we can compute a lower bound to $\beta_s$ when $1<s\le 3$.  
\begin{proposition} \label{ch5:4:lower_bnd_on_beta}
    Let $s\in (1,3]$ and let $\beta_s$ be defined as in \eqref{ch5:4:uniform_expression} then
    \begin{equation*}
         \beta_s \geq \min_{t \in [0,1]} \frac{1+t^s}{1+t^{s-1}} = \frac{s }{s-1} t_0 \eqqcolon b(s)^{-1} 
    \end{equation*}
    where $t_0 \in (0,1)$ is the unique root of $t \mapsto t^s+ st +1-s $ in $(0,1)$.
\end{proposition} 
\begin{proof}
    By definition and Lemma \ref{ch5:4:lemma_lax_milgram} we have 
    for any $s\in [1,3]$ 
    \begin{equation*}
        \beta_s 
          \geq 
            \inf \left\{ 
              \frac{\iint_{\R^3 \times \R^3} 
                \frac{\abs{x}^s + \abs{y}^s }{2\abs{x-y}}
                (d\overline{\rho}(x)d\overline{\rho}(y)}{\int_{\R^3} \abs{x}^{s-1} d\overline{\rho}(x)} : \overline{\rho} \in P_\text{rad}(\R^3) \cap H^{-1}(\R^3)
            \right\}
    \end{equation*}
    where the infimum is taken only over radial probability measures 
    $\overline{\rho}$. Using Newtons Theorem \cite[Theorem 5.2]{book:LS:2009} for fixed $\overline{\rho}$ yields
    \begin{equation*}
        \iint_{\R^3 \times \R^3} 
          \frac{\abs{x}^s + \abs{y}^s }{2\abs{x-y}}
          d\overline{\rho}(x)d\overline{\rho}(y) 
        =
          \frac{1}{2}\iint_{\R^3 \times \R^3} 
            \frac{\abs{x}^s + \abs{y}^s }{\max(\abs{x},\abs{y})} d\overline{\rho}(x)d\overline{\rho}(y) \,.
    \end{equation*}
    Therefore, with 
    $t(x,y)=\tfrac{\min(\abs{x},\abs{y})}{\max(\abs{x},\abs{y})}
    = \min\left(\tfrac{\abs{x}}{\abs{y}}, \tfrac{\abs{y}}{\abs{x}}\right)$, 
    \begin{equation*}
        \begin{split}
          \frac{1}{2} &\iint_{\R^3 \times \R^3} 
            \frac{\abs{x}^s + \abs{y}^s }{\abs{x-y}}
          d\overline{\rho}(x)d\overline{\rho}(y) \\
            &= 
              \frac{1}{2}\iint_{\R^3\times \R^3}  
                \frac{\abs{x}^s + \abs{y}^s }{\max(\abs{x},\abs{y})(\abs{x}^{s-1}+\abs{y}^{s-1})}
                (\abs{x}^{s-1}+\abs{y}^{s-1})
              d\overline{\rho}(x)d\overline{\rho}(y)\\
                        &= 
              \frac{1}{2}\iint_{\R^3\times \R^3}  
                \frac{1+ t(x,y)^s }{1+t(x,y)^{s-1}}
                (\abs{x}^{s-1}+\abs{y}^{s-1})
              d\overline{\rho}(x)d\overline{\rho}(y)\\ 
            &\ge 
              \min_{0\le t\le 1} 
              \frac{1+ t^s }{1+t^{s-1}}
              \int_{\R^3}  
                \abs{x}^{s-1}
              d\overline{\rho}(x)    \, .   
        \end{split}
    \end{equation*}
Consequently, 
\begin{equation} \label{ch5:4:beta_lower_bnd}
    \beta_s \geq \min_{t \in [0,1]} \frac{1+t^s}{1+t^{s-1}}.
\end{equation}
While this minimum can easily be computed for $s=2$ and $s=3$, it cannot be computed in a closed form for arbitrary $2<s<3$. 
The minimum is obviously not attained at the boundary 
$t \in \{0,1\}$ since
\begin{equation*}
    \frac{1+\left(\frac{1}{2}\right)^s}{1+\left(\frac{1}{2}\right)^{s-1}} = \frac{2^s+1}{2^s+2}<1.
\end{equation*}
We locate the position of the minimum by differentiating
\begin{equation*}
  \left( \frac{1+t^s}{1+t^{s-1}} \right)'=\frac{t^s(t^s+s(t-1)+1)}{(t^s+t)^2} 
\end{equation*}
which vanishes for $t_0\in (0,1)$ iff 
\begin{equation} \label{ch5:4:pos_of_min}
    t_0^s+ st_0 +1-s =0\, .
\end{equation}
The minimum in equation \eqref{ch5:4:beta_lower_bnd} is therefore attained at $t_0 \in (0,1)$. Combining the equation \eqref{ch5:4:pos_of_min} and \eqref{ch5:4:beta_lower_bnd} shows
\begin{equation*}
     \beta_s \geq \min_{t \in [0,1]} \frac{1+t^s}{1+t^{s-1}} = \frac{s}{s-1} t_0  = b(s)^{-1} \, .
\end{equation*}
\end{proof} 
\begin{remark}
Nam conjectured in \cite{N:2012} that $\beta_2$ could be calculated using only radial probability measures. 
Theorem \ref{ch5:4:thm:radial lower bound} shows that this is indeed correct. 
Because of this, our lower bound for $\beta_2$ is slightly better than the one in \cite{N:2012}. More importantly, the main improvement in our analysis of the ionization problem comes from the fact that 
the lower bound for $\beta_s$ is increasing in $s\in[2,3]$ and substantially bigger than $\beta_2$ when $s$ is close to $3$.  We would like to take $s$ much larger than $2$, however, we do not know whether the lower bound of Proposition 
\ref{ch5:4:lower_bnd_on_beta} extends to $s>3$. 

For $s=2$ and $s=3$, one can compute the value of  $\beta_s$ explicitly. One finds
\begin{equation*}
    \begin{split}
         &\beta_2 = 2(\sqrt{2}-1) \Rightarrow b(2) = \frac{1}{2}(\sqrt{2}+1)  \leq 1.2072 \\
         &\beta_3= \frac{3}{2} \frac{(1+\sqrt{2})^{2/3}-1}{\sqrt[3]{1+\sqrt{2}}} \Rightarrow b(3) = \frac{2}{3}\frac{\sqrt[3]{1+\sqrt{2}}}{(1+\sqrt{2})^{2/3}-1} \leq 1.1185
    \end{split}
\end{equation*}    
\end{remark}
\begin{figure}[!ht]
		\centering
            		\begin{tikzpicture}
			\begin{axis}[
				xlabel={\( s \)},
				legend pos=north east,
				grid=both,
				width=10cm,
				height=7cm
				]
				
				\addplot[blue, only marks, mark=*, mark size=2pt] table {
				1.500000 1.333333
				1.551724 1.313393
				1.603448 1.295754
				1.655172 1.280033
				1.706897 1.265928
				1.758621 1.253198
				1.810345 1.241648
				1.862069 1.231120
				1.913793 1.221482
				1.965517 1.212624
				2.017241 1.204455
				2.068966 1.196897
				2.120690 1.189882
				2.172414 1.183355
				2.224138 1.177264
				2.275862 1.171569
				2.327586 1.166230
				2.379310 1.161216
				2.431034 1.156497
				2.482759 1.152049
				2.534483 1.147847
				2.586207 1.143873
				2.637931 1.140108
				2.689655 1.136536
				2.741379 1.133143
				2.793103 1.129914
				2.844828 1.126840
				2.896552 1.123908
				2.948276 1.121109
				3.000000 1.118434
				};
				\addlegendentry{\( b(s) \)};
				
				\addplot[red, only marks, mark=x, mark size=2pt] table {
				1.500000 1.237325
				1.551724 1.223309
				1.603448 1.210936
				1.655172 1.199921
				1.706897 1.190043
				1.758621 1.181132
				1.810345 1.173045
				1.862069 1.165671
				1.913793 1.158917
				1.965517 1.152706
				2.017241 1.146974
				2.068966 1.141666
				2.120690 1.136736
				2.172414 1.132144
				2.224138 1.127856
				2.275862 1.123842
				2.327586 1.120076
				2.379310 1.116536
				2.431034 1.113202
				2.482759 1.110056
				2.534483 1.107082
				2.586207 1.104266
				2.637931 1.101597
				2.689655 1.099062
				2.741379 1.096652
				2.793103 1.094358
				2.844828 1.092171
				2.896552 1.090084
				2.948276 1.088090
				3.000000 1.086183
				};
				\addlegendentry{\( b_{num}(s) \) };
				
			\end{axis}
 		\end{tikzpicture}
		\caption{Values of $b(s)$ according to Proposition \ref{ch5:4:lower_bnd_on_beta}, where $t_0$ was computed numerically. The lower bounds $b_{num}(s)$ on $\beta_s^{-1}$ have been found by choosing explicit measures in \eqref{ch5:4:eq:radial lower bound} and numerical optimization. The exact value of $\beta_s^{-1}$ has to be between both lines. The values $b(3)$ and $b_{num}(3)$ differ by approximately $3 \, \%$.  }  \label{ch5:4:figure:01}
\end{figure}
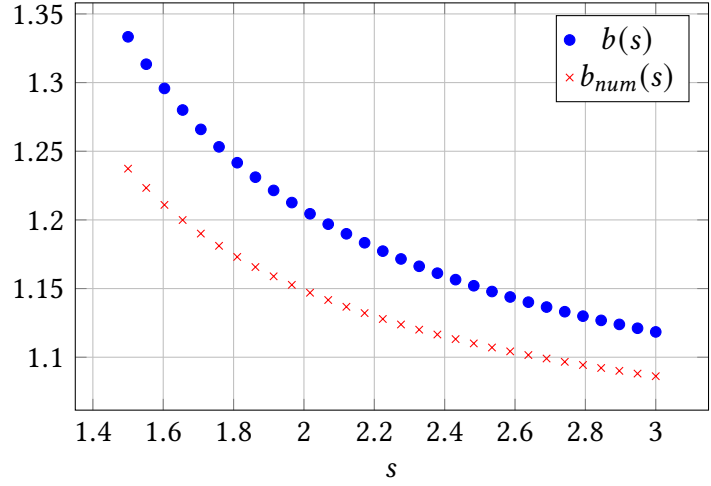
To find upper bounds on $\beta_s$, respectively lower bounds on $b(s)$, one can choose an explicit measure in \eqref{ch5:4:eq:radial lower bound}. To produce Figure \ref{ch5:4:figure:01} we have chosen
\begin{equation*}
    \overline{\rho}_{num}(x) =
    \begin{cases}
        A \abs{x}^{-p}, \quad &\abs{x} \in [1,n]\\
        0, \quad &\abs{x} \notin [1,n]
    \end{cases} \, .
\end{equation*}
and have optimized numerically in the parameters $p$ and $n$ and chose $A>0$ such that $\overline{\rho}_{num}$ is the density of a probability measure. The results of this study are plotted in Figure \ref{ch5:4:figure:01} where $b_{num}(s)$ are the numerically obtained lower bounds on $\beta_s^{-1}$ after the optimization explained above.

Next, we give the
\begin{proof}[Proof of Lemma \ref{ch5:4:lemma_lax_milgram}]
Before we give a detailed proof let us clarify the strategy. 
We would like to mimic the strategy for the Coulomb potential $I_0$ outlined in Remark \ref{ch5:4:rem:positive definiteness I_0}, but the weight $|x|^s$ seems to spoil the argument and, as far we know, no arguments using positive definiteness, as for $I_0$,  or conditional positive definiteness, i.e,  $I_s(\nu)\ge 0$ for all (suitable) signed measures $\nu$ with toal mass $\nu(\R^3)=0$, seem to be available. 
So we have to use a different route. 
Introduce again the bilinear version  
\begin{align}\label{ch5:4:eq:splitting informal}
  I_s(\rho_1,\rho_2) = \iint_{\R^3\times\R^3}\frac{|x|^s+|y|^s}{2|x-y|}d\rho_1(x)d\rho_2(y) \, . 
\end{align}
Then as for $I_0$ we have 
\begin{align*}
    I_s(\rho) = I_s(\overline\rho) + 2I_s(\overline{\rho},\nu) +I_s(\nu)
\end{align*}
where $\nu=\rho-\overline{\rho}$ is the non--radial part of $\rho$. 
Since $\overline{\rho}$ is a radial measure, Newton's theorem shows again that the functions 
\begin{align*}
 \int\frac{|x|^s}{|x-y|}d\overline{\rho}(x)=\int\frac{|x|^s}{\max(|x|,|y|)}d\overline{\rho}(x)
 &\eqqcolon V_1(|y|)\, , \\
 \int\frac{1}{|x-y|}d\overline{\rho}(x)=\int\frac{1}{\max(|x|,|y|)}d\overline{\rho}(x)
 &\eqqcolon V_2(|y|) 
\end{align*}
are radial. Thus 
\begin{align}\label{ch5:4:eq:bilinear term vanishes informal}
  2I_s(\overline{\rho},\nu)  
    = 
      \int_{\R^3} V_1(|y|) d\nu(y) + \int_{\R^3} V_2(|y|) |y|^sd\nu(y) 
      =0
\end{align}
since the measure $\nu$ is orthogonal to radial functions. 
Thus it is enough to show that $I_s(\nu)\ge 0$ for measures $\nu$ which are orthogonal to radial functions. Note that the potential 
$V_\mu= \int\frac{1}{|x-y|}d\mu(y)$ solves the equation 
\begin{align*}
    -\Delta V_\mu = 4\pi\mu 
\end{align*}
in the sense of distributions. Hence, at least informally, 
\begin{align*}
    I_s(\mu) 
    &= 
       \int_{\R^3}\abs{x}^s \int_{\R^3} \frac{1}{\abs{x-y}} d\mu(x)d\mu(y) 
       = 
         \int_{\R^3}\abs{x}^s V_\mu(x) d\mu(x) \nonumber \\
    &= \frac{1}{4\pi}\int |x|^s V_\mu(x) (-\Delta V_\mu(x))d x 
       = 
         \frac{1}{4\pi}\left\langle|x|^sV_\mu, -\Delta V_\mu \right\rangle
\end{align*}
where $\langle \cdot,\cdot\rangle$ is the usual scalar product on $L^2(\R^3)$ and we used symmetry for the first equality. 
Since $\mu$ is real-valued, so is its potential $V_\mu$, hence using the IMS formula and Hardy's inequality, there exists a $c_H>0$ such that 
\begin{align*}
    4\pi I_s(\mu) 
    &= 
      \Re\langle|x|^sV_\mu, -\Delta V_\mu \rangle 
      = \langle |x|^{s/2}V_\mu,-\Delta(|x|^{s/2}V_\mu)\rangle 
        - \langle V_\mu, \frac{s^2}{4}|x|^{s-2} V_\mu \rangle \nonumber \\
    &= 
      \Big\langle |x|^{s/2}V_\mu,\left[-\Delta -\frac{s^2}{4|x|^2}\right](|x|^{s/2}V_\mu)\Big\rangle
      \ge       \left( c_H-\frac{s^2}{4} \right) \langle V_\mu, \abs{\cdot}^{s-2} V_\mu\rangle
\end{align*}
Using $\mu=\nu=\rho-\overline{\rho}$ above and noting that 
$\nu$ is orthogonal to radial functions one sees that also its potential $V_\nu$ is orthogonal to radial functions. Hence  
\begin{align*}
    4\pi I_s(\nu) 
      \ge  
         \left( c_H-\frac{s^2}{4} \right) \langle V_\nu, \abs{\cdot}^{s-2} V_\nu\rangle
\end{align*}
with the improved Hardy constant $c_H=d^2/4$ in dimension $d\geq 2$, 
(see \cite[Lemma 2.4]{EF:2006}), since the potential $V$ is orthogonal to radial symmetric functions in the $L^2$-sense. Hence 
$I_s(\nu)= I_s(\rho-\overline{\rho})\ge 0$ whenever $s\leq d=3$. 
Of course, this is not a proof, the weight $|x|^s$ is not bounded, so the 
application of the IMS localization formula is informal. In addition, it 
is not clear that the potential $V_\nu$ is well--defined. 
\smallskip 

To make this argument rigorous, one has to be a bit more careful. 
For any $x\in \R^3$, $\varepsilon,\lambda>0$ and $\mu \in H^{-1}(\R^3)$ we define
\begin{equation} \label{ch5:4:sol_diff_eq}
   \varphi_{\varepsilon,s}(x) \coloneqq  \frac{\abs{x}^s}{1+\varepsilon \abs{x}^s},  \quad V_{\mu,\lambda}(x) \coloneqq \int_{\R^3} \frac{e^{-\lambda\abs{x-y}}}{4\pi \abs{x-y}} d\mu(y),  
\end{equation}
These are regularized versions of the weight $|x|^s$ and the potential 
$V_\nu$, respectively. 
Let us collect some properties of $\varphi_{\varepsilon,s}$,  $V_{\lambda,\mu}$ and $u[\mu]$ before we continue.
Note that $V_{\mu,\lambda}$ is the solution to the differential equation
\begin{equation} \label{ch5:4:helmholtz}
    (-\Delta + \lambda^2) V_{\mu,\lambda} = \mu.
\end{equation}
in the sense of distributions. It is elementary to verify that $\varphi_{\varepsilon,s} \in W^{1,\infty}$ since $x\mapsto\abs{x}^s$ is weakly differentiable and $\varphi_{\varepsilon,s}$ is bounded by construction. 
In the definition of $\beta_s$ in \eqref{ch5:4:uniform_expression} we also assumed that $\mu\in H^{-1}(\R^d)$. This ensures that 
$V_{\mu,\lambda} \in H^1$. Indeed, \eqref{ch5:4:helmholtz} shows that the Fourier transform of $V_{\lambda,\nu}$ is given by $\hatt{V}_{\lambda,\mu}(k)= (|k|^2+\lambda)^{-1}\hatt{\mu}(k)$. Hence the $H^{-1}$ norm of $V_{\lambda,\nu}$ 
is given by 
\begin{equation*}
    \int_{\R^3} (1+\abs{k}^2)\abs{\hatt{V}_{\mu,\lambda}(k)}^2 dk \lesssim \int_{\R^2} \frac{(1+\abs{k}^2)}{\lambda^2+\abs{k}^2 } \frac{\abs{\hatt{\mu}(k)}^2}{\lambda^2+\abs{k}^2 } dk \lesssim \int_{\R^3 }\frac{\abs{\hatt{\mu}(k)}^2}{\lambda^2+\abs{k}^2 } dk <\infty
\end{equation*}
The last integral is finite for any $\lambda>0$ since, by assumption,  $\mu\in H^{-1}$. In particular, we also have 
$\varphi_\varepsilon V_{\mu,\lambda} \in H^1$ for any $\varepsilon,\lambda>0$ as a product of an $W^{1,\infty}$ and an $H^1$ function. 
Note that if we split a probability measure $\rho\in H^{-1}(\R^3)$ as 
$\rho=\overline{\rho}+\nu$, with $\overline{\rho}$ the radial and $\nu$ the non--radial parts, the same holds for the potential $V_{\overline{\rho}}$ and $V_\nu$. Thus all potentials we need are in $H^1(\R^3)$. 
By monotone convergence  
\begin{equation} \label{ch5:4:monotn_conv}
    \begin{split}
        I_s(\rho)&=\iint_{\R^3 \times \R^3} \frac{\abs{x}^s + \abs{y}^s }{2\abs{x-y}}d\rho(x) d\rho(y) 
          = \lim\limits_{\lambda \to 0} \iint_{\R^3\times\R^3}      
              \frac{(\abs{x}^s+\abs{y}^s)e^{-\lambda\abs{x-y}}}{2\abs{x-y}} d\rho(y)d\rho(x) \\
        &=\lim\limits_{\lambda \to 0}  \lim\limits_{\varepsilon \to 0} \iint_{\R^3\times\R^3} \frac{(\varphi_{\varepsilon,s}(x) + \varphi_{\varepsilon,s}(y))e^{-\lambda\abs{x-y}}}{2\abs{x-y}} d\rho(y)d\rho(x)
    \end{split}
\end{equation}
Define 
\begin{align*}
    I_{s}^{\lambda,\varepsilon}(\rho)
    = \iint_{\R^3\times\R^3} 
         \frac{(\varphi_{\varepsilon,s}(x) 
           + \varphi_{\varepsilon,s}(y))e^{-\lambda\abs{x-y}}}
           {\abs{x-y}} 
      d\rho(y)d\rho(x)
\end{align*}
and its bilinear version 
\begin{align*}
    I_{s}^{\lambda,\varepsilon}(\rho_1,\rho_2)
    = \iint_{\R^3\times\R^3} 
         \frac{(\varphi_{\varepsilon,s}(x) 
           + \varphi_{\varepsilon,s}(y))e^{-\lambda\abs{x-y}}}
           {\abs{x-y}} 
      d\rho_1(y)d\rho_1(x)\, .
\end{align*}
 Split $\rho$ into its radial part $\overline{\rho}$ and its non--radial part $\nu=\rho-\overline{\rho}$. Then as in 
 \eqref{ch5:4:eq:splitting informal} one sees 
 \begin{align*}
    I_{s}^{\lambda,\varepsilon}(\rho) 
      =
        I_{s}^{\lambda,\varepsilon}(\overline{\rho}) + 2 I_{s}^{\lambda,\varepsilon}(\overline{\rho},\nu) + I_{s}^{\lambda,\varepsilon}(\nu)\, .
 \end{align*}
 Again we have that the potentials 
 \begin{align*}
     \wti{V}_1= (-\Delta_\lambda)^{-1} \overline{\rho}
     \quad \text{and } 
     \wti{V}_2= (-\Delta_\lambda)^{-1} \varphi_{\epsilon,s}\overline{\rho}
 \end{align*}
 are rotationally symmetric. Thus as in \eqref{ch5:4:eq:bilinear term vanishes informal} we have 
 \begin{align*}
      8\pi I_s^{\lambda,\varepsilon}(\overline{\rho},\nu)  
    = 
      \int_{\R^3} \wti{V}_1(|y|) d\nu(y) + \int_{\R^3} \wti{V}_2(|y|) |y|^sd\nu(y) 
      =0
 \end{align*}
 since $\nu$ is a bounded measure orthogonal to radial functions. 
 Thus we have 
 \begin{align}\label{ch5:4:eq:rigorous representation punchline 1}
   I_{s}^{\lambda,\varepsilon}(\rho) 
      =
        I_{s}^{\lambda,\varepsilon}(\overline{\rho}) + I_{s}^{\lambda,\varepsilon}(\nu) 
\end{align}
and the claim follows once we show that 
$I_{s}^{\lambda,\varepsilon}(\nu)\ge 0$ for $\nu=\rho-\overline{\rho}$. 
\smallskip

We claim that for any probability measure 
$\mu\in P(\R^3)\cap H^{-1}(\R^3)$ 
\begin{align}\label{ch5:4:eq:rigorous representation}
    \frac{1}{4\pi}I_s^{\lambda,\varepsilon}(\mu,\mu) 
      = 
        \left\langle 
          \nabla(\varphi_{\varepsilon,s}V_{\lambda,\mu}), 
          \nabla(V_{\lambda,\mu})
        \right\rangle 
        + \lambda \left\langle 
          \varphi_{\varepsilon,s}^{1/2}V_{\lambda,\mu}, 
          \varphi_{\varepsilon,s}^{1/2}V_{\lambda,\mu}
        \right\rangle 
\end{align}
where $V_{\lambda,\mu} = (-\Delta+\lambda)^{-1}\mu$ and $\varphi_{\varepsilon,s}V_{\lambda,\nu} \in H^1(\R^3)$.  
Assuming this representation allows to finish the proof since now we can apply the IMS localization formula. Since the right--hand side of \eqref{ch5:4:eq:rigorous representation} is real, we have 
\begin{align}\label{ch5:4:eq:rigorous representation 2}
    \frac{1}{4\pi}&I_s^{\lambda,\varepsilon}(\mu,\mu) 
      = 
        \Re\left\langle 
          \nabla(\varphi_{\varepsilon,s}V_\mu), 
          \nabla(V_\mu)
        \right\rangle 
        + \lambda \left\langle 
          \varphi_{\varepsilon,s}^{1/2}V_{\lambda,\mu}, 
          \varphi_{\varepsilon,s}^{1/2}V_{\lambda,\mu}
        \right\rangle \nonumber\\
      &= \left\langle 
          \nabla(\varphi_{\varepsilon,s}^{1/2}V_\mu), 
          \nabla(\varphi_{\varepsilon,s}^{1/2}V_\mu)
        \right\rangle
        - \left\langle 
          V_\mu, |\nabla \varphi_{\varepsilon,s}^{1/2}|^2 V_\mu
        \right\rangle
        + \lambda \left\langle 
          \varphi_{\varepsilon,s}^{1/2}V_{\lambda,\mu}, 
          \varphi_{\varepsilon,s}^{1/2}V_{\lambda,\mu}
        \right\rangle 
\end{align}
from the IMS localization formula. 
Computing the derivative shows 
\begin{equation*}
    |\nabla \varphi_{\varepsilon,s}^{1/2}|^2 
    = \varphi_{\varepsilon,s}\abs{\frac{\nabla\varphi_{\varepsilon,s}(x)}{2\varphi_{\varepsilon,s}(x)}}^2 
    = \varphi_{\varepsilon,s}\frac{s^2}{4\abs{x}^2} \, ,
\end{equation*}
so, since $\lambda\ge 0$,  
\begin{align*}
    \frac{1}{4\pi}I_s^{\lambda,\varepsilon}(\mu,\mu) 
      &\ge \left\langle 
          \nabla(\varphi_{\varepsilon,s}^{1/2}V_\mu), 
          \nabla(\varphi_{\varepsilon,s}^{1/2}V_\mu)
        \right\rangle
        - \left\langle 
          \varphi_{\varepsilon,s}^{1/2} V_\mu, \frac{s^2}{4\abs{x}^2}  \varphi_{\varepsilon,s}^{1/2} V_\mu
        \right\rangle
\end{align*}
We use this for $\mu=\nu=\rho-\overline{\rho}\in H^{-1}(\R^d)$. Note  
that $\nu$ being orthogonal to radial functions implies that its potential 
$V_\nu=(\Delta+\lambda)^{-1}\nu$ is orthogonal to radial functions in $L^2(\R^3)$. 
Also, since both $\rho$ and $\overline{\rho}\in H^{-1}(\R^3)$ we also have that $\nu\in H^{-1}(\R^3)$, so we can use 
the representation \eqref{ch5:4:eq:rigorous representation} for 
$\mu=\nu$. 

Hence with the improved Hardy inequality $\wti{C}^H_d= d^2/4 $ (see \cite[Lemma 2.4]{EF:2006}), valid for functions orthogonal to radial functions, we get 
\begin{align}\label{ch5:4:eq:rigorous representation punchline}
    \frac{1}{4\pi}I_s^{\lambda,\varepsilon}(\nu,\nu) 
      &\ge \left(\wti{C}^H_3- s^2/4\right)\left\langle 
          \varphi_{\varepsilon,s}^{1/2}V_\nu, |x|^{-2} 
          \varphi_{\varepsilon,s}^{1/2}V_\nu
        \right\rangle\, .
\end{align} 
In our case $d=3$, so 
\eqref{ch5:4:eq:rigorous representation punchline} shows that 
$I_s(\nu,\nu) \ge 0$ as long as $s\le 3$. 
Together with \eqref{ch5:4:eq:rigorous representation punchline 1} we get 
 \begin{align}\label{ch5:4:eq:rigorous representation punchline final}
   I_{s}(\rho) 
     = \lim_{\varepsilon\to 0} \lim_{\lambda\to 0} 
         I_{s}^{\lambda,\varepsilon}(\rho) 
     \ge 
        \lim_{\varepsilon\to 0} \lim_{\lambda\to 0} I_{s}^{\lambda,\varepsilon}(\overline{\rho}) 
     =  I_{s}(\overline{\rho}) 
\end{align}
for any measure $\rho\in P(\R^3)\cap H^{-1}(\R^3)$ and all 
$0\le s\le 3 $. 

It remains to prove the representation \eqref{ch5:4:eq:rigorous representation}. 
First note that by symmetry we have 
\begin{align*}
    \frac{1}{4\pi}I_S^{\lambda\varepsilon}(\mu,\mu) 
     = \iint \frac{\varphi_{\varepsilon,s}(x)e^{-\lambda\abs{x-y}}}{4\pi \abs{x-y}} d\mu(y)\, d\mu(x)
     = \int \varphi_{\varepsilon,s}(x) V_{\lambda,\mu}(x) d\mu(x)
\end{align*}
To show that this leads to \eqref{ch5:4:eq:rigorous representation} we 
use the the Lax-Milgram theorem \cite[Chapter 6.2.1, Theorem 1]{book:E:2022}. For any $\lambda>0$ define the bilinear, or better sesquilinear, form  
\begin{equation*}
    B[\cdot,\cdot]: H^1\times H^1\to \R, \quad (u,v) \mapsto B[u,v] \coloneqq \langle \nabla u, \nabla v\rangle_{L^2} + \lambda^2 \langle u, v\rangle_{L^2} .
\end{equation*}
Then $B$ is coercive and bounded (with respect to the $H^1$ norm) 
and thus fulfills the conditions of the Lax-Milgram Theorem. Let
\begin{equation*}
    f(\cdot):H^1\to \C, \quad g\mapsto f(g) \coloneqq \int_{\R^3} g(x) d\mu(x)
\end{equation*}
for $\mu \in H^{-1}$ a bounded measure.  
Then $f$ is a bounded linear functional on $H^1$ and by the Lax-Milgram theorem there exists a unique $v_0\in H^1$ such that
\begin{equation} \label{ch5:4:from_lax_milgram}
   \int_{\R^3} g d\mu = f(g) = B[g,v_0] = \langle \nabla g, \nabla v_0\rangle_{L^2} + \lambda^2 \langle g, v_0\rangle_{L^2}.
\end{equation}
Recall that this solution $v_0$ is the weak solution of Equation 
\eqref{ch5:4:helmholtz}, hence 
\begin{equation} \label{ch5:4:from_lax_milgram 2}
   \int_{\R^3} g d\mu =  \langle \nabla g, \nabla V_{\mu,\lambda}\rangle_{L^2} + \lambda^2 \langle g, V_{\mu,\lambda} \rangle_{L^2}.
\end{equation}
for all $g\in H^1(\R^3)$. Using 
$g=\varphi_{\varepsilon,s}V_{\lambda,\mu}$ proves 
\begin{align*}
  \frac{1}{4\pi}I_s^{\lambda,\varepsilon}(\mu,\mu) 
     = \int \varphi_{\varepsilon,s} V_{\lambda,\mu} d\mu
     = \left\langle 
         \nabla(\varphi_{\varepsilon,s} V_{\lambda,\mu} ),
         \nabla V_{\lambda,\mu} 
       \right\rangle
       +\lambda \left\langle 
            \varphi_{\varepsilon,s} V_{\lambda,\mu} ,  V_{\lambda,\mu} 
         \right\rangle
\end{align*}
which is \eqref{ch5:4:eq:rigorous representation}. 
\end{proof}


\section{Mean--Field type bounds} \label{ch5:5:ch5:5:sec:lwr_bnds}
In this section we analyze $\alpha_{N,s}$ defined in equation \eqref{ch5:3:def_alpha} and derive lower bounds for $\alpha_{N,s}$ in terms of its mean--field version  $\beta_s$. Clearly, the quotient in the definition of $\alpha_{N,s}$ 
is infinite when two points coincide. So let   
\begin{equation*}
    A\coloneqq \{ (x_1,x_2,\dots x_N) \in \R^{3N}: x_i \neq x_j \text{ whenever } i\neq j \}
\end{equation*}
then
\begin{equation} \label{ch5:5:geometric_exp}
    \alpha_{N,s} = \inf \left\{ \frac{\sum_{\substack{1\leq i,k\leq N \\ i\neq k}} \frac{\abs{x_k}^s + \abs{x_i}^s}{|x_{i}-x_{k}|}}{2(N-1)\sum_{k=1}^N \abs{x_k}^{s-1} } : (x_1,x_2,\dots x_N) \in A\right\} \, .
\end{equation}
Note that the mapping
\begin{equation*}
    (x_1,x_2,\dots x_N) \mapsto \frac{\sum_{\substack{1\leq i,k\leq N \\ i\neq k}} \frac{\abs{x_k}^s + \abs{x_i}^s}{|x_{i}-x_{k}|}}{2(N-1)\sum_{k=1}^N \abs{x_k}^{s-1} }
\end{equation*}
 is continuous on $A$. The set 
\begin{equation}\label{ch5:5:def:A_0}
    A_0 \coloneqq \{ (x_1,x_2,\dots x_N) \in A:  x_k\neq 0  \text{ for } 1\leq k\leq N\}
\end{equation}
is dense in $A$ thus 
\begin{equation} \label{ch5:5:without_zero}
        \alpha_{N,s} = \inf \left\{ \frac{\sum_{\substack{1\leq j,k\leq N \\ i\neq k}} \frac{\abs{x_j}^s + \abs{x_k}^s}{|x_{i}-x_{k}|}}{2(N-1)\sum_{k=1}^N \abs{x_k}^{s-1} } : (x_1,x_2,\dots x_N) \in A_0 \right\} \, .
\end{equation}
For completeness we prove this in the Appendix in Lemma \ref{app:A:lem_a_0}. In order to prove that $\alpha_{N,s} \to  \beta_s$ for $N\to \infty$, with $\beta_s$ given by \eqref{ch5:4:uniform_expression} we need some  preparations. 
\begin{lemma} \label{ch5:5:is_in_dual}
    Let $r>0$, $x\in \R^3\setminus\{0\}$ and $\mu$ be the measure defined by 
    \begin{equation}\label{ch5:5:eq:admiisable measure 1}
        \int_{\R^3} f d\mu \coloneqq \int_{S^2} f(x + r\abs{x} \omega)\frac{d\omega}{4\pi}
    \end{equation}
    for any measurable function $f$. Then $\mu \in H^{-1}(\R^3)$.
\end{lemma}
\begin{remark}
  This lemma shows that convex combinations of probability 
  measures of the form \eqref{ch5:5:eq:admiisable measure 1} are 
  allowed in the computation of upper bounds for $\beta_s$ since 
  they clearly are in $P(\R^3)\cap M_{s-1}(\R^3)$ and the lemma shows that they are also in $H^{-1}(\R^3)$.  
\end{remark}
\begin{proof}
    Let $r>0$, $\Omega \coloneqq \{x\in \R^3: \abs{x}<r\}$ and  
    $T:H^1(\Omega) \to L^2(\partial \Omega)$ the trace operator as for example 
    defined in \cite[Chapter 5.5]{book:E:2022}. 
    Furthermore, let $f\in H^1(\R^3)$ then
    \begin{equation*}
        \abs{\int_{\R^3} f d\mu} = \abs{ \int_{S^2} f(x + r\abs{x} \omega)\frac{d\omega}{4\pi} } \leq \frac{1}{4\pi} \int_{\partial \Omega} \abs{(Tf)(\omega)} d\omega \leq r^{1/2}\norm{Tf}_{L^2(\partial\Omega)}. 
    \end{equation*}
    The Trace Theorem \cite[Chapter 5.5 Theorem 1]{book:E:2022} shows that there exists a $C>0$ such that 
    \begin{equation*}
        \norm{Tf}_{L^2(\partial\Omega)} \leq C\norm{f}_{H^1(\Omega)} < \infty \, .
    \end{equation*}
    Consequently, $\mu$ is in the dual space of $H^{1}(\R^3)$ by definition of the dual space. That is, $\mu\in H^{-1}(\R^3)$.
\end{proof}
We continue by comparing  $\alpha_{N,s}$ to $\beta_s$.  Let $r>0$ and $(x_1,\dots, x_N) \in \R^{3N}$ with $x_j\neq x_k$ for $j\neq k$. Following \cite{N:2012} we define
\begin{equation}\label{ch5:5:ansatz_measure}
    \mu \coloneqq  \frac{1}{N} \sum_{i=1}^N d\mu_i, \, \int_{\R^3} f d\mu_i \coloneqq \int_{S^2} f(x_j + r\abs{x_j} \omega)\frac{d\omega}{4\pi}
\end{equation}
for any measurable function $f$.  By Lemma \ref{ch5:5:is_in_dual} $x_i \neq 0$ implies that $ \mu_i \in H^{-1}$. 
In our analysis, we will use a refinement of the following Lemma \ref{ch5:5:beta_alpha_first_lemma}. 
We provide this bound, since it allows to compare 
 $\alpha_{N,s}$ and $\beta_s$ for all $s>0$ and large $N$, while the refinements only work for $s\ge 2$.  
\begin{lemma}[Comparison of $\alpha_{N,s}$ with $\beta_s$, all $s\ge 0$]\label{ch5:5:beta_alpha_first_lemma}
    Let $\alpha_{N,s}$ and $\beta_s$ be defined as in \eqref{ch5:4:uniform_expression} and \eqref{ch5:5:geometric_exp} then for every $N\geq2$, 
    $r >0$, and $s\ge 0$
    \begin{equation} \label{ch5:5:first_ineq_a_b}
        \frac{(r+1)^{s+1}-(1-r)^{s+1}}{2r(s+1)}  N\beta_s \leq (1+r)^s\left( \alpha_{N,s}(N-1)+ \frac{1}{r}\right)
    \end{equation}
\end{lemma}
\begin{remark}\label{ch5:5:convergence alpha for large N}
  Note that the prefactors in front of $N\beta_s $  
  and $\alpha_{N,s}(N-1)$ in 
  \eqref{ch5:5:first_ineq_a_b}  converges to one as $r\to 0$. 
  Thus \eqref{ch5:5:first_ineq_a_b} shows that for all $r>0$ 
\begin{equation*}
  \liminf_{N\to\infty} \alpha_{N,s} 
    \ge         
      \frac{(1+r)^{s+1}-(1-r)^{s+1}}
           {2r(s+1)(1+r)^s} \, \beta_s \, .
\end{equation*}
Taking the limit $r\to 0$ yields
$\liminf\limits_{N\to\infty} \alpha_{N,s} \ge \beta_s$ and with Lemma \ref{ch5:3:lem:alpha bounded by beta} this proves 
\begin{align*}
    \lim_{N\to\infty} \alpha_{N,s} = \beta_s
\end{align*}
for all $s> 0$.     
\end{remark}
\begin{proof}[Proof of Lemma \ref{ch5:5:beta_alpha_first_lemma}]
 Let $\mu = \sum_{j=1}^N \mu_j$ for points $x_1,\ldots,x_N\in A_0$, 
 where $A_0$ is defined in \eqref{ch5:5:def:A_0}, 
 be the measure given by \eqref{ch5:5:ansatz_measure}. By Lemma \ref{ch5:5:is_in_dual} we know that 
 $\mu \in H^{-1}(\R^3)$. Recall the definition of 
 $I_s$ in Lemma \ref{ch5:4:lemma_lax_milgram}. Then
    \begin{equation} \label{ch5:5:hard_to_est}
        \begin{split}
          N^2I_s(\mu) 
            &=  
              \sum_{j,k=1}^N 
              \iint_{\R^3 \times \R^3} 
                \frac{\abs{x}^s + \abs{y}^s }{2\abs{x-y}}
                d\mu_j(x) d\mu_k(y) \\
            &= 
              \sum_{j\neq k}^N
                \iint_{\R^3 \times \R^3} 
                  \frac{\abs{x}^s + \abs{y}^s }{2\abs{x-y}}
                d\mu_j(x) d\mu_k(y) \\
            & + \sum_{j=1}^N
              \iint_{\R^3 \times \R^3} 
                \frac{\abs{x}^s + \abs{y}^s }{2\abs{x-y}}
              d\mu_j(x) d\mu_j(y) \\
        \end{split}
    \end{equation}
    Note that by construction of the measure $\mu_i$ 
    \begin{equation*}
        \abs{x-x_i} = r \abs{x_i}
    \end{equation*}
    and hence
    \begin{equation} \label{ch5:5:simple_est_on_dist}
        \abs{x} \leq \abs{x-x_i} + \abs{x_i} \leq (1+r)\abs{x_i}.
    \end{equation}
    for any $x\in \supp(\mu_i)$. 
    We first bound the diagonal terms with $j=k$. 
    Note that
    \begin{equation} \label{ch5:5:equal_measures}
    \begin{split}
      \iint_{\R^3 \times \R^3} \frac{\abs{x}^s + \abs{y}^s }{2\abs{x-y}}d\mu_j(x) d\mu_j(y)&= \int_{\R^3} \abs{x}^s \int_{\R^3} \frac{1 }{\abs{x-y}}d\mu_j(y) d\mu_j(x) \\
      &\leq (1+r)^s\abs{x_i}^s\int_{\R^3}  \int_{S^2} \frac{1 }{\abs{x-x_j-r \abs{x_i}\omega}}\frac{d\omega}{4\pi} d\mu_i(x) \\
      &=(1+r)^s \abs{x_j}^s  \int_{S^2} \int_{S^2} \frac{1 }{\abs{r \abs{x_j}\eta -r \abs{x_j}\omega}}\frac{d\omega}{4\pi} \frac{d\eta}{4\pi} \\
      &=\frac{\abs{x_j}^{s-1}}{r} (1+r)^s.
    \end{split}
    \end{equation}
Hence
\begin{equation} \label{ch5:5:equal_measures_sol}
   \sum_{j=1}^N\iint_{\R^3 \times \R^3} \frac{\abs{x}^s + \abs{y}^s }{2\abs{x-y}}d\mu_j(x) d\mu_j(y) \leq \frac{(1+r)^s}{r} \sum_{j=1}^N\abs{x_j}^{s-1}.
\end{equation}
Similarly, we can bound the off--diagonal terms  
$j\neq k$. We have
\begin{equation} \label{ch5:5:unequal_measures}
    \begin{split}
      \iint_{\R^3 \times \R^3} &
        \frac{\abs{x}^s + \abs{y}^s }{2\abs{x-y}}
      d\mu_j(x) d\mu_k(y) \\
      &= 
        \int_{\R^3} \abs{x}^s \int_{\R^3} 
          \frac{1 }{\abs{x-y}}d\mu_k(y) d\mu_j(x) \\
      &\leq 
        (1+r)^s\abs{x_j}^s\int_{\R^3}  \int_{S^2} \frac{1 }{\abs{x-x_k-r \abs{x_k}\omega}}\frac{d\omega}{4\pi} d\mu_j(x) \\
      &= 
        (1+r)^s\abs{x_j}^s\int_{\R^3}  
          \frac{1 }{\max \{ \abs{x-x_j}, r \abs{x_j} \}} 
        d\mu_j(x) \\
      &\leq (1+r)^s\abs{x_i}^s\int_{\R^3}  \frac{1 }{ \abs{x-x_j}} d\mu_j(x) 
        \leq 
          (1+r)^s \frac{ \abs{x_i}^s }{\abs{x_i-x_j}} 
          \, .
    \end{split}
\end{equation}
Thus 
\begin{equation} \label{ch5:5:unequal_measures_sol}
    \sum_{j\neq k}\iint_{\R^3 \times \R^3}
      \frac{\abs{x}^s + \abs{y}^s }
           {2\abs{x-y}} 
    d\mu_j(x)d\mu_k(y) 
      \leq  
        (1+r)^s\sum_{j\neq k}^N 
        \frac{\abs{x_j}^s + \abs{x_k}^s}
             {2\abs{x_j-x_k}} \, .
\end{equation}
Combining \eqref{ch5:5:hard_to_est}, \eqref{ch5:5:equal_measures} and \eqref{ch5:5:unequal_measures_sol} shows
\begin{equation} \label{ch5:5:est_on_num}
    N^2I_s(\mu) 
      \leq 
        \frac{(1+r)^s}{r} 
          \sum_{j=1}^N \abs{x_j}^{s-1} 
        +  (1+r)^s\sum_{j\neq k}^N 
          \frac{\abs{x_j}^s + \abs{x_k}^s}
               {2\abs{x_j-x_k}}  \, .
\end{equation}
Let $t>-2$ then with Lemma \ref{app:A:int_001} we have 
    \begin{equation} \label{ch5:5:n_th_moment}
        \begin{split}
            N\int_{\R^3} \abs{x}^t d\mu 
              &= 
                \sum_{j=1}^N \int_{S^2} 
                  \abs{x_j+r \abs{x_j}\omega}^t \frac{d\omega}{4\pi}  
            = 
              \frac{(1+r)^{t+2} - (1-r)^{t+2}}
                   {2r(t+2)} 
              \sum_{j=1}^N \abs{x_j}^t \,.
        \end{split}
    \end{equation}
Applying \eqref{ch5:5:n_th_moment} for $t=s-1$ yields
\begin{equation} \label{ch5:5:s-1_th_moment}
        \begin{split}
            N\int_{\R^3} \abs{x}^{s-1} d\mu(x)     
              &= 
                \frac{(1+r)^{s+1} - (1-r)^{s+1}}
                     {2r(s+1)} 
                \sum_{j=1}^N \abs{x_j}^{s-1} .
        \end{split}
\end{equation}
Recall the definitions of $\beta_s$ and $I_s(\mu)$ in \eqref{ch5:4:uniform_expression} and \eqref{ch5:4:interaction} 
then $\beta_s\le I_s(\mu)/ \int\abs{x}^{s-1}d\mu$ or 
\begin{equation} \label{ch5:5:closing_lemma_argument}
  N\int_{\R^3} \abs{x}^{s-1} d\mu(x) \,  \beta_sN 
    \leq  N^2 I_s(\mu) \, ,
\end{equation}
which together with \eqref{ch5:5:est_on_num} and \eqref{ch5:5:s-1_th_moment} implies the inequality 
\begin{equation*}
  \frac{(1+r)^{s+1} - (1-r)^{s+1}}
       {2r(s+1)(1+r)^s} 
  \beta_s N 
    \leq 
      \frac{1}{r} 
      +\frac{ 
              \sum_{j\neq k} 
              \frac{\abs{x_j}^s + \abs{x_k}^s}
                   {2\abs{x_j-x_k}} 
             }{\sum_{j=1}^N \abs{x_j}^{s-1} }
\end{equation*}
Taking the infimum in the positions $(x_1,x_2,\dots,x_N) \in A_0$ together with the definition of $\alpha_{N,s}$ and applying Lemma \ref{app:A:lem_a_0} we conclude
\begin{equation*}
      \frac{(1+r)^{s+1} - (1-r)^{s+1}}{2r(s+1)(1+r)^s} \beta_s N
        \leq \frac{1}{r} 
        + \alpha_{N,s} (N-1) \, .
\end{equation*}
This proves \eqref{ch5:5:first_ineq_a_b}.
\end{proof}

\begin{lemma}[Refined comparison I of 
$\alpha_{N,s}$ with $\beta_s$, $s\ge 2$]
\label{ch5:5:lem:bound on alpha for large s}
  For $s\ge 2$ and all $N\in \N$ and $r>0$ we have 
  \begin{equation}\label{ch5:5:eq:large s}
       \left(\frac{s+2}{s+1}\right) \frac{(1+r)^{s+1}-\abs{1-r}^{s+1}}{(1+r)^{s+2}-\abs{1-r}^{s+2}}N\beta_s - \frac{1}{r} \leq \frac{2r(s+2)}{(1+r)^{s+2}-\abs{1-r}^{s+2}} g(r) \alpha_{N,s}(N-1) \, .
\end{equation}
where
\begin{equation}
    g(r) = (1+r^2)^{s/2}\left(\frac{(1+q)^{s/2}+(1-q )^{s/2}}{2} -\frac{s\left(s-2\right)}{15} q^2(1+q)^{\frac{s-4}{2}} \right),  \quad q = \frac{2r}{1+r^2} \, .
\end{equation}
\end{lemma}
\begin{remark}\label{ch5:5:remark_alpha_beta}  
For $s = 2$, the bound \eqref{ch5:5:eq:large s} is similar to the refined bound (27) in \cite{N:2012}. The main challenge in establishing a relationship between $\alpha_{N,s}$ and $\beta_s$ arises from the weighted Coulomb interaction term in \eqref{ch5:5:unequal_measures}. In the proof of Lemma \ref{ch5:5:lem:bound on alpha for large s}, we improve the estimate of these terms using a convexity argument.  

Since obtaining optimal estimates between $\alpha_{N,s}$ and $\beta_s$ 
is technically challenging, we present the proof of 
Lemma \ref{ch5:5:lem:bound on alpha for large s} here, relying on 
convexity. 
A further refinement is provided in Lemma \ref{ch5:5:refined_lower_bound}. 
This improvement, which is more difficult to prove as it involves 
a multipole expansion and estimates for all multipole moments using 
certain nontrivial properties of Legendre polynomials, yields a 
better constant in front of $\beta_s N$.  
\end{remark}

Before we give the proof of Lemma 
\ref{ch5:5:lem:bound on alpha for large s}, we state 
and prove a result which 
is extremely helpful  dropping certain terms when 
deriving a bound on $\alpha_{N,s}$ in terms of 
$\beta_s$ when $s\ge 2$. 

\begin{lemma}
\label{ch5:5:Lem:positivity}
  Let $\gamma:(0,\infty)\times(0,\infty)\to [0,\infty)$ 
  a function such that $\gamma(u,v)$ is increasing in $v$ 
  for any fixed $u>0$. 
  Then for any $N$  dinstinct points  $x_1, \ldots, x_N$ in 
  $\R^3\setminus\{0\}$ we have 
  \begin{align}\label{ch5:5:eq:positivity}
      \sum_{j\neq k} \frac{\gamma(\abs{x_j - x_k}, r\abs{x_j})}{\abs{x_j}} \, x_j\cdot(x_j - x_k) 
      \ge 0
  \end{align}
  for all $r>0$
\end{lemma}
\begin{proof}
    \  Since the sum is over pais $j\neq k$ it is enough to consider the case $N=2$ and $i=1, j=2$.  
    Set $a=x_1-x_2$. 
    We have $x_1\cdot(x_1-x_2) = \abs{x_1}^2 - x_1\cdot x_2 \ge 
    \abs{x_1}^2 - \abs{x_1}\abs{x_2}$ and, similarly, 
    $x_2\cdot(x_2-x_1)\ge \abs{x_2}^2 - \abs{x_2}\abs{x_1}$. 
    Since $\gamma\ge 0$  $\gamma(\abs{a},v_1)\ge \gamma(\abs{a},v_2)$ if $v_1\ge v_2$, 
  by assumption this implies
  \begin{equation*}
      \begin{split}
          &\frac{\gamma(\abs{x_1-x_2}, r\abs{x_1})}
               {\abs{x_1}} \, x_1\cdot (x_1-x_2) 
          + 
          \frac{\gamma(\abs{x_2-x_1}, r\abs{x_2})}
               {\abs{x_2}} \, x_2\cdot (x_2-x_1) \\
        &\ge 
           \gamma(\abs{a}, r\abs{x_1})
                \, \big(\abs{x_1}-\abs{x_2}\big) 
          + 
           \gamma(\abs{a}, r\abs{x_2})
                \, \big(\abs{x_2}-\abs{x_1}\big) \\
        &= 
           \big(\gamma(\abs{a}, r\abs{x_1}) 
                - \gamma(\abs{a}, r\abs{x_2})
            \big)
            \big( \abs{x_1}-\abs{x_2}) \big) \ge 0 \, .
      \end{split}
  \end{equation*} 
\end{proof}
\begin{remark}\label{ch5:5:rem:monotonicity}
   We note that unlike the proof in \cite{N:2012}, we do not 
   need the explicit form of $\gamma(u, v)$ in \eqref{ch5:5:def:gamma} 
   for \eqref{ch5:5:eq:positivity}. Our argument shows that it is enough that $\gamma(u,v)\ge 0$ is increasing in $v>0$ for fixed $u>0$. 
\end{remark}
\begin{proof}[Proof of Lemma \ref{ch5:5:lem:bound on alpha for large s}]
The diagonal terms are easy to calculate. 
  Without loss of generality, let $j=1$. Then  
  by symmetry, the definition of the measures 
  $\mu_j$, and Newton's theorem  we have  
\begin{equation} \label{ch5:5:eq:diag_part_refinement}
    \begin{split}
         \iint_{\R^3\times\R^3} 
      &\frac{\abs{x}^s+\abs{y}^s}{2\abs{x-y}} 
    d\mu_1(x)d\mu_1(y) 
       = 
        \iint_{\R^3\times\R^3} 
          \frac{\abs{x}^s}{\abs{x-y}} 
        d\mu_1(x)d\mu_1(y)  \\
      &= 
        \frac{1}{(4\pi)^2} \int_{S^2} \int_{S^2}
          \frac{|x_1+r\abs{x_1}\omega_1|^s}{r\abs{x_1}\abs{\omega_1-\omega_2}} 
        d\omega_1 d\omega_2 
        = 
          \frac{1}{4\pi r\abs{x_1}}\abs{x_1}^{s-1} 
          \int_{S^2}
            |\hatt{x}_1+r\omega|^s 
          d\omega    \\
     &= 
       \frac{\abs{x_1}^{s-1}}{r} 
       \frac{(1+r)^{s+2}-\abs{1-r}^{s+2}}
            {2r(s+2)}\, .
    \end{split}
\end{equation}
See Lemma \ref{app:A:int_001} for the explicit calculation of the last integral in \eqref{ch5:5:eq:diag_part_refinement}. Thus the diagonal sum 
is  given by 
\begin{equation} \label{ch5:5:eq:diagonal terms}
  \sum_{j=1}^N 
    \iint_{\R^3\times\R^3} 
      \frac{\abs{x}^s+\abs{y}^s}{2\abs{x-y}} 
    d\mu_j(x)d\mu_k(y) 
    = 
    \frac{(1+r)^{s+2}-\abs{1-r}^{s+2}}
            {2r^2(s+2)}
    \sum_{j=1}^N \abs{x_j}^{s-1}\, .
\end{equation}
For the off--diagonal sum we use symmetry 
and Newton's theorem -- the measure $\mu_j$ is 
radially symmetric around the point $x_j$ -- to 
see that 
\begin{equation}\label{ch5:5:eq:off-diagonal with Newton}
 \begin{split}
  \sum_{j\neq k}
  \iint_{\R^3 \times \R^3} 
    &\frac{\abs{x}^s + \abs{y}^s }{2\abs{x-y}}
  d\mu_j(x) d\mu_k(y) 
    = 
      \sum_{j\neq k}
      \iint_{\R^3 \times \R^3} 
        \frac{\abs{x}^s} {\abs{x-y}}
      d\mu_k(y) d\mu_j(x)   \\
    &\le 
       \sum_{j\neq k} 
       \int_{\R^3} 
         \frac{\abs{x}^s }{\abs{x-x_j}}
       d\mu_j(x)    
      = 
        \sum_{j\neq k} 
        \frac{1}{4\pi}\int_{S^2} 
         \frac{\abs{x_j + r\abs{x_j}\omega}^s }
              {\abs{x_j - x_k + r\abs{x_j}\omega}}
       d\omega       
       \end{split}
\end{equation}
To bound the integral in the last sum  for $s\neq 2$ the estimates presented in \cite[Section 4]{N:2012} can not easily be applied, we will proceed differently. With $\hatt{x}= x/\abs{x}$ for $x\in\R^3\setminus\{0\}$ and $q=2r/(1+r^2)$ we have for the numerator in the last term of \eqref{ch5:5:eq:off-diagonal with Newton}
\begin{equation}\label{ch5:5:numerator_extended}
    \abs{x_j + r\abs{x_j}\omega}^s = \abs{x_j}^s(\hat{x}_j + r\omega)^s=\abs{x_j}^s (1+r^2+2r \hat{x}_j\cdot \omega )^{s/2} = \abs{x_j}^s (1+r^2)^{s/2} (1+q \hat{x}_j\cdot \omega )^{s/2} \, .
\end{equation}
Set for $t\in [-1,1]$ and $d\in \R$
\begin{equation}
    F(t) \coloneqq (1+q t )^{s/2}, \quad H_d(t) \coloneqq F(t) - d(t^2-1)
\end{equation}
We determine $d\in \R$ depending on $q,s \in \R$ such that $H$ is convex. Note that $H_d'' = F'' - 2d$. Consequently to ensure convexity of $H_d$ we need
\begin{equation*}
    d\leq \frac{1}{2}F''(t) = \frac{s}{4}\left(\frac{s}{2}-1\right)q^2(1+qt)^{\frac{s-4}{2}} \leq \frac{s}{4}\left(\frac{s}{2}-1\right)q^2(1+q)^{\frac{s-4}{2}}
\end{equation*}
We fix
\begin{equation}\label{ch5:5:def_d0}
    d_0 \coloneqq \frac{s}{4}\left(\frac{s}{2}-1\right)q^2(1+q)^{\frac{s-4}{2}}
\end{equation}
such that $H_{d_0}$ is convex. Due to the convexity of $H_{d_0}$ we have
\begin{equation*}
   H_{d_0}(t) \leq \frac{H_{d_0}(1)-H_{d_0}(-1)}{2}t +  \frac{H_{d_0}(1)+H_{d_0}(-1)}{2} \, .
\end{equation*}
Since $ H_{d_0}=F(t) - d_0(t^2-1)$ this yields
\begin{equation}\label{ch5:5:convexity_applied}
    F(t) \leq \frac{F(1)-F(-1)}{2}t +  \frac{F(1)+F(-1)}{2} + d_0(t^2-1)
\end{equation}
Inserting $F(t) = (1+q t )^{s/2}$ and $d_0$ from \eqref{ch5:5:def_d0} into \eqref{ch5:5:convexity_applied} we arrive at
\begin{equation}\label{ch5:5:main_ineq_after_convex}
    (1+q t )^{s/2} \leq \frac{(1+q)^{s/2}-(1-q )^{s/2}}{2}t + \frac{(1+q)^{s/2}+(1-q )^{s/2}}{2} + \frac{s}{4}\left(\frac{s}{2}-1\right)q^2(1+q)^{\frac{s-4}{2}}(t^2-1)
\end{equation}
Combining \eqref{ch5:5:eq:off-diagonal with Newton}, \eqref{ch5:5:numerator_extended} and \eqref{ch5:5:main_ineq_after_convex} yields
\begin{equation}\label{ch5:5:everything_combined}
 \begin{split}
  \sum_{j\neq k}
  \iint_{\R^3 \times \R^3} 
    &\frac{\abs{x}^s + \abs{y}^s }{2\abs{x-y}}
  d\mu_j(x) d\mu_k(y) \\
    &\le(1+r^2)^{s/2}\frac{(1+q)^{s/2}-(1-q )^{s/2}}{2}  \sum_{j\neq k} 
       \abs{x_j}^s\int_{S^2} 
         \frac{\hat{x}_j\cdot \omega}
              {\abs{x_j - x_k + r\abs{x_j}\omega}}
       \frac{d\omega}{4\pi}       \\
       &\phantom{\le~}+(1+r^2)^{s/2}\frac{(1+q)^{s/2}+(1-q )^{s/2}}{2}  \sum_{j\neq k} 
       \abs{x_j}^s\int_{S^2} 
         \frac{1}
              {\abs{x_j - x_k + r\abs{x_j}\omega}}
       \frac{d\omega}{4\pi} \\
       &\phantom{\le~}+(1+r^2)^{s/2}\frac{s}{4}\left(\frac{s}{2}-1\right)q^2(1+q)^{\frac{s-4}{2}}\sum_{j\neq k} 
       \abs{x_j}^s\int_{S^2} 
         \frac{(\hat{x}_j\cdot \omega)^2-1}
              {\abs{x_j - x_k + r\abs{x_j}\omega}}
       \frac{d\omega}{4\pi} 
       \end{split}
\end{equation}
We proceed by estimating each of the summands in the right--hand side of \eqref{ch5:5:everything_combined} independently. We begin by showing that the first summand is negative. Let $a_{jk}\coloneqq (x_k-x_j)/\abs{x_j}$ then
\begin{equation} \label{ch5:5:to_be_expanded}
    \abs{x_j}^s\int_{S^2} 
         \frac{\hat{x}_j\cdot \omega}
              {\abs{x_j - x_k + r\abs{x_j}\omega}}
       \frac{d\omega}{4\pi}  =\abs{x_j}^{s-1}\int_{S^2} 
         \frac{\hat{x}_j\cdot \omega}
              {\abs{a_{jk}-r\omega}} 
       \frac{d\omega}{4\pi} 
\end{equation}
which either can be solved in polar coordinates directly or using multipol expansion, that is expanding the Coulomb--kernel in terms of the Legendre Polynomials $P_l(t)$, $l\in \N_0$. 
Using the generating function 
\begin{equation}\label{ch5:5:eq:generating function}
    (1+\delta^2- 2\delta t)^{-1/2} = \sum_{n=0}^\infty \delta^n P_n(t)
\end{equation}
which is valid for $\abs{t}\le 1$ and $\abs{\delta}<1$. We can always assume that $r\neq \abs{a_{jk}}$ since otherwise we replace $r$ with $r_\varepsilon=r+\varepsilon$ and take the limit $\varepsilon\to 0$ after solving the integral.  Expanding $\abs{a_{jk}-r\omega}^{-1}$ yields
\begin{equation} \label{ch5:5:multipol_expansion_001}
    \abs{a_{jk}-r\omega}^{-1} = \sum_{n=0}^\infty \frac{\min\{\abs{a_{jk}},r\}^n}{\max\{\abs{a_{jk}},r\}^{n+1}} P_n(\omega \cdot \hat{a}_{jk})
\end{equation}
Using $P_1(\hat{x}_j \cdot \omega) = \hat{x}_j \cdot \omega$ and inserting \eqref{ch5:5:multipol_expansion_001} into \eqref{ch5:5:to_be_expanded} we arrive at
\begin{equation} \label{ch5:5:use_twisted_orthogonality}
    \begin{split}
        \abs{x_j}^s\int_{S^2} 
         \frac{\hat{x}_j\cdot \omega}
              {\abs{x_j - x_k + r\abs{x_j}\omega}}
       \frac{d\omega}{4\pi}  &=\abs{x_j}^{s-1}\int_{S^2} 
        \sum_{n=0}^\infty \frac{\min\{\abs{a_{jk}},r\}^n}{\max\{\abs{a_{jk}},r\}^{n+1}} P_n(\omega \cdot \hat{a}_{jk}) P_1(\hat{x}_j \cdot \omega)
       \frac{d\omega}{4\pi}  \\
       &=\abs{x_j}^{s-1} 
        \sum_{n=0}^\infty \frac{\min\{\abs{a_{jk}},r\}^n}{\max\{\abs{a_{jk}},r\}^{n+1}} \int_{S^2} P_n(\omega \cdot \hat{a}_{jk}) P_1(\hat{x}_j \cdot \omega)
       \frac{d\omega}{4\pi}  \
    \end{split}
\end{equation}
Legendre Polynomials are orthogonal in the following sense 
\begin{equation}
    \int_{-1}^1 P_n(t) P_m(t) \, dt = \frac{2\delta_{mn}}{2n+1},
\end{equation}
and consequently by the Funk-Hecke formula in Lemma \ref{app:A:eq:atkinson han formula} we arrive at
\begin{equation} \label{ch5:5:twisted_orthogonality}
    \int_{S^2} P_n(\hat{x}_j \cdot \omega) P_m(\omega \cdot \hat{a}_{jk}) \, \frac{d\omega}{4\pi} = \frac{\delta_{mn}}{2n+1} \, .
\end{equation}
Inserting \eqref{ch5:5:twisted_orthogonality} into \eqref{ch5:5:use_twisted_orthogonality} we arrive at
\begin{equation*} 
    \begin{split}
        \abs{x_j}^s\int_{S^2} 
         \frac{\hat{x}_j\cdot \omega}
              {\abs{x_j - x_k + r\abs{x_j}\omega}}
       \frac{d\omega}{4\pi}   
       &=\abs{x_j}^{s-1}  \frac{\min\{\abs{a_{jk}},r\}}{\max\{\abs{a_{jk}},r\}^{2}} \frac{\hat{a}_{jk}\cdot \hat{x}_j}{3}  \\
       &= -\abs{x_j}^{s-1}  \frac{\min\{\abs{a_{jk}},r\}}{\max\{\abs{a_{jk}},r\}^{2}} \frac{(x_j-x_k)\cdot \hat{x}_j}{3\abs{x_k-x_j}} \\
       &= -\abs{x_j}^{s}\frac{\min\{\abs{x_j-x_k},r\abs{x_j}\}}{\max\{\abs{x_j-x_k},r\abs{x_j}\}^{2}}\frac{1}{3\abs{x_k-x_j}}(x_j-x_k)\cdot \hat{x}_j \\
       &= -r^{-s} \gamma\left(\abs{x_j-x_k},r\abs{x_j}\right) (x_j-x_k)\cdot \hat{x}_j
    \end{split}
\end{equation*}
with 
\begin{align}
\label{ch5:5:def:gamma}
    \gamma(u,v) 
      =
        \frac{v^s\min(u, v)}{3u\max(u, v)^2}   \, .
\end{align}
for $u,v>0$. Summing \eqref{ch5:5:use_twisted_orthogonality} over $j\neq k$ yields
\begin{equation}
     \sum_{j\neq k} 
       \abs{x_j}^s\int_{S^2} 
         \frac{\hat{x}_j\cdot \omega}
              {\abs{x_j - x_k + r\abs{x_j}\omega}}
       \frac{d\omega}{4\pi}   = -r^{-s} \sum_{j\neq k} \gamma\left(\abs{x_j-x_k},r\abs{x_j}\right) (x_j-x_k)\cdot \hat{x}_j
\end{equation}
Applying Lemma \ref{ch5:5:Lem:positivity} and noting Remark \ref{ch5:5:rem:monotonicity}  we find
\begin{align*}
\sum_{j\neq k} 
          \gamma(\abs{x_j-x_k}, r\abs{x_j}) \, 
          \hatt{x}_j\cdot \big(x_j - x_k\big) \ge 0\, .    
\end{align*}
Thus the first summand in the right--hand side of \eqref{ch5:5:everything_combined} is not positive and consequently
\begin{equation}\label{ch5:5:fist_summand_neglect}
 \begin{split}
  \sum_{j\neq k}
  \iint_{\R^3 \times \R^3} 
    &\frac{\abs{x}^s + \abs{y}^s }{2\abs{x-y}}
  d\mu_j(x) d\mu_k(y)      \\
       &\leq \phantom{+}(1+r^2)^{s/2}\frac{(1+q)^{s/2}+(1-q )^{s/2}}{2}  \sum_{j\neq k} 
       \abs{x_j}^s\int_{S^2} 
         \frac{1}
              {\abs{x_j - x_k + r\abs{x_j}\omega}}
       \frac{d\omega}{4\pi} \\
       &\phantom{\le~}+(1+r^2)^{s/2}\frac{s}{4}\left(\frac{s}{2}-1\right)q^2(1+q)^{\frac{s-4}{2}}\sum_{j\neq k} 
       \abs{x_j}^s\int_{S^2} 
         \frac{(\hat{x}_j\cdot \omega)^2-1}
              {\abs{x_j - x_k + r\abs{x_j}\omega}}
       \frac{d\omega}{4\pi} 
       \end{split}
\end{equation}
The first integral in the right--hand side of \eqref{ch5:5:fist_summand_neglect} can be solved since due to Newton's theorem 
\begin{equation} \label{ch5:5:Newton_applied}
    \int_{S^2} 
         \frac{1}
              {\abs{x_j - x_k + r\abs{x_j}\omega}}
       \frac{d\omega}{4\pi} = \frac{1}{\max\{\abs{x_j-x_k},r\abs{x_j}\} } \, .
\end{equation}
Inserting \eqref{ch5:5:Newton_applied} into \eqref{ch5:5:fist_summand_neglect} yields
\begin{equation}\label{ch5:5:one_int_remains}
 \begin{split}
  \sum_{j\neq k}
  \iint_{\R^3 \times \R^3} 
    &\frac{\abs{x}^s + \abs{y}^s }{2\abs{x-y}}
  d\mu_j(x) d\mu_k(y)      \\
       &\leq \phantom{+}(1+r^2)^{s/2}\frac{(1+q)^{s/2}+(1-q )^{s/2}}{2}  \sum_{j\neq k} 
         \frac{\abs{x_j}^s}
              {\max\{\abs{x_j-x_k},r\abs{x_j}\}} \\
       &\phantom{\le~}+(1+r^2)^{s/2}\frac{s}{4}\left(\frac{s}{2}-1\right)q^2(1+q)^{\frac{s-4}{2}}\sum_{j\neq k} 
       \abs{x_j}^s\int_{S^2} 
         \frac{(\hat{x}_j\cdot \omega)^2-1}
              {\abs{x_j - x_k + r\abs{x_j}\omega}}
       \frac{d\omega}{4\pi} \, .
       \end{split}
\end{equation}
Next we estimate the remaining integral in the right--hand side of \eqref{ch5:5:one_int_remains}, in particular we aim to solve
\begin{equation}\label{ch5:5:third_moment}
    \abs{x_j}^s\int_{S^2} 
         \frac{(\hat{x}_j\cdot \omega)^2-1}
              {\abs{x_j - x_k + r\abs{x_j}\omega}}
       \frac{d\omega}{4\pi} =  \abs{x_j}^{s-1}\int_{S^2} 
         \frac{(\hat{x}_j\cdot \omega)^2-1}
              {\abs{a_{jk}-r\omega}} 
       \frac{d\omega}{4\pi} 
\end{equation}
We use 
\begin{equation}\label{ch5:5:2nd_order_legendre}
    \frac{2}{3} P_2(t) - \frac{2}{3} = t^2-1, \, \quad  \forall t\in [-1,1],
\end{equation}
where $P_2$ is the second--order Legendre polynomial. Inserting \eqref{ch5:5:2nd_order_legendre} into \eqref{ch5:5:third_moment} we find
\begin{equation}\label{ch5:5:third_moment_with_legendre}
    \begin{split}
    \abs{x_j}^s\int_{S^2} 
         \frac{(\hat{x}_j\cdot \omega)^2-1}
              {\abs{x_j - x_k + r\abs{x_j}\omega}}
       \frac{d\omega}{4\pi} &= \frac{2}{3}  \abs{x_j}^{s-1}\int_{S^2} 
         \frac{P_2(\hat{x}_j\cdot \omega)-1}
              {\abs{a_{jk}-r\omega}} 
       \frac{d\omega}{4\pi} \\
       &= \frac{2}{3}  \abs{x_j}^{s-1} \left( \int_{S^2} 
         \frac{P_2(\hat{x}_j\cdot \omega)}
              {\abs{a_{jk}-r\omega}} 
       \frac{d\omega}{4\pi} -  \frac{1}{\max\{\abs{a_{jk}}, r \}} \right) \, .\\
    \end{split}
\end{equation}
where we have used Newton's theorem. To solve the integral involving the Legendre polynomial $P_2$ we use the multipol expansion in \eqref{ch5:5:multipol_expansion_001} and \eqref{ch5:5:twisted_orthogonality} to find
\begin{equation} \label{ch5:5:2nd_order_multipol}
    \begin{split}
        \int_{S^2} 
         \frac{P_2(\hat{x}_j\cdot \omega)}
              {\abs{a_{jk}-r\omega}} 
       \frac{d\omega}{4\pi}&=
        \sum_{n=0}^\infty \frac{\min\{\abs{a_{jk}},r\}^n}{\max\{\abs{a_{jk}},r\}^{n+1}} \int_{S^2} P_n(\omega \cdot \hat{a}_{jk}) P_2(\hat{x}_j \cdot \omega)
       \frac{d\omega}{4\pi} \\
       &= \frac{\min\{\abs{a_{jk}},r\}^2}{\max\{\abs{a_{jk}},r\}^{3}} \frac{P_2(\hat{a}_{jk}\cdot \omega)}{5} \leq \frac{1}{5\max\{\abs{a_{jk}},r\}}
    \end{split}
\end{equation}
Using $\abs{a_{jk}} = \abs{x_j-x_k}/\abs{x_j}$ and inserting \eqref{ch5:5:2nd_order_multipol} into \eqref{ch5:5:third_moment_with_legendre} shows
\begin{equation}\label{ch5:5:third_integral_estimated}
    \begin{split}
    \abs{x_j}^s\int_{S^2} 
         \frac{(\hat{x}_j\cdot \omega)^2-1}
              {\abs{x_j - x_k + r\abs{x_j}\omega}}
       \frac{d\omega}{4\pi} &\le \frac{-8}{15} \frac{\abs{x_j}^s}{\max\{\abs{x_j-x_k},r\abs{x_j}\} } \, .
    \end{split}
\end{equation}
Combining \eqref{ch5:5:one_int_remains} and \eqref{ch5:5:third_integral_estimated} we arrive at
\begin{equation*}
 \begin{split}
  \sum_{j\neq k}
  &\iint_{\R^3 \times \R^3} 
    \frac{\abs{x}^s + \abs{y}^s }{2\abs{x-y}}
  d\mu_j(x) d\mu_k(y)      \\
       &\leq \phantom{+}(1+r^2)^{s/2}\frac{(1+q)^{s/2}+(1-q )^{s/2}}{2}  \sum_{j\neq k} 
         \frac{\abs{x_j}^s}
              {\max\{\abs{x_j-x_k},r\abs{x_j}\}} \\
       &\phantom{\le~}-\frac{8}{15} (1+r^2)^{s/2}\frac{s}{4}\left(\frac{s}{2}-1\right)q^2(1+q)^{\frac{s-4}{2}} \sum_{j\neq k} 
        \frac{\abs{x_j}^s}{\max\{\abs{x_j-x_k},r\abs{x_j}\} }\\
        &=\underbrace{(1+r^2)^{s/2}\left(\frac{(1+q)^{s/2}+(1-q )^{s/2}}{2} -\frac{s\left(s-2\right)}{15} q^2(1+q)^{\frac{s-4}{2}} \right)}\limits_{\eqqcolon g(r)} \sum_{j\neq k} 
         \frac{\abs{x_j}^s}
              {\max\{\abs{x_j-x_k},r\abs{x_j}\}}
       \end{split}
\end{equation*}
Using $q=2r/(1+r^2)$ one checks by direct computations that $g(r)\geq 0$. Consequently we find
\begin{equation} \label{ch5:5:off_diaognal_refined}
    \sum_{j\neq k}
  \iint_{\R^3 \times \R^3} 
    \frac{\abs{x}^s + \abs{y}^s }{2\abs{x-y}}
  d\mu_j(x) d\mu_k(y) \leq g(r)\sum_{j\neq k} 
         \frac{\abs{x_j}^s+\abs{x_k}^s}
              {2\abs{x_j-x_k}}
\end{equation}
Combining the estimates of the diagonal terms \eqref{ch5:5:eq:diag_part_refinement} and the off--diagonal terms \eqref{ch5:5:off_diaognal_refined} together with \eqref{ch5:5:hard_to_est} yields
\begin{equation} \label{ch5:5:numerator_for_beta}
    N^2 I_s(\mu) \leq  
       \frac{(1+r)^{s+2}-\abs{1-r}^{s+2}}
            {2r^2(s+2)} \sum_{j=1}^N \abs{x_j}^{s-1} + g(r)\sum_{j\neq k} 
         \frac{\abs{x_j}^s+\abs{x_k}^s}
              {2\abs{x_j-x_k}}
\end{equation}
Applying Lemma \ref{app:A:int_001} we also get 
\begin{align}\label{ch5:5:denominator_for_beta}
    N\int \abs{x}^{s-1} d\mu(x) 
    = \sum_{j=1}^N \abs{x}^{s-1} d\mu_j(x) 
    = \frac{(1+r)^{s+1}-\abs{1-r}^{s+1}}{2r(s+1)} 
      \sum_{j=1}^N \abs{x_j}^{s-1} \, .
\end{align}
Combining \eqref{ch5:5:numerator_for_beta} and \eqref{ch5:5:denominator_for_beta} shows
\begin{equation} \label{ch5:5:beta_alpha_refined_almost}
    \begin{split}
        N\beta_s &\leq \frac{ N^2 I_s(\mu)}{ N\int \abs{x}^{s-1} d\mu(x)} \\
        &\leq \frac{1}{r} \left(\frac{s+1}{s+2}\right) \frac{(1+r)^{s+2}-\abs{1-r}^{s+2}}{(1+r)^{s+1}-\abs{1-r}^{s+1}} + \frac{2r(s+1)g(r)}{(1+r)^{s+1}-\abs{1-r}^{s+1}} \frac{\sum\limits_{1\le j<k\le N}\frac{\abs{x_j}^s+\abs{x_k}^s}{2\abs{x_j-x_k}}}{\sum_{j=1}^N \abs{x_j}^{s-1}} \, .
    \end{split}
\end{equation}
Taking the infimum over the positions $x_1,x_2,\dots x_N\in A_0$ together with the definition of $\alpha_{N,s}$ in \eqref{ch5:5:without_zero} we conclude from \eqref{ch5:5:beta_alpha_refined_almost} 
\begin{equation}\label{ch5:5:beta_alpha_refined}
      N\beta_s \leq \frac{1}{r} \left(\frac{s+1}{s+2}\right) \frac{(1+r)^{s+2}-\abs{1-r}^{s+2}}{(1+r)^{s+1}-\abs{1-r}^{s+1}} + \frac{2r(s+1)g(r)}{(1+r)^{s+1}-\abs{1-r}^{s+1}} \alpha_{N,s}(N-1)
\end{equation}
and equivalently
\begin{equation*}
       \left(\frac{s+2}{s+1}\right) \frac{(1+r)^{s+1}-\abs{1-r}^{s+1}}{(1+r)^{s+2}-\abs{1-r}^{s+2}}N\beta_s - \frac{1}{r} \leq \frac{2r(s+2)g(r)}{(1+r)^{s+2}-\abs{1-r}^{s+2}} \alpha_{N,s}(N-1)
\end{equation*}
This finishes the proof of Lemma \ref{ch5:5:lem:bound on alpha for large s}.
\end{proof}
For small $r$ one can find a better bound than the one in Lemma \ref{ch5:5:lem:bound on alpha for large s} by estimating more carefully and not using convexity.  But the argument is much more involved. 
However, since we are interested in bounds for small $r>0$, in order to make the prefactor in \eqref{ch5:5:eq:large s} close to one, 
we will give this improved bound now. 
\begin{lemma}[Refined comparison II of 
$\alpha_{N,s}$ with $\beta_s$, $4\ge s\ge 2$] \label{ch5:5:refined_lower_bound}
    Let $\alpha_{N,s}$ and $\beta_s$ be defined as in \eqref{ch5:5:geometric_exp} and \eqref{ch5:4:uniform_expression} then for every $N\geq2$, 
    $r >0$, and $4\ge s\ge 2$
\begin{equation} \label{ch5:5:refined_inequality}
    \begin{split}
      \frac{(1+r)^{s+1}-(1-r)^{s+1}}{2r(s+1)}  N\beta_s 
        &\leq   
           \frac{(1+r)^{s+2}-(1-r)^{s+2}}{2r(s+2)} 
           \left( \alpha_{N,s}(N-1)  + \frac{1}{r} \right) \\
        &\phantom{+~}+r^2 f(r,s)\alpha_{N,s} (N-1)
    \end{split}
\end{equation}
with
\begin{equation} \label{ch5:5:remainder_f}
    f(r,s) \coloneqq \frac{s}{2}\left(\frac{s}{2}-1\right)\left( \frac{4}{15} + \left(2-\frac{s}{2}\right) \frac{8}{105}r + \left(2-\frac{s}{2}\right) \frac{448}{625} r^2 \right)\, .
\end{equation}
In particular, for $s=3$ this gives  
\begin{equation} \label{ch5:5:refined_inequality_s=3}
    \begin{split}
      (1+r^2)N\beta_3 - \frac{1+2r^2+r^4/5}{r} \leq  \alpha_{N,3}(N-1) \left(1+ \frac{r^2}{5} + \frac{r^3}{35} + \frac{168}{625} r^4 \right)\, .
    \end{split}
\end{equation}
\end{lemma}
\begin{remark}
Note that $f(r,2)$ vanishes for all $r>0$ and for $s\in (2,3]$ it adds a positive correction to the leading order term in 
the prefactor of $\alpha_{N,s}$.  For later usage, we note that one has the rough estimate $f(r,s)\leq f(r,3)< \frac{1}{2}$ for $r\in [0,1]$ any $s\in (2,3]$.
\end{remark} 

\begin{proof}
  We again use the bounds \eqref{ch5:5:hard_to_est} together with \eqref{ch5:5:eq:diagonal terms} and  \eqref{ch5:5:eq:off-diagonal with Newton}. 
  In order to improve on Lemma \ref{ch5:5:lem:bound on alpha for large s}, we have to bound the integral in the last sum of 
  \eqref{ch5:5:eq:off-diagonal with Newton} more carefully. 
  Recalling  $a_{jk} = (x_k-x_j)/\abs{x_j}$  and $\hatt{x}_j= x_j/\abs{x_j}$ we can rewrite  
  \begin{equation}\label{ch5:5:go_on_from_here}
     \int_{S^2} 
         \frac{\abs{x_j + r\abs{x_j}\omega}^s }
              {\abs{x_j - x_k + r\abs{x_j}\omega}}
       \frac{d\omega}{4\pi} =\abs{x_j}^{s-1} \int_{S^2} 
         \frac{(1+r^2+2r\hat{x}_j \cdot \omega)^{s/2} }
              {\abs{a_{jk}-r\omega}}
       \frac{d\omega}{4\pi}
  \end{equation}
Such integrals can be solved by multipole expansion, that is expanding the Coulomb--kernel in terms of the Legendre Polynomials $P_l(t)$, $l\in \N_0$. 
Using the generating function 
\begin{equation*}
    (1+\delta^2- 2\delta t)^{-1/2} = \sum_{n=0}^\infty \delta^n P_n(t)
\end{equation*}
which is valid for $\abs{t}\le 1$ and $\abs{\delta}<1$, and 
expanding $\abs{a_{jk}-r\omega}^{-1}$ 
in \eqref{ch5:5:go_on_from_here} 
with the help of \eqref{ch5:5:eq:generating function} 
and using Lemma \ref{app:A:lem:atkinson han formula}  
yields
\begin{equation} \label{ch5:5:multipol_expansion}
  \begin{split}
   \abs{x_j}^{s-1} \int_{S^2} 
         &\frac{(1+r^2+2r\hat{x}_j \cdot \omega)^{s/2} }
              {\abs{a_{jk}-r\omega}}
       \frac{d\omega}{4\pi}\\
   & = 
      \abs{x_j}^{s-1} \sum_{l=0}^\infty  
        \frac{\min(\abs{a_{jk}},r)^l}
             {\max(\abs{a_{jk}},r)^{l+1}} 
        \int_{S^2} 
          (1+r^2+2r\hat{x}_j \cdot \omega)^{s/2} 
          P_l(\langle \hat{a}_{jk}, \omega \rangle) 
        \frac{d\omega}{4\pi}    \\
  & = 
      \abs{x_j}^{s-1} \sum_{l=0}^\infty  
       \frac{\min(\abs{a_{jk}},r)^l}
             {\max(\abs{a_{jk}},r)^{l+1}} 
             \lambda_{l,s}(r) P_l\big(  \hat{a}_{jk} \cdot \hat{x}_j\big)     
  \end{split}
\end{equation}
with 
\begin{equation*}
    \lambda_{l,s}(r) 
      = \frac{1}{2}\int_{-1}^1 
          (1+r^2 +2rt)^{s/2} P_l(t) 
        dt\, .
\end{equation*}
  We willl see shortly that the sum $\sum_{l=0}^\infty \abs{\lambda_{l,s}(r)}$ converges --  see \eqref{ch5:5:eq:convergence} -- so 
  the series in the last line of  \eqref{ch5:5:multipol_expansion} 
  converges for all $r>0$ since $-1\le P_l(t)\le 1$ for  
  $-1\le t\le 1$.
Using $P_0(t)\equiv 1$ the first multipole moment $l=0$ is easy to compute,  
\begin{equation} \label{ch5:5:multipol_moment 0} 
    \begin{split}
        \lambda_{0,s}(r)
         &=
           \frac{(1+r)^{s+2} - (1-r)^{s+2}}{2(s+2)r} \,. 
    \end{split}
\end{equation}
Note that with $P_1(t)=t$ and consequently the second multipole moment is positive non--negative since
\begin{equation}\label{ch5:5:multipol_moment 1} 
   \begin{split} 
        \lambda_{1,s}(r)
         &= 
           \frac{1}{2} \int_{-1}^1 (1+r^2+2rt)^{s/2} t dt 
           >0 
           \,. 
    \end{split}
\end{equation}
The calculation for higher moments is a bit involved. 
Before we embark on this, let us note that if 
$\sum_{l\ge 0}\lambda_{l,s}(r)$ converges absolutely, we can 
further bound \eqref{ch5:5:multipol_expansion} as follows. 
\begin{equation*}
  \begin{split}
    \abs{x_j}^{s-1} \sum_{l=0}^\infty  
        &\frac{\min(\abs{a_{jk}},r)^l}
             {\max(\abs{a_{jk}},r)^{l+1}}  
             \lambda_{l,s}(r)P_l\big( \hat{a}_{jk} \cdot \hat{x}_j \big)    \\
        &\le  
          \lambda_{0,s}(r)
          \frac{\abs{x_j}^{s-1}}{\max(\abs{a_{jk}},r)} 
          + 
          \lambda_{1,s}(r) 
          \frac{\abs{x_j}^{s-1}\min(\abs{a_{jk}},r) }
             {\max(\abs{a_{jk}},r)^{2}}
          \, (\hat{a}_{jk} \cdot \hat{x}_j) \\
          &\phantom{\le  
          	\lambda_{0,s}(r)
          	\frac{\abs{x_j}^{s-1}}{\max(\abs{a_{jk}},r)} }
          + \frac{\abs{x_j}^{s-1}}{\max(\abs{a_{jk}},r)} 
            \sum_{l=2}^\infty \abs{\lambda_{l,s}(r)}  
  \end{split}
\end{equation*}
Using   $a_{jk} = (x_k-x_j)/\abs{x_j}$  we arrive at
\begin{equation*}
  \begin{split}
    \abs{x_j}^{s-1} \sum_{l=0}^\infty  
        &\frac{\min(\abs{a_{jk}},r)^l}
             {\max(\abs{a_{jk}},r)^{l+1}}  
             \lambda_{l,s}(r)P_l\big( \hat{a}_{jk} \cdot \hat{x}_j \big)    \\
        &\le  
          \left( \lambda_{0,s}(r)+\sum_{l=2}^\infty \abs{\lambda_{l,s}(r)}\right)
          \frac{\abs{x_j}^{s}}{\abs{x_j-x_k}} 
          \\
          &\phantom{\le} \, \, - 
          \lambda_{1,s}(r) 
          \frac{\abs{x_j}^{s}\min(\abs{x_j-x_k},r\abs{x_j}) }
             {\abs{x_j-x_k}\max(\abs{x_j-x_k},r\abs{x_j})^{2}}
          \,  \hat{x}_j \cdot(x_j-x_k)
  \end{split}
\end{equation*}
Let $C_{s}(r)= \sum_{l=2}^\infty \abs{\lambda_{l,s}(r)} $ 
and 
\begin{align*}
    \wti{\gamma}(u,v) 
      = 
        \frac{v^s\min(u, v)}
               {u\max(u, v)^2}  ,
\end{align*}
then
\begin{equation*}
  \begin{split}
    \abs{x_j}^{s-1} \sum_{l=0}^\infty  
        &\frac{\min(\abs{a_{jk}},r)^l}
             {\max(\abs{a_{jk}},r)^{l+1}}  
             \lambda_{l,s}(r)P_l\big( \hat{a}_{jk} \cdot \hat{x}_j \big)    \\
        &\le  
          \left( \lambda_{0,s}(r)+C_{s}(r)\right)
          \frac{\abs{x_j}^{s}}{\abs{x_j-x_k}} 
          - 
          \lambda_{1,s}(r)r^{-s} \wti{\gamma}(\abs{x_j-x_k},r\abs{x_j})
          \,  \hat{x}_j \cdot(x_j-x_k)
  \end{split}
\end{equation*}
Note that  $\wti{\gamma}(u,v)$ is increasing in $v$ for fixed $u>0$. Applying Lemma \ref{ch5:5:Lem:positivity} and noting Remark \ref{ch5:5:rem:monotonicity}  we find
\begin{align}\label{ch5:5:apply_gamma_argument}
\sum_{j\neq k} 
          \wti{\gamma}(\abs{x_j-x_k}, r\abs{x_j}) \, 
          \hatt{x}_j\cdot \big(x_j - x_k\big) \ge 0\, .    
\end{align}
As in \eqref{ch5:5:eq:off-diagonal with Newton} one sees  
\begin{align*}
  \sum_{j\neq k} &\iint_{\R^3\times\R^3} 
    \frac{\abs{x}^s+ \abs{x}^s}{2\abs{x-y}}
  d\mu_j(x)d\mu_k(y) \\
    &\le 
     \left( \lambda_{0,s}(r)+C_{s}(r)\right)
         \sum_{j\neq k} \frac{\abs{x_j}^{s}}{\abs{x_j-x_k}} 
          - 
          \lambda_{1,s}(r)r^{-s}\sum_{j\neq k}\wti{\gamma}(\abs{x_j-x_k},r\abs{x_j})
          \,  \hat{x}_j \cdot(x_j-x_k) \\
    &\le 
      \big(\lambda_{0,s}(r) +C_{s}(r)\big)\sum_{j\neq k}
          \frac{\abs{x_j}^s + \abs{x_j}^s}{2\abs{x_j-x_k} }
\end{align*}
where we used \eqref{ch5:5:apply_gamma_argument} to drop the second sum. In the last line we symmetrized the remaining expression. Thus we get a similar bound as \eqref{ch5:5:off_diaognal_refined} with $g(r)$ replaced by $\lambda_{0,s}(r) +C_{s}(r)$. 

\smallskip 

Thus, as in the proof of Lemma 
\ref{ch5:5:lem:bound on alpha for large s}, a bound similar 
to \eqref{ch5:5:numerator_for_beta}, but with 
$g(r)$ in \eqref{ch5:5:numerator_for_beta} replaced by 
$\lambda_{0,s}(r) +C_{s}(r)$, follows from this. In particular it follows
\begin{equation}\label{ch5:5:impoved_g}
       \left(\frac{s+2}{s+1}\right) \frac{(1+r)^{s+1}-\abs{1-r}^{s+1}}{(1+r)^{s+2}-\abs{1-r}^{s+2}}N\beta_s - \frac{1}{r} \leq \left(1 + \frac{2r(s+2)C_{s}(r)}{(1+r)^{s+2}-\abs{1-r}^{s+2}} \right) \alpha_{N,s}(N-1)
\end{equation}
Hence the claimed bound \eqref{ch5:5:refined_inequality} follows as 
soon as we can show that 
\begin{equation}\label{ch5:5:eq;punchline 2}
    C_s(r)\le r^2  f(r,s)
\end{equation}
with $f$ given in \eqref{ch5:5:remainder_f}. We will do this in the rest of this proof. 

\smallskip 
To get a grip on the higher order moments $\lambda_{l,s}(r)$ for $l\ge 2$  we expand $\abs{\hat{x}_j+r \omega}^s $ in a 
binomial series. 
With $t= \hatt{x}_j\cdot\omega$ 
and 
\begin{equation}\label{ch5:5:def:q}
    q=2r/(1+r^2)\le 1
\end{equation}
 we have   
\begin{equation}\label{ch5:5:eq:binomial series}
  \begin{split}
    \big(1+r^2 + 2r t\big)^{s/2} 
    = \big(1+r^2\big)^{s/2} \big(1+qt\big)^{s/2} 
  & = 
      \big(1+r^2\big)^{s/2} 
      \sum_{n=0}^\infty \binom{s/2}{n} q^n t^n     
  \end{split}
\end{equation}
According to \cite[Satz 22.8]{book:F:2023} the binomial series converges absolutely and uniformly for $-1\le t\le 1$  
and all $0\le q\le 1$. 
Hence we can interchange the summation and 
integration in \eqref{ch5:5:multipol_expansion} to see that 
\begin{equation} \label{ch5:5:multipol_moments step 0} 
    \begin{split}
      \lambda_{l,s}(r) 
          = (1+r^2)^{s/2}\sum_{n=0}^\infty 
              \binom{s/2}{n} q^n \frac{1}{2}\int_{-1}^1 t^n P_l(t) dt 
              \quad l\geq2 \, .
    \end{split}
\end{equation}
Write 
\begin{equation} \label{ch5:5:ext_monomial_002}
    t^n = \sum_{m=0}^n c_{n,m} P_m(t) \quad \text{for } t\in [-1,1] \, .
\end{equation}
Using that the Legendre polynomials are orthogonal in $L^2([-1,1])$ and normalized by $$\frac{1}{2}\int_{-1}^1 P_l(t)^2 dt = \frac{1}{2l+1}\, ,$$ and $P_l $ has degree $l$,  
one sees that $\int_{-1}^1 t^n P_l(t) dt =0$ if $n<l$ and for 
$n\ge l$ we have 
$\frac{1}{2}\int_{-1}^1 t^n P_l(t) dt = \frac{c_{n,l}}{1l+1}$. 
Thus 
\begin{equation} \label{ch5:5:multipol_moments} 
    \begin{split}
       \lambda_{l,s}(r)  
        = (1+r^2)^{s/2}\sum_{n=l}^\infty \binom{s/2}{n}  \frac{q^n c_{n,l}}{2l+1} , \quad l\geq2 \, .
    \end{split}
\end{equation}
We will use that the coefficients $c_{n,l}$ for 
$n,l\in \N_0$ are non--negative, see \eqref{app:A:coeff} in Appendix \ref{AppendixA}. 
Moreover, 
\begin{equation*}
    \sum_{m=0}^n c_{n,m} =1\, ,
\end{equation*}
which follows from setting $t=1$ in \eqref{ch5:5:ext_monomial_002} and using $P_l(1)=1$ for $l\in\N_0$. Together with $c_{n,m}\ge 0$, this also shows $c_{n,m}\le 1$. From \cite[Hilfssatz 22.8a]{book:F:2023} we have the bound 
\begin{align}\label{ch5:5:eq:hilfssatz}
    \left| \binom{s/2}{n}\right| \le \frac{c}{n^{1+s/2}}\, .
\end{align}
This implies that the series in the right--hand--side of 
\eqref{ch5:5:multipol_moments} converges absolutely for all 
$0\le q\le 1$, since $0\le c_{n,l}\le 1$ and hence 
\begin{equation} \label{ch5:5:est_on_sums}
  \sum_{n=l}^\infty 
        \abs{\binom{s/2}{n}}
        \frac{q^n c_{n,l}}{2l+1} 
    \lesssim  
      \sum_{n=l}^\infty 
        \frac{1}{2l+1} 
        \frac{1}{n^{1+s/2}} 
        <\infty
\end{equation}
for any $s>0$. Moreover, we also have 
\begin{equation} \label{ch5:5:eq:convergence} 
    \begin{split}
       \sum_{l\ge 2} \abs{\lambda_{l,s}(r)}   
        &\le  
          (1+r^2)^{s/2}\sum_{l=2}^\infty \sum_{n=l}^\infty \abs{\binom{s/2}{n}}  \frac{q^n c_{n,l}}{2l+1} 
        \lesssim 
          \sum_{l=2}^\infty \sum_{n=l}^\infty 
            \frac{1}{n^{1+s/2}} \frac{1}{2l+1}  \\
          &= 
            \sum_{n=2}^\infty \sum_{l=2}^n 
            \frac{1}{n^{1+s/2}} \frac{1}{2l+1}   
        \lesssim 
          \sum_{n=2}^\infty 
            \frac{\ln(2+n)}{n^{1+s/2}}           
            <\infty\, .
    \end{split}
\end{equation}
For $n\in\N$ define
\begin{equation*}
    A_n 
      \coloneqq   
        \abs{\binom{s/2}{n}} 
        \sum_{l=2}^n  \frac{ c_{n,l}}{2l+1}\, .
\end{equation*}
We have
\begin{equation} \label{ch5:5:sum_in_terms_of_A_k}
    \sum_{l=2}^\infty\abs{\lambda_{l,s}(r)} 
      \le 
        (1+r^2)^{s/2}\sum_{l=2}^\infty \sum_{n=l}^\infty 
        \abs{\binom{s/2}{n}} \frac{ c_{n,l}}{2l+1} 
        \,q^{n}  
      = 
        (1+r^2)^{s/2}\sum_{n=2}^\infty A_n q^n  \, .
\end{equation}
Since $q=  2r/(1+r^2)$ (see \eqref{ch5:5:def:q}) we get  
\begin{equation} \label{ch5:5:truncated}
    \sum_{n=2}^\infty A_n q^n \leq \frac{4r^2}{\left(1+r^2\right)^2} A_2    + \frac{8r^3}{\left(1+r^2\right)^3} A_3 + \frac{16r^4}{\left(1+r^2\right)^4} \sum_{n=4}^\infty A_n \, .
\end{equation}
In  Lemma \ref{app:A:remainder_of_sum} in Appendix \ref{AppendixA} we show that 
\begin{equation*}
    \begin{split}
        A_2=\abs{\binom{s/2}{2}} \frac{2}{15}, \quad A_3=\abs{\binom{s/2}{3}} \frac{2}{35}, \quad \sum_{k=4}^\infty A_k  \leq \frac{s}{2}\left(\frac{s}{2}-1\right)\left(2-\frac{s}{2}\right) \frac{28}{625} 
        \, .
    \end{split}
\end{equation*}
 It follows that
\begin{equation} \label{ch5:5:series_expansion}
    \sum_{k=2}^\infty A_k q^k \leq \frac{r^2}{\left(1+r^2\right)^2} f(r,s) .
\end{equation}
with $f$ defined in \eqref{ch5:5:remainder_f}. Combining \eqref{ch5:5:series_expansion} with \eqref{ch5:5:sum_in_terms_of_A_k} proves for $s\leq 4$
\begin{equation}
   C_{s}(r) = \sum_{l=2}^\infty\abs{\lambda_{l,s}(r)} 
      \le  (1+r^2)^{(s-4)/2}r^2 f(r,s) \leq r^2 f(r,s) \,.
\end{equation}

This proves \eqref{ch5:5:eq;punchline 2}, which finishes the proof of 
Lemma \ref{ch5:5:refined_lower_bound}. 
\end{proof} 
\begin{remark}
We truncate the series in \eqref{ch5:5:truncated} at the fourth power because we will later choose $r \lesssim Z^{-1/3}$. As a result, even terms like $Zr^4$ become negligible as $Z$ increases. For further refinements at small $Z$ respectively large $r$, it is more appropriate to evaluate the inequality using computational methods.    
\end{remark} 
\section{Upper bounds on the weighted kinetic energy} \label{ch5:6:sec_upper}
In this section, we derive an upper bound on
\begin{equation} \label{ch5:6:need_upper_bound}
    Z-\frac{1}{2}\frac{\left\langle \abs{x_1}^s \psi_{N,Z},  P_1^2\psi_{N,Z} \right\rangle}{\left\langle \abs{x_1}^{s-1} \psi_{N,Z}, \psi_{N,Z} \right\rangle} \, 
\end{equation}
which is the right--hand side of \eqref{ch5:3:main_inequality}. Note that 
the denominator has to be interpreted in the quadratic form sense 
\begin{align*}
  \left\langle \abs{x_1}^s \psi_{N,Z},  P_1^2\psi_{N,Z} \right\rangle 
    &=
      \left\langle \nabla_1(\abs{x_1}^s \psi_{N,Z}),  \nabla_1\psi_{N,Z} \right\rangle_{L^2(\R^3N)} \\
    &=  
    \int_{\R^{3(N-1)}} \langle \abs{x_1}^s \psi_{N,Z}, P_1^2 \psi_{N,Z} \rangle_{L^2(dx_1)} dx_2 \dots dx_N
\end{align*}
In the case $s=1$ Lieb used in \cite{L:1984:b} the fact that
\begin{equation} \label{ch5:6:superharmonic_prop}
    -2\re \langle \abs{x}\varphi, -\Delta \varphi \rangle_{L^2(\R^3)} 
    = 
    \langle\varphi,  (\abs{x} \Delta + \Delta \abs{x}) \varphi \rangle_{L^2(\R^3)} \leq 0\, .
\end{equation}
Thus the second term in \eqref{ch5:6:need_upper_bound}
can be dropped when $s=1$. Together with $\alpha_{N,1} \geq 1/2$ 
this recovers Lieb's bound $N_c<2Z+1$. 
In \cite{CS:2013} Chen and Siedentop showed that in dimension $d=3$ for any $b\in[0,1]$
\begin{equation}\label{ch5:6:chen_siedentop}
    \langle \varphi,(\abs{x}^b \Delta + \Delta \abs{x}^b)\varphi \rangle \leq  0
\end{equation}
for any $\varphi \in L^2(\R^3)$. For $b>1$ \eqref{ch5:6:chen_siedentop} does not hold in general.
 Before we proceed let us clarify in what sense we understand the inner product in \eqref{ch5:6:need_upper_bound}. Recall that $\psi_{N,Z}\in \mathcal{H}^f_N$ is the normalized many particle ground-state of $H_{N,Z}$ in \eqref{ch5:2:hamilton} and does depend on the positions of particles $(x_1,x_2,\dots, x_n)$ and the spin degrees of freedom $(\sigma_1,\sigma_2, \dots, \sigma_N)$ with $\sigma_i \in \{1/2,-1/2\}$ . Following \cite[Chapter 3]{book:LS:2009} we define the one-particle density by
\begin{equation*} 
    \rho_{\psi_{N,Z}}(x) \coloneqq  \sum_{i=1}^N \rho^i_{\psi_{N,Z}}(x)
\end{equation*}
where
\begin{equation*}
    \rho^i_{\psi_{N,Z}}(x) \coloneqq \int_{\R^{3(N-1)}} \abs{\psi(x_1,\dots x_{i-1},x,x_{i+1}, \dots x_N)}^2 dx_1 \dots \hat{dx_i}  \dots dx_N  \, .
\end{equation*}
where $\hat{dx_i} $ means that the integration of $x_i$ is omitted. Remember that we ignore any degrees of freedom related to spin. Due to \eqref{ch5:2:pauli_princ} we have $\rho^i = \rho^1$ for any $i \in \{1,2,\dots, N\}$ and thus
\begin{equation} \label{ch5:6:one-particle-density}
    \rho_{\psi_{N,Z}}(x_1) \coloneqq N \rho^1_{\psi_{N,Z}}(x_1)
\end{equation}
with
\begin{equation*}
    \int_{\R^3}\rho_{\psi_{N,Z}}(x_1) \, dx_1 = N \, .
\end{equation*}
Consequently for any $p\in \R$
\begin{equation}\label{ch5:6:eq:with_one_part_dens}
    \left\langle \abs{x_1}^{p} \psi_{N,Z}, \psi_{N,Z} \right\rangle = \frac{1}{N} \int_{R^3} \abs{x_1}^{p} \rho_{\psi_{N,Z}}(x_1) \, dx_1 
\end{equation}
As a substitute for \eqref{ch5:6:superharmonic_prop} we prove
\begin{lemma} \label{ch5:6:IMS+Hardy}
    For any $s\geq 2$ 
    \begin{equation} \label{ch5:6:upper_bound_001}
        \frac{-1}{2}\frac{\left\langle \abs{x_1}^s \psi_{N,Z},  P_1^2 \psi_{N,Z} \right\rangle}{\left\langle \abs{x_1}^{s-1} \psi_{N,Z}, \psi_{N,Z} \right\rangle} \leq \frac{s^2-1}{8 }\left\langle \abs{x_1}^{s-1} \psi_{N,Z}, \psi_{N,Z} \right\rangle^{\frac{-1}{s-1}} \, .
    \end{equation}
\end{lemma}
\begin{proof}
  For readability we drop the indices $N,Z$ and write $\psi = \psi_{N,Z}$. 
  Applying the IMS-localization formula, see for example 
  \cite[Theorem 3.2]{CFKS:1987}, yields
    \begin{equation} \label{ch5:6:used_hardy_and_ims}
        \begin{split}
            \Re \langle \abs{x_1}^s \psi,P_1^2 \psi \rangle_{L^2(dx_1)} &= \left\langle \abs{x_1}^{s/2} \psi,  \left[ P_1^2- \abs{\frac{\nabla_1 \abs{x_1}^s}{2\abs{x_1}}}^2 \right] \abs{x_1}^{s/2} \psi \right\rangle_{L^2(dx_1)} \\
            &=\left\langle \abs{x_1}^{s/2} \psi,  \left[P_1^2 - \frac{s^2}{4}\abs{x_1}^{-2}\right] \abs{x_1}^{s/2} \psi \right\rangle_{L^2(dx_1)} \\
            &\geq \frac{1-s^2}{4}\left\langle \abs{x_1}^{s-2} \psi,   \psi \right\rangle_{L^2(dx_1)}
        \end{split}
    \end{equation}
    Due to \eqref{ch5:6:eq:with_one_part_dens} we have
    \begin{equation*}
         \left\langle \abs{x_1}^{s-2} \psi, \psi \right\rangle = \frac{1}{N} \int_{R^3} \abs{x_1}^{s-2} \rho_{\psi}(x_1) \, dx_1 \, .
    \end{equation*}
    Applying H\"older's inequality for 
    \begin{equation*}
        p=\frac{s-1}{s-2}, \, q =s-1
    \end{equation*}
    yields
    \begin{equation*}
        \left\langle \abs{x_1}^{s-2} \psi,   \psi \right\rangle_{L^2(dx_1)} \leq \left( \frac{1}{N}\int_{\R^3} \abs{x_1}^{s-1} \rho_{\psi}(x_1) \, dx_1  \right)^{\frac{s-2}{s-1}} = \left(\left\langle \abs{x_1}^{s-1} \psi,   \psi \right\rangle_{L^2(dx_1)}\right)^{\frac{s-2}{s-1}} \, .
    \end{equation*}
    Note that we used $s>2$ in this step. If one wants to cover the cases $s\in (1,2)$ one needs to estimate the expression above differently. Consequently
    \begin{equation} \label{ch5:6:used_hoelder}
        \frac{\left\langle \abs{x_1}^{s-2} \psi,   \psi \right\rangle_{L^2(dx_1)}}{\left\langle \abs{x_1}^{s-1} \psi,   \psi \right\rangle_{L^2(dx_1)}}  \leq \left(\left\langle \abs{x_1}^{s-1} \psi,   \psi \right\rangle_{L^2(dx_1)} \right)^{\frac{-1}{s-1}} \, .
    \end{equation}
    Combining \eqref{ch5:6:used_hardy_and_ims} and \eqref{ch5:6:used_hoelder} proves Lemma \eqref{ch5:6:IMS+Hardy}. The case $s=2$ follows in the limit $s\to 2$.
\end{proof}
\begin{remark} \label{ch5:6:jensen_remark} There exists a straightforward simplification of the inequality in Lemma \ref{ch5:6:IMS+Hardy} since by Jensen's Inequality
\begin{equation} \label{ch5:6:jensen_001}
    \left\langle \abs{x_1}^{s-1} \psi_{N,Z}, \psi_{N,Z} \right\rangle^{\frac{-1}{s-1}} \leq \left\langle \abs{x_1}^{-1} \psi_{N,Z}, \psi_{N,Z} \right\rangle  \
\end{equation}
The right--hand side of \eqref{ch5:6:upper_bound_001} is growing quadratic in $s$ which is unfortunate since the bound on $b(s)$ in Theorem \ref{ch5:2:main_thm} is decreasing. Note that the right--hand side of \eqref{ch5:6:jensen_001} can be interpreted as the inverse expectation of the radius of the atom which in Thomas--Fermi Theory grows as $Z^{-1/3}$ (see \cite[p. 560]{L:1976}) but should be bounded in $Z$ for real atoms. In \cite{N:2012} Nam did control the right--hand side of \eqref{ch5:6:jensen_001}. We proceed similarly to his proof. 
\end{remark}
We continue by estimating the right--hand side of \eqref{ch5:6:upper_bound_001}. We want to apply the following inequality introduced by Lieb in \cite[p. 563]{L:1976}
\begin{align} \label{ch5:6:p-inequ}
	\left(\int_{\R^3} f(x)^{\frac53} dx \right)^{\frac{p}{2}} \int_{\R^3} \abs{x}^p f(x)dx  \geq C_p \left(\int_{\R^3} f(x) dx \right)^{1+\frac{5p}{6}} 
\end{align}
which holds for any non negative measurable function $f$ and $p\geq 0$ where the sharp constant $C_p$ is attained for 
\begin{equation} \label{ch5:6:sharp_const}
	f_p(x) \coloneqq \begin{cases}
	\left(1-\abs{x}^p\right)^\frac32 &\abs{x}\leq 1 \\
	0 &\text{elsewise}
	\end{cases} .
\end{equation}
We give the explicit constant $C_p$ in the Appendix in equation \eqref{app:A:explicit_constant}. We prove
\begin{lemma} \label{ch5:6:upper_bound_002}
    Let $p\geq 0$ then 
    \begin{equation*}
         \left( \frac{1}{N} \int_{\R^3} \abs{x_1}^{p} \rho_{\psi_{N,Z}}(x_1) \, dx_1 \right)^{-1/p} \leq \kappa C_p^{-1/p}  Z N^{-2/3}
    \end{equation*}
    where $\kappa =  \sqrt{5}\left(\frac{2}{9\pi^2} 1.456  \right)^{1/3} $ and $C_p$ the constant in \eqref{ch5:6:p-inequ}. 
\end{lemma}
\begin{proof}
Applying \eqref{ch5:6:p-inequ} for $f= \rho_{\psi_{N,Z}}$ yields
\begin{equation*}
    \left( \frac{1}{N} \int_{\R^3} \abs{x_1}^p \rho_{\psi_{N,Z}}(x_1) dx_1 \right)^{-1/p}  \leq C_p^{-1/p} N^{-5/6} \left(\int_{\R^3} \rho_{\psi_{N,Z}}(x_1)^{\frac53} dx_1 \right)^{\frac{1}{2}} \, .
\end{equation*}
By the fermionic kinetic energy inequality in \cite[Theorem 1]{FNvdB:2018} 
\begin{equation} \label{ch5:6:kinetic_inequality}
    \frac{u^{\frac{-2}{3}}}{2} K_3  \int_{\R^3} \rho_{\psi_{N,Z}}(x)^{\frac53} dx \leq \sum_{i=1}^N\frac12\langle \psi_{N,Z}, [-\Delta_{i}] \psi_{N,Z} \rangle 
\end{equation}
with $K_3 = \frac{3}{5}\left( \frac{1.456}{6 \pi^2} \right)^{-2/3} \approx 7.096$. Here $u$ denotes the degrees of freedom in the spin components. We consider spin $1/2$ particles (for example electrons) and thus $u=2$. By the quantum mechanic virial theorem (see \cite{JW:1967}, \cite{AHK:2023})
\begin{equation} \label{ch5:6:virial_thrm}
    -E_{N,Z} =  \sum_{i=1}^N\frac12\langle \psi_{N,Z}, [-\Delta_{i}] \psi_{N,Z} \rangle \, .
\end{equation}
Combining \eqref{ch5:6:kinetic_inequality} and \eqref{ch5:6:virial_thrm} for $u=2$ we arrive at
\begin{equation*}
    \int_{\R^3} \rho_{\psi_{N,Z}}(x)^{\frac53} dx_1 \leq  -\frac{2^{5/3}}{K_3} E_{N,Z} \, .
\end{equation*}
Together with 
\begin{equation*}
-E_{N,Z} \leq A Z^2 N^{1/3}
\end{equation*}
for $A=(3/2)^{1/3}$ (see Lemma \ref{app:A:bnd_on_atom_energy}) this yields
\begin{equation*} 
    \begin{split}
          \left( \frac{1}{N} \int_{\R^3} \abs{x_1}^p \rho_{\psi_{N,Z}}(x_1) dx_1 \right)^{-1/p}  &\leq C_p^{-1/p}  \left( \frac{2^{5/3}}{K_3}A\right)^{1/2}  Z N^{-2/3} \\  
          &= C_p^{-1/p} \sqrt{5}\left(\frac{2}{9\pi^2} 1.456  \right)^{1/3} ZN^{-2/3} \, .
    \end{split}
\end{equation*}
\end{proof}
Combining Lemma \ref{ch5:6:upper_bound_001} and Lemma \ref{ch5:6:upper_bound_002} we can prove
\begin{lemma} \label{ch5:6:upper_bound_003}
    Let $s\geq 2$ then
    \begin{equation} \label{ch5:6:final_upper_bound_ferm}
        Z+\frac{1}{2}\frac{\left\langle \abs{x_1}^s \psi_{N,Z},  \Delta_{1}\psi_{N,Z} \right\rangle}{\left\langle \abs{x_1}^{s-1} \psi_{N,Z}, \psi_{N,Z} \right\rangle}  \leq Z + \frac{s^2-1}{8 } C_{s-1}^{-1/(s-1)} \kappa ZN^{-2/3}
    \end{equation}
    with $\kappa=  \sqrt{5}\left(\frac{2}{9\pi^2} 1.456  \right)^{1/3} $ and $C_{s-1}$ the constant in \eqref{ch5:6:p-inequ}. 
\end{lemma}
\begin{proof}
The inequality \eqref{ch5:6:final_upper_bound_ferm} follows directly by combining Lemma \ref{ch5:6:IMS+Hardy} and Lemma \ref{ch5:6:upper_bound_002}. The fact that we can either apply Lemma \ref{ch5:6:upper_bound_002} for $p=1$ or $p=s-1$ is due to Jensen's inequality as explained in Remark \ref{ch5:6:jensen_remark}. An explicit calculation shows
\begin{equation*}
\begin{split}
    &C_{1}^{-1} =  \left( 3^\frac53 5^\frac56\frac{ \left(\frac{7}{\pi}\right)^\frac{1}{3}}{22\sqrt{11}} \right)^{-1} \approx 2.341\dots \\
    &C_{2}^{-1/2} = 4 \frac{\pi^{2/3}}{\sqrt{15}} \approx 2.215\dots
\end{split}
\end{equation*}
and $p\mapsto C_p^{-1/p}$ is decreasing. We give in Appendix Lemma \ref{app:A:nasty_constant} the explicit constant $C_p$ for any $p\in [1,2]$.
\end{proof} 
\section{Bounds on maximal excess charge}  \label{ch5:7:ch5:7:proof_main_thrm}
From Lemma \ref{ch5:5:refined_lower_bound} and Lemma \ref{ch5:6:upper_bound_003} it is straightforward to prove the inequality in Theorem \ref{ch5:2:main_thm}. We begin with the general inequality for $s\in [2,3]$ before we discuss some refinements in the cases $s=2$ and $s=3$.
\subsection{Proof of the Main Theorem} 
\begin{proof}[Proof of Theorem \ref{ch5:2:main_thm}]
We aim to solve \eqref{ch5:5:refined_inequality} namely
\begin{equation} \label{ch5:7:eq:main_thrm_start}
    \begin{split}
      \frac{(1+r)^{s+1}-(1-r)^{s+1}}{2r(s+1)} N\beta_s \leq &\frac{(1+r)^{s+2}-(1-r)^{s+2}}{2r(s+2)}\left(  \alpha_{N,s}(N-1) + \frac{1}{r} \right) \\
        &+  r^2 f(r,s)  \alpha_{N,s} (N-1)
    \end{split}
\end{equation}
for $N$. Note that the fraction on the left--hand side of \eqref{ch5:7:eq:main_thrm_start} is positive for all $r>0$ and $s>0$ and consequently we can use the lower bound $b(s)^{-1}$ from Proposition \ref{ch5:4:lower_bnd_on_beta} to bound the left--hand side $\beta_s$. 

Direct computations show for $r\in [0,1]$ and $p \geq 2$
\begin{equation*}
    2rp \leq (1+r)^p - (1-r)^p \, .
\end{equation*}
Consequently
\begin{equation*}
    \frac{2r(s+2)}{(r+1)^{s+2}-(1-r)^{s+2}} \leq 1
\end{equation*}
together with $f(r,s)<1/2$
\begin{equation*}
    \frac{2r(s+2)}{(r+1)^{s+2}-(1-r)^{s+2}} r^2 f(r,s) \leq r^2 f(r,s) \leq \frac{r^2}{2} \, .
\end{equation*}
Consequently from \ref{ch5:5:refined_lower_bound} we conclude
\begin{equation} \label{ch5:7:main_for_main}
     \left( \frac{s+2}{s+1} \frac{(r+1)^{s+1}-(1-r)^{s+1}}{(r+1)^{s+2}-(1-r)^{s+2}}  \frac{N}{b(s)} - \frac{1}{r}\right) \leq \alpha_{N,s}(N-1)\left(1+\frac{r^2}{2}\right)
\end{equation}
We prove in the appendix as Lemma \ref{app:A:appendix:lemma:technichal_est} that for any $r\in (0,1)$ and $s \geq 0$
\begin{equation} \label{ch5:7:binomial}
    \frac{s+2}{s+1} \frac{(r+1)^{s+1}-(1-r)^{s+1}}{(r+1)^{s+2}-(1-r)^{s+2}} \geq 1- \frac{s}{3}r^2 \, .
\end{equation}
 Combining the \eqref{ch5:7:main_for_main}  and \eqref{ch5:7:binomial}  we conclude
\begin{equation} \label{ch5:7:here_we_can_choose_r}
    \left( 1- \frac{s}{3}r^2 \right) \frac{N}{b(s)}- \frac{1}{r}< \alpha_{N,s}(N-1)\left(1+\frac{r^2}{2}\right)
\end{equation}
We minimize the left-hand side of \eqref{ch5:7:here_we_can_choose_r} and therefore we choose
\begin{equation}\label{ch5:7:so_i_choose_r}
    r= \left( \frac{3}{2s} \right)^{1/3}  \left(\frac{N}{b(s)}\right)^{-1/3} \eqqcolon \lambda N^{-1/3} \, .
\end{equation}
Combining \eqref{ch5:7:here_we_can_choose_r} and  \eqref{ch5:7:so_i_choose_r} to find
\begin{equation*}
    \frac{N}{b(s)} \leq \alpha_{N,s}(N-1)\left(1+\frac{\lambda^2}{2} N^{-2/3}\right) + \left( \lambda^{-1}  + \frac{s}{3}  \frac{\lambda^2}{b(s)} \right) N^{1/3} \, .
\end{equation*}
Applying Lemma \ref{ch5:6:final_upper_bound_ferm} shows
\begin{equation*}
    \begin{split}
        \frac{N}{b(s)} &\leq \ Z\left(1 + A  N^{-2/3}\right)(1+(\lambda^2/2) N^{-2/3}) + \left( \lambda^{-1}  + \frac{s}{3} \frac{\lambda^2}{b(s)}\right) N^{1/3} \\
        &= Z +\left( A + (\lambda^2/2) \right)ZN^{-2/3} +  \left( \lambda^{-1}  + \frac{s}{3} \frac{\lambda^2}{b(s)}\right) N^{1/3} + \frac{\lambda^2 A}{2} ZN^{-4/3}
    \end{split}
\end{equation*}
where 
\begin{equation*}
    A \coloneqq \frac{s^2-1}{8} C_{s-1}^{-1/(s-1)} \kappa
\end{equation*}
is the parameter in Lemma \ref{ch5:6:final_upper_bound_ferm}. Note that $Z\leq N \leq 3Z$ and thus there exists some $c(s)>0$ such that
\begin{equation*}
    N_c < b(s)\, Z + c(s)Z^{1/3}
\end{equation*}
Since the calculations hold for any lower bound $b(s)^{-1} < \beta_s$ the statement of Theorem \ref{ch5:2:main_thm} follows.
\end{proof}

\subsection{The Case of a Quadratic Weight}  
\begin{proof}[Proof of Proposition \ref{ch5:2:case_s=2}]
Applying Lemma \ref{ch5:5:refined_lower_bound} yields
\begin{equation} \label{ch5:7:starting_point_s=2}
    \frac{r^2/3 + 1}{r^2 + 1} N \beta_2  - \frac{1}{r} \leq  \alpha_{N,2}(N-1)<Z + \frac{3}{8 } C_{1}^{-1} \kappa ZN^{-2/3}
\end{equation}
Note that by a straightforward calculation
\begin{equation} \label{ch5:7:eq:prop2.2_direct}
     N \beta_2\left(1-\frac{2r^2}{3} \right)   \leq \frac{r^2/3 + 1}{r^2 + 1} N \beta_2 
\end{equation}
and consequently by inserting \eqref{ch5:7:eq:prop2.2_direct} into \eqref{ch5:7:starting_point_s=2} we arrive at
\begin{equation} \label{ch5:7:eq:refined_starting_point_s=2}
    N \beta_2\left(1-\frac{2r^2}{3} \right) - \frac{1}{r} \leq Z + \frac{3}{8 } C_{1}^{-1} \kappa ZN^{-2/3}
\end{equation}
Optimizing the left--hand side of \eqref{ch5:7:eq:refined_starting_point_s=2} in $r>0$ gives
\begin{equation} \label{ch5:7:choice_for_r_s_2}
    r= \left(\frac{3}{4}\right)^{1/3} \left(N\beta_2\right)^{-1/3} \, .
\end{equation}
Note that $r<1$ for $N>1$. Inserting \eqref{ch5:7:choice_for_r_s_2} into \eqref{ch5:7:eq:refined_starting_point_s=2} yields
\begin{equation*}
    N\beta_2 - \left( \frac{9}{2} \right)^{1/3} (N\beta_2)^{1/3} \leq  Z + \frac{3}{8 } C_{1}^{-1} \kappa ZN^{-2/3} \, .
\end{equation*}
Applying Lemma \ref{ch5:6:upper_bound_003} we arrive at
\begin{equation*}
    N\beta_2  \leq Z + \frac{3}{8 } C_{1}^{-1} \kappa Z N^{-2/3}+ \left( \frac{9}{2} \beta_2 \right)^{1/3}  N^{1/3} \, .
\end{equation*}
We define 
\begin{equation*}
    \lambda \coloneqq  \frac{3}{8 } C_{1}^{-1}   \kappa \approx 0.6284 \, .
\end{equation*}
Then
\begin{equation} \label{ch5:7:go_on_from_here001}
     N\beta_2  \leq Z + \lambda Z N^{-2/3}+ \left( \frac{9}{2} \beta_2 \right)^{1/3}  N^{1/3} \, .
\end{equation}
Let $a>0$ and assume that
\begin{equation} \label{ch5:7:assumption_s=2}
    N\beta_2 \geq Z + \beta_2 a Z^{1/3} \, .
\end{equation}
Combining \eqref{ch5:7:go_on_from_here001} and \eqref{ch5:7:assumption_s=2} yields
\begin{equation} \label{ch5:7:poly}
     Z + \beta_2 a Z^{1/3} \leq  Z + \lambda Z N^{-2/3}+ \left( \frac{9}{2} \beta_2 \right)^{1/3}  N^{1/3} \, .
\end{equation}
Dividing by $Z^{1/3}$ gives
\begin{equation*}
    a \leq  \beta_2^{-1}  \lambda \left( \frac{N}{Z}\right)^{-2/3}+ \beta_2^{-1}\left( \frac{9}{2} \beta_2 \right)^{1/3}  \left( \frac{N}{Z}\right)^{1/3} \, .
\end{equation*}
From Lieb's bound, we conclude $N/Z<5/2$ for any $Z\geq2$. Maximizing the right hand side of \eqref{ch5:7:poly} for $N/Z \in [1,5/2]$ yields
\begin{equation*}
    a \leq  \beta_2^{-1}  \lambda \left( \frac{N}{Z}\right)^{-2/3}+ \beta_2^{-1}\left( \frac{9}{2} \beta_2 \right)^{1/3}  \left( \frac{N}{Z}\right)^{1/3} \leq 2.953 \, .
\end{equation*}
Thus for $a\coloneqq 2.96$ assumption \eqref{ch5:7:assumption_s=2} cannot hold and thus
\begin{equation*}
    N \leq \frac{1}{\beta_2} Z  + 2.96 Z^{1/3} < b(2)Z  + 2.96 Z^{1/3}
\end{equation*}
for any $Z\geq 2$. The assertion in Proposition \ref{ch5:2:case_s=2} follows.    
\end{proof}

\subsection{The Case of a Cubic Weight}
\begin{proof}[Proof of Proposition \ref{ch5:2:corol_s=3}]
By application of Lemma \ref{ch5:5:refined_lower_bound} we find
\begin{equation} \label{ch5:7:deduced_from_lem5.9}
    \begin{split}
      \frac{(1+r)^{s+1}-(1-r)^{s+1}}{2r(s+1)}  N\beta_s 
        &\leq   
           \frac{(1+r)^{s+2}-(1-r)^{s+2}}{2r(s+2)} 
           \left( \alpha_{N,s}(N-1)  + \frac{1}{r} \right) \\
        &\phantom{+~}+r^2 f(r,s)\alpha_{N,s} (N-1) 
    \end{split}
\end{equation}
or equivalently
\begin{equation}\label{ch5:7:starting_reformed}
       \left(\frac{s+2}{s+1}\right) \frac{(1+r)^{s+1}-\abs{1-r}^{s+1}}{(1+r)^{s+2}-\abs{1-r}^{s+2}}N\beta_s - \frac{1}{r} \leq \left(1 + \frac{2r(s+2)r^2 f(r,s)}{(1+r)^{s+2}-\abs{1-r}^{s+2}} \right) \alpha_{N,s}(N-1)
\end{equation}
In Appendix \ref{AppendixA} as Lemma \ref{app:A:appendix:lemma:technichal_est} we show that for any $r\in (0,1)$ and $s \geq 0$
\begin{equation} \label{ch5:7:auto}
   1- \frac{s}{3}r^2  \leq \frac{s+2}{s+1} \frac{(r+1)^{s+1}-(1-r)^{s+1}}{(r+1)^{s+2}-(1-r)^{s+2}} \, .
\end{equation}
Combining \eqref{ch5:7:auto} and \eqref{ch5:7:starting_reformed} with $s=3$ proves
\begin{equation}\label{ch5:7:insert_here}
       (1-r^2)N\beta_3 - \frac{1}{r} \leq \left(1 +\frac{5}{r^2(r^2+10)+5} r^2 f(r,3) \right) \alpha_{N,3}(N-1) \, .
\end{equation}
where
\begin{equation}\label{ch5:7:taylor01}
    f(r,3) = \frac{1}{5} + \frac{1}{35} r +  \frac{168}{625} r^2\, .
\end{equation}
By direct computations one shows for any $r\geq 0$
\begin{equation}\label{ch5:7:taylor02}
    \frac{5}{r^2(r^2+10)+5} \leq 1-2r^2+\frac{19}{5}r^4 \, .
\end{equation}
Assume $r<0.5$ then by combining \eqref{ch5:7:taylor01} and \eqref{ch5:7:taylor02} we find 
\begin{equation}\label{ch5:7:taylor03}
    \begin{split}
        \left(1 +\frac{5}{r^2(r^2+10)+5} r^2 f(r,3) \right) &\leq 1+\frac{r^2}{5} + \frac{r^3}{35} \underbrace{- \frac{82 r^4}{625} - \frac{2 r^5}{35} + \frac{139 r^6}{625} + \frac{19 r^7}{175} + \frac{3192 r^8}{3125}}\limits_{\leq 0, \text{ for } r<0.53} \\
        &\leq 1+\frac{r^2}{5} + \frac{r^3}{35} \, .
    \end{split}
\end{equation}
Inserting \eqref{ch5:7:taylor03} into the right--hand side of \eqref{ch5:7:insert_here} we arrive at
\begin{equation}\label{ch5:7:refined_starting_simple}
       (1-r^2)N\beta_3 - \frac{1}{r} \leq \left( 1+\frac{r^2}{5} + \frac{r^3}{35} \right) \alpha_{N,3}(N-1) \, .
\end{equation}
for any $r\leq 0.5$. Applying Lemma \ref{ch5:6:upper_bound_003} for $s=3$ yields
\begin{equation*}
   (1-r^2)N\beta_3 - \frac{1}{r} \leq \left( 1+\frac{r^2}{5} + \frac{r^3}{35} \right) \left(Z + c Z N^{-2/3} \right), \quad c=C_2^{-1/2} \kappa\,.
\end{equation*}
We continue by choosing $r \in (0,0.5]$. As in the previous cases let
\begin{equation} \label{ch5:7:choice_of_r_s3}
    r= \lambda (N\beta_3)^{-1/3}, \quad \lambda>0\, .
\end{equation}
Inserting \eqref{ch5:7:choice_of_r_s3} into \eqref{ch5:6:upper_bound_003} yields
\begin{equation}
    \begin{split}
        N\beta_3 \leq \phantom{+~}Z  &+ \lambda^{-1}(N\beta_3)^{1/3} +\lambda^2(N\beta_3)^{1/3} + \frac{\lambda^2}{5}(N\beta_3)^{-2/3}Z +  c Z N^{-2/3} \\
        &+ \frac{\lambda^3}{35}(N\beta_3)^{-1}Z+c\frac{\lambda^2}{5}(N\beta_3)^{-2/3}  Z N^{-2/3} \\
        &+c\frac{\lambda^3}{35}(N\beta_3)^{-1} Z N^{-2/3}
    \end{split}
\end{equation}
We can always assume $Z<N\beta_3$ (since otherwise $N\leq \beta_3^{-1}Z$ already proves an inequality than the statement) and consequently we find
\begin{equation} \label{ch5:7:we_determine_lambda_here}
    \begin{split}
        N\beta_3 \leq &\phantom{+~}Z  + \left(\lambda^{-1} + \frac{6\lambda^2}{5} \right)(N\beta_3)^{1/3} +  c Z N^{-2/3} \\
        &+ \frac{\lambda^3}{35} +c\frac{\lambda^2}{5}(\beta_3)^{1/3}  N^{-1/3} +c\frac{\lambda^3}{35}N^{-2/3}
    \end{split}
\end{equation}
To optimize the leading correction term that grows as $N^{1/3}$ we minimize
\begin{equation*}
    \lambda \mapsto \lambda^{-1}+\frac{6}{5} \lambda^2,
\end{equation*}
and consequently, we choose
\begin{equation} \label{ch5:7:choice_of_lambda_s=3}
    \lambda = \left( \frac{5}{12} \right)^{1/3}, \quad \text{ such that } \quad \lambda^{-1}+\frac{6}{5} \lambda^2 = 3 \left( \frac{3}{10} \right)^{1/3} \, .
\end{equation}
To ensure $r<0.5$ as assumed after \eqref{ch5:7:choice_of_lambda_s=3} we need to have
\begin{equation}
    N > \frac{10}{3 \beta_3} > \frac{10}{3}  \, .
\end{equation}
We always assume $N\geq Z$ and consequently the result will hold for $Z\geq 4$. Inserting \eqref{ch5:7:choice_of_lambda_s=3} into \eqref{ch5:7:we_determine_lambda_here} yields
\begin{equation} \label{ch5:7:lambda_inserted}
    \begin{split}
        N\beta_3 \leq &\phantom{+~}Z  + 3 \left( \frac{3}{10} \right)^{1/3} (N\beta_3)^{1/3} +  c Z N^{-2/3} \\
        &+ \frac{1}{84} + \frac{c}{5} \left( \frac{5}{12} \right)^{2/3}(\beta_3)^{1/3}  N^{-1/3} +c\frac{1}{84}N^{-2/3}
    \end{split}
\end{equation}
We can always assume $N \geq \beta_3^{-1}Z $ because otherwise $N\leq \beta_3^{-1}Z$ and we are done. Inserting $N \geq \beta_3^{-1}Z $ into the last two summands in the right--hand side of  \eqref{ch5:7:lambda_inserted} yields
\begin{equation*} 
    \begin{split}
        N\beta_3 \leq &\phantom{+~}Z  + 3 \left( \frac{3}{10} \right)^{1/3} (N\beta_3)^{1/3} +  c Z N^{-2/3} \\
        &+ \frac{1}{84} + \frac{c}{5} \left( \frac{5}{12} \right)^{2/3}(\beta_3)^{2/3}  Z^{-1/3} +c\frac{\beta_3^{2/3}}{84}Z^{-2/3} \, .
    \end{split}
\end{equation*}
To prove the desired inequality
\begin{equation*}
    N\leq \beta_3^{-1} Z +  a_1 Z^{1/3} + a_2 +  a_3 Z^{-1/3} +  a_4 Z^{-2/3} \, .
\end{equation*}
for optimal $a_1,a_2,a_3,a_4 \geq 0$ and all $N\geq 3$ we assume that for any arbitrary but fixed $N,Z$ with $N\geq 3$ holds
\begin{equation} \label{ch5:7:use_for_contra}
     N\beta_3 \geq Z + \beta_3 a_1 Z^{1/3} + \beta_3 a_2 + \beta_3 a_3 Z^{-1/3} + \beta_3 a_4 Z^{-2/3} 
\end{equation}
and bring this to a contradiction by choosing $a_1,a_2,a_3,a_4 \geq 0$ and comparing \eqref{ch5:7:use_for_contra} with \eqref{ch5:7:lambda_inserted}. We do this now to finish the proof. Combining  \eqref{ch5:7:use_for_contra} with \eqref{ch5:7:lambda_inserted} yields
\begin{equation}\label{ch5:7:choose_a_from_here}
    \begin{split}
          \beta_3 a_1 Z^{1/3}& + \beta_3 a_2 + \beta_3 a_3 Z^{-1/3} + \beta_3 a_4 Z^{-2/3} 
          \\
          \leq &\phantom{+~} 3 \left( \frac{3}{10} \right)^{1/3} (N\beta_3)^{1/3} +  c Z N^{-2/3} + \frac{1}{84} \\
          &\phantom{\leq}+ \frac{c}{5} \left( \frac{5}{12} \right)^{2/3}(\beta_3)^{2/3}  Z^{-1/3} +c\frac{\beta_3^{2/3}}{84}Z^{-2/3}
    \end{split}
\end{equation}
After comparing both sides of \eqref{ch5:7:choose_a_from_here} we choose
\begin{equation}
    \begin{split}
       a_2 = \beta_3^{-1}/84, \quad   a_3 =  \frac{c}{5} \left( \frac{5}{12} \right)^{2/3}\beta_3^{-1/3}, \quad a_4 = c\frac{\beta_3^{-1/3}}{84} \, .
    \end{split}
\end{equation}
Using $\beta_3^{-1}< 1.1185$ and $c<1.5855$ this gives
\begin{equation}
    a_2 \leq 0.0134, \quad a_3 \leq 0.184, \quad a_4 \leq 0.0196 \,.
\end{equation}
For this choice of $a_2,a_3,a_4$ we arrive at
\begin{equation}\label{ch5:7:to_be_failed_02}
    \begin{split}
          \beta_3 a_1 Z^{1/3}  
          \leq &\phantom{+~} 3 \left( \frac{3}{10} \right)^{1/3} (N\beta_3)^{1/3} +  c Z N^{-2/3} 
    \end{split}
\end{equation}
Dividing \eqref{ch5:7:to_be_failed_02} by $\beta_3 Z^{1/3}$ we find 
\begin{equation}\label{ch5:7:to_be_failed_03}
    \begin{split}
          a_1 
          \leq &\phantom{+~} 3\left( \frac{3}{10} \right)^{1/3} \beta_3^{-2/3} \left( \frac{N}{Z} \right)^{1/3}  +  c \beta_3^{-1}\left( \frac{N}{Z} \right)^{-2/3} 
    \end{split}
\end{equation}
We can always assume $N\geq \beta_3^{-1}Z$ as explained earlier and $N<2Z+1$ due to Lieb's result and consequently we can assume $N/Z \in [\beta_3^{-1},5/2]$ for $Z\geq 2$. Thus
\begin{equation}\label{ch5:7:supremum}
   a_1 \leq \sup \left\{ 3\left( \frac{3}{10} \right)^{1/3} \beta_3^{-2/3} x^{1/3}  +  c \beta_3^{-1} \, x^{-2/3} : x\in [\beta_3^{-1},5/2] \right\}
\end{equation}
Following Lemma \ref{ch5:6:upper_bound_003} we have
\begin{equation}
   c=  \sqrt{5}\left(\frac{2}{9\pi^2} 1.456  \right)^{1/3} 4 \frac{\pi^{2/3}}{\sqrt{15}},
\end{equation}
and since $\beta_3^{-1} \in [1.0,  1.1185]$ one can show that the supremum in the right--hand side of \eqref{ch5:7:supremum} is attained at $x=\beta_3^{-1}$ and consequently
\begin{equation}
    a_1 \leq 3\left( \frac{3}{10} \right)^{1/3} \beta_3^{-1}  +  c \beta_3^{-1/3} < 3.893 \, .
\end{equation}
For the choice $a_1=3.90$ and $a_2,a_3,a_4$ as in \eqref{ch5:7:choose_a_from_here} the inequality \eqref{ch5:7:use_for_contra} fails and therefore we find
\begin{equation*}
    N\leq \beta_3^{-1} Z +  3.90 Z^{1/3} + 0.0134 +  0.184 Z^{-1/3} +  0.0196 Z^{-2/3}, \quad Z\geq 4 \, .
\end{equation*}
this proves the statement of of Proposition \ref{ch5:2:corol_s=3}.
\end{proof}


\begin{appendix}

\section{Technical Details}
\label{AppendixA} 
\subsection{Density Argument for $\alpha_{N,s}$}
\begin{lemma} \label{app:A:lem_a_0}
    Let $\alpha_{N,s}$ be defined as in equation \eqref{ch5:5:geometric_exp} and $A\subset \R^{3N}$ be defined as
    \begin{equation*}
        A_0=\{ (x_1,x_2,\dots x_N) \in A:  x_k\neq 0  \text{ for } 1\leq k\leq N\}
    \end{equation*}
    then
    \begin{equation} \label{app:A:without_zero}
        \alpha_{N,s}= \inf \left\{ \frac{\sum_{\substack{1\leq i,k\leq N \\ i\neq k}} \frac{\abs{x_k}^s + \abs{x_i}^s}{|x_{i}-x_{k}|}}{2(N-1)\sum_{k=1}^N \abs{x_k}^{s-1} } : (x_1,x_2,\dots x_N) \in A_0 \right\}
    \end{equation}
\end{lemma}
\begin{proof}  We define
    \begin{equation} \label{app:A:defF}
        F(x_1,x_2,\dots,x_N) \coloneqq  \frac{\sum_{\substack{1\leq i,k\leq N \\ i\neq k}} \frac{\abs{x_k}^s + \abs{x_i}^s}{|x_{i}-x_{k}|}}{2(N-1)\sum_{k=1}^N \abs{x_k}^{s-1} } 
    \end{equation}
Note that $F$ is continuous in $A$ and $A_0$ is dense in $A$. Since any $(x_1,x_2,\dots, x_N)\in A$ consists of arbitrary but distinct point in $\R^3$ we may always assume that $\abs{x_1} \leq \abs{x_2} \leq \dots \leq \abs{x_N}$ by relabeling the indices. Since the vectors are distinct only $x_1$ may vanish. The set $A_0$ is a subset of $A$ and thus 
\begin{equation*}
    \alpha_{N,s}  \leq  \inf \{ F(x_1,x_2,\dots,x_N) : (x_1,x_2,\dots,x_N) \in A_0\} \eqqcolon \zeta_{N,s}
\end{equation*}
The claim follows if we can show $\zeta_{N,s}\leq \alpha_{N,s}$. We show for arbitrary $\varepsilon>0$ that $\zeta_{N,s} \leq \alpha_{N,s}+\varepsilon $ and conclude the statement in the limit $\varepsilon\to 0$. Let $\varepsilon>0$ arbitrary then by the definition of the infimum, there exists some $r_\varepsilon \in A $ such that $F(r_\varepsilon) \leq \alpha_{N,s}+\varepsilon/2 $. If $r_\varepsilon\in A_0$ then $F(r_\varepsilon)\geq \zeta_{N,s}$ and the inequality follows directly. If $r_\varepsilon \in A\setminus A_0$ then $(r_\varepsilon)_1 = 0$. By continuity of $F$ in $r_\varepsilon$ we can find $u_\varepsilon \in A_0$ such that $\abs{F(r_\varepsilon) - F(u_\varepsilon) } \leq \varepsilon/2$ and thus 
\begin{equation*}
    \zeta_{N,s} \leq F(u_\varepsilon) \leq \alpha_{N,s} + \varepsilon.
\end{equation*}
Thus we have shown $\zeta_{N,s} = \alpha_{N,s}$ and the statement follows immediately.
\end{proof}
\subsection{Funk-Hecke formula and Multipol Moments}
\begin{lemma}\label{app:A:lem:atkinson han formula}
  For any function $f\in L^1(-1,1)$ and any Legendre polynomial 
  $P_l$ we have 
  \begin{equation}
  \label{app:A:eq:atkinson han formula}
      \frac{1}{4\pi}
        \int_{S^2} 
          f(\xi\cdot\omega) P_l(\zeta\cdot\omega) 
        d\omega 
      =
        \lambda P_l(\xi\cdot\zeta) 
  \end{equation}
  for all $\xi,\zeta\in S^2$ where 
  \begin{equation*}
      \lambda= \frac{1}{2} \int_{-1}^1 P_l(t) f(t) dt
  \end{equation*}
\end{lemma}
\begin{proof}
  This is a direct consequence of the Funk--Hecke formula in three dimensions, see equation (2.66) just after Theorem 2.22 in Chapter 2.6 of \cite{AH:2012}.  
\end{proof}
\begin{lemma} \label{app:A:int_001}
    Let $\lambda>-2$,$a\in\R^3$ and $r, \abs{a}> 0$, then
    \begin{equation*}
         \int_{S^2} \abs{a+r\omega}^\lambda \frac{d\omega}{4\pi} =\frac{(\abs{a}+r)^{\lambda+2} - (\abs{a}-r)^{\lambda+2}}{2r\abs{a}(\lambda+2)} \\
    \end{equation*}
\end{lemma}
\begin{proof}
 Since the Legendre polynomial $P_0=1$ and 
 $\abs{a+r\omega}^2 = \abs{a}^2 +r^2 +2r\abs{a}\hatt{a}\cdot\omega$
 we see from Lemma \ref{app:A:lem:atkinson han formula} that 
 \begin{align*}
   \int_{S^2} \abs{a+r\omega}^t\lambda \frac{d\omega}{4\pi} 
     = 
       \frac{1}{2}\int_{-1}^1 
         (\abs{a}^2+r\omega +2r\abs{a}t)^{\lambda/2} 
       \frac{d\omega}{4\pi} 
     =\frac{(\abs{a}+r)^{\lambda+2} - (\abs{a}-r)^{\lambda+2}}{2r\abs{a}(\lambda+2)}  
 \end{align*}
\end{proof}

\begin{lemma} \label{app:A:int_002}
    Let $s\geq -2$,  $\zeta\in S^2$, $w\in\R^3\setminus \{0\}$, 
    and $r>0$. Then 
    \begin{equation*} 
    \begin{split}
        &\int_{S^2} \abs{w+r \omega}^s \frac{d\omega}{4\pi} =\frac{(\abs{w}+r)^{s+2} - \abs{\abs{w}1-r}^{s+2}}{2r\abs{w}(s+2)} \\
        &\int_{S^2} 
           \abs{w+r \omega}^s \langle \zeta, \omega \rangle 
         \frac{d\omega}{4\pi} 
           = c(|r|) \langle w, \zeta\rangle\\
        &\int_{S^2} 
           \abs{w+r \omega}^s P_l(\langle\hat{w},\omega\rangle)
         \frac{d\omega}{4\pi} 
           = 
             \frac{1}{2}\int_{-1}^1
               (w^2+r^2 +2r\abs{w}t)^{s/2} P_l(t) 
             dt 
             %
    \end{split}
\end{equation*}
where $c(r) \coloneqq \int_{-1}^1 (1+r^2+2rt)^{s/2}t \, dt $ and $ c_{k,n} \coloneqq (2n+1)\int_{-1}^1 t^k P_n(t) \, dt$.
\end{lemma}
\begin{proof}
The zero-order moment is easy to and follows directly from Lemma \ref{app:A:int_001}. Similar
\begin{equation*}
    \begin{split}
        \int_{S^2} \abs{\hat{x}_j+r \omega}^s P_1(\langle \hat{a}, \omega \rangle) \frac{d\omega}{4\pi} &= \int_{S^2} \abs{\hat{x}_j+r \omega}^s \langle \hat{a},\omega \rangle \frac{d\omega}{4\pi}  \\
        &=\langle \hat{a}, \int_{S^2} \abs{\hat{x}_j+r \omega}^s \omega \frac{d\omega}{4\pi } \rangle \\
        &= \langle U^{-1} \hat{a}, \int_{S^2} \abs{U^{-1}\hat{x}_j+r \omega}^s \omega \frac{d\omega}{4\pi } \rangle 
    \end{split}
\end{equation*}
for any $U\in SO(3)$. Choose $U\in SO(3)$ such that $U^{-1}\hat{x}_j = \hat{e}_3$ then 
\begin{equation*}
    \begin{split}
        \int_{S^2} \abs{U^{-1}\hat{x}_j+r \omega}^s \omega \frac{d\omega}{4\pi } &=\int_{S^2} \abs{\hat{e}_3+r \omega}^s \omega \frac{d\omega}{4\pi }  \\
        &= 2\pi \hat{e}_3 \int_{0}^\pi (1+r^2+2r\cos{\theta})^{s/2} \cos{\theta} \frac{\sin{\theta}d\theta}{4\pi} \\
        &= \frac{\hat{e}_3}{2} \int_{-1}^1 (1+r^2+2rt)^{s/2}t dt 
        \eqqcolon \hat{e}_3 c(r)
    \end{split}
\end{equation*}
and thus
\begin{equation} \label{app:A:second_moment}
    \int_{S^2} \abs{\hat{x}_j+r \omega}^s P_1(\langle \hat{a}, \omega \rangle) \frac{d\omega}{4\pi} = c(r) \langle \hat{a}, U \hat{e}_3\rangle 
\end{equation}
for some $c(r) \geq 0$.
To compute higher multipole moments we need to extend $\abs{\hat{x_j}+r\omega}^s$ in terms of Legendre polynomials. Note, that
\begin{equation} \label{app:A:to_be_extended}
   \abs{\hat{x}_j+r\omega}^s = \left(1+r^2+2r\langle\hat{x_j}, \omega\rangle  \right)^{s/2} = (1+r^2)^{s/2} \left(1+\frac{2r}{1+r^2} \langle\hat{x}_j, \omega\rangle \right)^{s/2} 
\end{equation}
For convenience we define 
\begin{equation*}
    q \coloneqq \frac{2r}{1+r^2} \, .
\end{equation*}
By Newton's generalized binomial theorem, we can extend the right--hand side of equation \eqref{app:A:to_be_extended} 
\begin{equation} \label{app:A:binomial_series}
    (1+r^2)^{s/2}\left(1+\frac{2r}{1+r^2} \langle\hat{x}_j, \omega\rangle \right)^{s/2} = (1+r^2)^{s/2}\sum_{k=0}^\infty \binom{s/2}{k} q^k \left(\langle\hat{x}_j, \omega\rangle\right)^k
\end{equation}
which converges absolutely since $q\langle\hat{x}_j, \omega\rangle \leq 1$ by assumption. If $s/2\in \N$ then the generalized binomial coefficients are identical to the normal binomial coefficients with 
\begin{equation} \label{app:A:normal_binomial}
    \binom{s/2}{k} = 0, \quad k> s/2\in \N
\end{equation}
and hence the series above is only a finite sum in that case. If $s/2\notin \N$ then for any $k\in \N_0$ the generalized binomial coefficients are defined as
\begin{equation*}
    \binom{s/2}{k} = \frac{\frac{s}{2}(\frac{s}{2}-1) \cdots (\frac{s}{2}-(k-1))}{k!} = \frac{ \prod_{n=0}^{k-1} (\frac{s}{2}-n)}{k!}.
\end{equation*}
We extend monomials in terms of Legendre polynomials. From Rodrigou's Formula, one can derive the explicit representation
\begin{equation*}
    P_n(t)= 2^n \sum_{m=0}^n t^m \binom{n}{m} \binom{\frac{n+m-1}{2}}{n} 
\end{equation*}
(See \cite{book:AS:1964}[Chapter 8]). Let
\begin{equation}\label{app:A:coeff}
    \begin{split}
        c_{k,n} &\coloneqq \frac{(2n+1)}{2}\int_{-1}^1 t^k P_n(t) \, dt  \\
        &= (2n+1)2^{n-1}\sum_{m=0}^n \binom{n}{m} \binom{\frac{n+m-1}{2}}{n} 
 \int_{-1}^1 t^{k+m} \, dt  \\
    \end{split}
\end{equation}
then its clear that $c_{k,n}=0$ for $n>k$. We differ the cases for which $k$ is even and for which $k$ is odd. When $k$ is even then $c_{k,n}=0$ whenever $n$ is odd as one easily concludes from equation \eqref{app:A:coeff}. Analogous when $k$ is odd $c_{k,n}=0$ whenever $n$ is even. Hence for $n\leq k$,
\begin{equation}\label{app:A:coeff_explicit}
        c_{k,n} = \begin{cases}
            (2n+1)2^{n-1} \sum_{m=0}^n \binom{n}{m} \binom{\frac{n+m-1}{2}}{n} 
            \frac{2}{k+m+1}, &k+m \text{ is even},\\
            0, &k+m \text{ is odd},\\
            1, &k=m=0
        \end{cases} \, .
\end{equation}
By the orthogonality of Legendre Polynomials
\begin{equation*}
    \int_{-1}^1 P_n(t) P_m(t) \, dt = \frac{2\delta_{mn}}{2n+1} \, 
\end{equation*}
it then follows
\begin{equation} \label{app:A:ext_monomial}
    t^k = \sum_{l=0}^k c_{k,l} P_l(t) \, .
\end{equation}
By the usual definition of the double factorial, one can derive by elementary steps that
\begin{equation} \label{app:A:coeff_better}
    t^k = \sum_{l= k,k-2,\dots} \frac{(2l+1)k!}{2^{(k-l)/2} \left( \frac{k-l}{2} \right)!(l+k+1)!!} P_k(t).
\end{equation}
For even numbers of $k=2n$ with $n\in \N_0$ this means
\begin{equation} \label{app:A:coeff_even}
    t^{2n} = \sum_{u=0}^n \frac{(4u+1)(2n)!}{2^{(n-u)} \left( n-u\right)!(2(n+u)+1)!!} P_{2u}(t).
\end{equation}
and for odd numbers of $k=2n+1$ with $n\in \N_0$  this is
\begin{equation} \label{app:A:coeff_odd}
    t^{2n+1} = \sum_{u=0}^n \frac{(4u+3)(2n+1)!}{2^{(n-u)} \left( n-u\right)!(2(n+u)+3)!!} P_{2u+1}(t).
\end{equation}
Combining the equation \eqref{app:A:ext_monomial}, \eqref{app:A:binomial_series} and \eqref{app:A:to_be_extended} shows
\begin{equation} \label{app:A:extended}
   \abs{\hat{x}_j+r\omega}^s = (1+r^2)^{s/2}\sum_{k=0}^\infty \binom{s/2}{k} q^k  \sum_{n=0}^k c_{k,n} P_n(\langle\hat{x}_j, \omega\rangle) \, .
\end{equation}
Using this representation we can compute the remaining multipole moments. For $l\geq 2$ we compute
\begin{equation} \label{app:A:higher_multipole_moments}
    \begin{split}
        &\int_{S^2} \abs{\hat{x}_j+r \omega}^s P_l(\langle \hat{a}, \omega \rangle) \frac{d\omega}{4\pi} \\
        &= (1+r^2)^{s/2}\sum_{k=0}^\infty \binom{s/2}{k} q^k  \sum_{n=0}^k c_{k,n} \int_{S^2}  P_n(\langle\hat{x}_j, \omega\rangle)  P_l(\langle\hat{a}, \omega\rangle) \frac{d\omega}{4\pi} . \\
    \end{split}
\end{equation}
The remaining integral follows from the orthogonality of Legendre Polynomials. By extending the Legendre Polynomials $P_n$ into the spherical harmonics $Y_{nm}$ one easily shows the following orthogonality relation for $a,b\in \R^3$ with $\norm{a}=\norm{b}=1$,
\begin{equation} \label{app:A:twisted_orthogonality}
	\begin{split}
	&\int_{ S^{2}} P_l(\langle \omega,a\rangle) P_n(\langle \omega , b \rangle)              \frac{d\omega}{\abs{ S^{2}}} \\
        &= \sum_{m'=-l}^l\sum_{m=-n}^n  \frac{4\pi}{2l+1}\frac{4\pi}{2n+1} \int_{S^{2}} Y^*_{lm}(\omega)Y_{nm'}(\omega)\frac{d\omega}{\abs{S^{2}}}Y^*_{lm}(a)Y_{nm'}(b) \\
		&=\sum_{m'=-l}^l\sum_{m=-n}^n \frac{4\pi}{2l+1}\frac{4\pi}{2n+1} \frac{ \delta_{mm'} \delta_{ln}}{ \abs{S^{2}}} Y_{lm}(a)Y^*_{nm'}(b) \\
		&= \frac{\delta_{ln}}{2l+1} P_l(\langle a, b\rangle).
	\end{split}
\end{equation}
Thus combining equation \eqref{app:A:twisted_orthogonality} and \eqref{app:A:higher_multipole_moments} shows for $l\geq 2$,
\begin{equation} \label{app:A:higher_multipole_moments_solution}
    \begin{split}
        &\int_{S^2} \abs{\hat{x}_j+r \omega}^s P_l(\langle \hat{a}, \omega \rangle) \frac{d\omega}{4\pi} \\
        &= (1+r^2)^{s/2}\sum_{k=2}^\infty \binom{s/2}{k} q^k  \sum_{n=0}^k c_{k,n} \frac{\delta_{ln}}{2l+1} P_l(\langle \hat{a}, \hat{x}_j \rangle )  \\
        &= (1+r^2)^{s/2}\sum_{k=l}^\infty \binom{s/2}{k}  \frac{q^k c_{k,l}}{2l+1} P_l(\langle \hat{a}, \hat{x}_j \rangle ) 
    \end{split}
\end{equation}
where the series still converges absolutely as mentioned after equation \eqref{app:A:binomial_series}. Note that in the case $s=2$ the expression vanishes due to equation \eqref{app:A:normal_binomial}.
\end{proof}
\begin{lemma} \label{app:A:remainder_of_sum}
For $s\in [2,3], k\in \N$ with $k\geq 2$ let
\begin{equation*}
    A_k =  \abs{\binom{s/2}{k}} \sum_{l=2}^k  \frac{ c_{k,l}}{2l+1}
\end{equation*}
with $c_{k,l}$ defined in equation \eqref{app:A:coeff}. Then
\begin{equation*}
    \sum_{k=4}^\infty A_k \leq \frac{s}{2}\left(\frac{s}{2}-1\right)\left(2-\frac{s}{2}\right) \frac{28}{625} \, .
\end{equation*}
\end{lemma}
\begin{proof}
    Note that 
    \begin{equation*}
        \sum_{l=0}^k c_{k,l} = 1 
    \end{equation*}
    due to its defining equation
    \begin{equation*}
    t^k = \sum_{l=0}^k c_{k,l} P_l(t) \, .
    \end{equation*}
    by choosing $t=1$ and noting $P_l(1) = 1$ for any $l\in \N$. Consequently
    \begin{equation*}
        A_k \leq \frac{1}{5}  \abs{\binom{s/2}{k}} \sum_{l=2}^k  c_{k,l} \leq \frac{1}{5}  \abs{\binom{s/2}{k}} \sum_{l=0}^k  c_{k,l} = \frac{1}{5} \abs{\binom{s/2}{k}}
    \end{equation*}
    where we used $c_{k,l} \geq 0$ (see their explicit form in \eqref{app:A:coeff}). For the generalized binomial coefficients the inequality
    \begin{equation*}
        \abs{\binom{s/2}{k}} \leq \frac{\frac{s}{2}(\frac{s}{2}-1)}{k(k-1)} \leq \frac{s}{2}\left(\frac{s}{2}-1\right)\left(\frac{1}{k-1} - \frac{1}{k}\right)
    \end{equation*}
    holds for any $s\in [2,3]$ and $k\geq 2$, $k\in \N$. Thus for $n_0\in \N$
    \begin{equation*}
        \sum_{k=n_0}^\infty A_k \leq  \sum_{k=n_0}^N A_k +  \frac{s}{2}\left(\frac{s}{2}-1\right)\frac{1}{5(N-1)}
    \end{equation*}
    and consequently
    \begin{equation*}
        \sum_{k=N}^\infty A_k < \frac{s}{2}\left(\frac{s}{2}-1\right)\left(2-\frac{s}{2}\right)\frac{2}{5(N-1)} \eqqcolon \delta(N)
    \end{equation*}
    is arbitrary small for $N$ large enough. Thus we can estimate the series over $A_k$ to arbitrary precision. Let $n_0=4$ then
    \begin{equation*}
        \sum_{k=n_0}^\infty A_k \leq  \sum_{k=4}^2N A_k + \delta(2N) = \sum_{k=2}^{N} A_{2k} + \sum_{k=2}^{N-1} A_{2k+1} +\delta(2N) 
    \end{equation*}
    With the explicit expression of $c_{k,l}$ in \eqref{app:A:coeff_even} we find
    \begin{equation*}
        A_{2k} \leq \frac{s}{2}\left(\frac{s}{2}-1\right)\left(2-\frac{s}{2}\right) \sum_{l=1}^k \frac{(2k-2)!}{2^{k-l}(k-l)!(2(k+l)+1)!!}
    \end{equation*}
    and analogous with \eqref{app:A:coeff_odd}
    \begin{equation*}
        A_{2k-1} \leq \frac{s}{2}\left(\frac{s}{2}-1\right)\left(2-\frac{s}{2}\right) \sum_{l=1}^k \frac{(2k-1)!}{2^{k-l}(k-l)!(2(k+l)+3)!!}
    \end{equation*}
    Choosing $N=1000$ and computing those terms on a computer explicitly we find
    \begin{equation*}
        \sum_{k=4}^{1001} A_{k} \leq \frac{s}{2}\left(\frac{s}{2}-1\right)\left(2-\frac{s}{2}\right) \left(0.0242 + 0.0203 \right)
    \end{equation*}
    Adding the error estimate
    \begin{equation*}
        \delta(2000) \leq \frac{s}{2}\left(\frac{s}{2}-1\right)\left(2-\frac{s}{2}\right) 0.0003 
    \end{equation*}
    such that
    \begin{equation*}
        \sum_{k=4}^{\infty} A_{k} \leq  \frac{s}{2}\left(\frac{s}{2}-1\right)\left(2-\frac{s}{2}\right) \frac{448}{10000} 
    \end{equation*}
\end{proof}
\subsection{Explicit constant in an inequality due to Lieb}
\begin{lemma} \label{app:A:nasty_constant}
    For any $p\in [1,2]$ the constant $C_p$ in equation \eqref{ch5:6:p-inequ} is given by
    \begin{equation} \label{app:A:explicit_constant}
        C_p = \frac{3 \sqrt{\pi}}{4} \frac{(4\pi)^{-p/3}}{p^{1+p/2}} \frac{\left(\frac{15\sqrt{\pi}}{8} \frac{\Gamma(3/p)}{\Gamma(7/2+3/p)} \right)^{p/2} \frac{\Gamma(3/p+1)}{\Gamma(3/p+7/2)}}{\left( \frac{\sqrt{\pi}}{4} \frac{\Gamma(3/p+1)}{\Gamma(3/p+5/2)}\right)^{1+5p/6}}
    \end{equation}
    where $\Gamma$ is the Gamma function defined by the improper integral
    \begin{equation*}
        \Gamma(r) \coloneqq  \int_{0}^\infty x^{r-1}e^{-x} dx \, .
    \end{equation*}
\end{lemma}
\begin{proof}
    We give only a sketch of the calculations. Following \cite{L:1976}[p. 563] one needs to solve the following three integrals
    \begin{equation*}
        \int_{\R^3} f_p(x) dx, \, \int_{\R^3} \abs{x}^p f_p(x)dx \text{ and } \int_{\R^3} [f_p(x)]^{5/3}dx
    \end{equation*}
    for 
    \begin{equation} \label{app:A:sharp_const}
	f_p(x) \coloneqq \begin{cases}
	\left(1-\abs{x}^p\right)^\frac32 &\abs{x}\leq 1 \\
	0 &\text{elsewise}
	\end{cases} .
\end{equation}
the first integral can be solved by a straightforward calculation and the latter two can be solved by substituting $u = \abs{x}^p$. The first integral reduces to
\begin{equation*}
    \int_{0}^1 r^2(1-r^p)^{3/2} \, dr = \frac{\sqrt{\pi}}{4} \frac{\Gamma(3/p+1)}{\Gamma(3/p+5/2)} \, .
\end{equation*}
After substituting the second integral reduces to
\begin{equation*}
    \frac{1}{p}\int_0^1 u^{3/p}(1-u)^{3/2} du = \frac{3\sqrt{\pi}}{4p} \frac{\Gamma(3/p+1)}{\Gamma(3/p+7/2)} \, .
\end{equation*}
The third integral reduces to
\begin{equation*}
    \frac{1}{p}\int_0^1 u^{(3-p)/p}(1-u)^{5/2} du = \frac{15\sqrt{\pi}}{8p} \frac{\Gamma(3/p)}{\Gamma(3/p+7/2)} \, .
\end{equation*}
Combining these three integrals gives the desired constant.
\end{proof}
\subsection{Bound on the Groundstate Energie of the Bohr Atom}
The following Lemma is a well-known fact. We include it here for convenience. 
\begin{lemma}\label{app:A:bnd_on_atom_energy} Given the operator in \eqref{ch5:2:hamilton} (in atomic units) for $N$ fermions with $u\in \N$ spin-degrees of freedom and nuclear charge $Z$ then the ground state energy is bound by
\begin{equation*}
    -E_{N,Z} \leq A Z^2 N^{1/3} 
\end{equation*}
with $A= \frac12 u^{2/3}3^{1/3} $.
\end{lemma}
\begin{proof}
    The energy levels of hydrogen with nuclear charge $Z$
    \begin{equation}\label{app:A:hydrogen_like} 
    h \coloneqq -\frac{1}{2} \Delta -\frac{Z}{|x|}
    \end{equation}
    are
    \begin{equation}\label{app:A:rydbergs}
        E_n = \frac{-Z^2}{n^2} Ry
    \end{equation}
    where $Ry$ is the Rydberg energy. 
    In atomic units 
    \begin{equation} \label{app:A:rydberg_units}
        Ry = \frac{m_e e^2}{2(4\pi \varepsilon_0)^2 \hbar}  = \frac{1}{2} 
    \end{equation}
    Each of these energy levels $n^2$-times degenerated. Note that the interelectronic repulsion is a positive contribution to the operator in \eqref{ch5:2:hamilton} and thus
    \begin{equation}
    \label{app:A:hamilton_without_inerelectronic}
    H_{N,Z} \geq  \sum_{i=1}^{N}\left(-\frac{1}{2} \Delta_{i}-\frac{Z }{|x_{i}|}\right)
    \end{equation}
    The ground state of the right--hand side can be computed explicitly in terms of the energies in equation \eqref{app:A:rydbergs}. Due to the fermionic symmetry, the particles occupy the energy levels starting from the lowest counting the degeneration in the ground state. Let $E_0(N,Z)$ the ground-state energy of the right hand side of equation \eqref{app:A:hamilton_without_inerelectronic} and Denote by $L\in \N$ the last completely filled energy Level then there exists $c\in [0,1)$ such that
    \begin{equation*}
        E_{0}(N,Z) =cu(L+1)^2E_{L+1} + u\sum_{j=1}^L j^2 E_j  = - u Z^2 R_y (c+L) \,.
    \end{equation*}
    Here we have used that due to the spin of the particles, each state can be $u$-times occupied. It remains to compare $L$ with $N$.
    \begin{equation*}
        \begin{split}
            \frac{N}{u}&=  c(L+1)^2  +  \sum_{j=1}^L j^2 = c(L+1)^2 +  \frac{L^3}{3} + \frac{L^2}{2} + \frac{L}{6} 
        \end{split}
    \end{equation*}
    A straightforward calculation using $c\in[0,1]$ shows
    \begin{equation*}
        (L+c)^3 \leq \frac{3N}{u}
    \end{equation*}
    We conclude
    \begin{equation*}
        -E_{N,Z} \leq -E_0(N,Z) \leq uZ^2R_y \left( \frac{3N}{u} \right)^{1/3} = \frac12 u^{2/3}3^{1/3}Z^2 N^{1/3} \, .
    \end{equation*}
\end{proof}
\subsection{An Useful Inequality}
\begin{lemma} \label{app:A:appendix:lemma:technichal_est}
    Let $ 0<\abs{r}\le 1 $ and $p\geq 1$ then
    \begin{equation} \label{app:A:eq:technical_appendix}
    \begin{split}
       \frac{(1+r)^{p-1}- (1-r)^{p-1}}{2r(p+1)} 
        &\ge 
        \frac{(1+r)^{p}- (1-r)^{p}}{2rp} \\
        &\geq \left(1- \frac{p-1}{3} r^2 \right)  
        		\frac{(1+r)^{p+1} - (1-r)^{p+1}}{2r{(p+1)}} \, . 
    \end{split}
    \end{equation}
\end{lemma}
\begin{proof} 
Since the inequalities in \eqref{app:A:eq:technical_appendix} are symmetric under changing $r$ to $-r$, it is enough to prove them for $0<r\le 1$.  
In this case \eqref{app:A:eq:technical_appendix} is 
equivalent to 
\begin{equation} \label{app:A:eq:technical_appendix-2}
  \begin{split}
     \frac{(1+r)^{p+1}- (1-r)^{p+1}}{p+1} 
        &\ge 
          \frac{(1+r)^{p}- (1-r)^{p}}{p} \\
        &\geq 
          \left(
            1- \frac{p-1}{3} r^2 
          \right)    
          \frac{(1+r)^{p+1} - (1-r)^{p+1}}{{p+1}} \, . 
    \end{split}
    \end{equation}
The first bound, 
\begin{equation} \label{app:A:eq:technical_appendix-left-1}
    h(r)\coloneqq \frac{(1+r)^{p+1}- (1-r)^{p+1}}{(p+1)} 
        \ge 
          \frac{(1+r)^{p}- (1-r)^{p}}{p}  
        \eqqcolon k(r) 
    \end{equation}
for $r\ge 0$ is easy to show. Since 
$h(0)= 0 = k(0)$ it follows as soon as  
$h'(r)\ge k'(r)$ for $r\ge 0$. This is 
equivalent to 
\begin{align}\label{app:A:eq:technical_appendix-left-2}
    (1+r)^{p} + (1-r)^{p} 
        \ge 
          (1+r)^{p-1}- (1-r)^{p-1}
          \quad \text{for } 0 \le r\le 1 \, .
\end{align}
Expanding 
\begin{equation}\label{app:A:eq:expansion}
 \begin{split}
  (1+r)^p \pm (1-r)^p 
    &=  
      (1+r)^{p-1}(1+r) \pm (1-r)^{p-1}(1-r) \\
    &= 
      (1+r)^{p-1} \pm (1-r)^{p-1} 
      +r\big( (1+r)^{p-1} \mp (1-r)^{p-1} \big)
 \end{split}
\end{equation}
in the left--hand--side of 
\eqref{app:A:eq:technical_appendix-left-2} 
we see that \eqref{app:A:eq:technical_appendix-left-2} is 
equivalent to 
\begin{align*}
  r\left((1+r)^{p-1} - (1-r)^{p-1}\right) 
    &\ge 0 \,
\end{align*} 
for $0\le r\le 1$, which is true if $p\ge 1$. 
This proves the first bound  in \eqref{app:A:eq:technical_appendix-2}. 

We prove the second inequality in 
\eqref{app:A:eq:technical_appendix-2} by determining 
$c\in \R$ such that  
\begin{equation}\label{app:A:eq:technical_est_01}
   f_1(r)\coloneqq  \frac{(1+r)^{p}- (1-r)^{p}}{p} \geq \left(1- c r^2 \right)  \frac{(1+r)^{p+1} - (1-r)^{p+1}}{{(p+1)}} \eqqcolon g_1(r)\, , 
\end{equation}
for $0\le r\le 1$. At $r=0$ we have $f_1(0)= g_1(0)$, so 
\eqref{app:A:eq:technical_est_01} is true for $r\ge 0$ as soon as $f_1'(r)\ge g_1'(r)$ for $0\le r\le 0$. 
This is equivalent to 
\begin{equation}\label{app:A:eq:technical_est_02}
    \begin{split}
        (1+r)^{p-1} + (1-r)^{p-1} 
          &\geq      
             (1 - c r^2) \left((1 + r)^{p} + (1 - r)^{p}\right) \\
          &\phantom{\ge ~~~~} 
            \frac{2 c r \left( (1 + r)^{p+1}-(1 - r)^{p+1}\right)}{p+1}
    \end{split}
\end{equation}
Using \eqref{app:A:eq:expansion} this ie equivalent to 
\begin{equation}\label{app:A:eq:technical_est_03}
    \begin{split}
       f_2(r) 
         &\coloneqq  
           - \big(  (1+r)^{p-1} - (1-r)^{p-1}  \big) 
           + \frac{2 c}{p+1} 
             \left( 
               (1 + r)^{p+1}-(1 - r)^{p+1}
             \right) \\
         &\phantom{=~~~} 
           + c r \left((1 + r)^{p} + (1 - r)^{p}\right) 
          \ge 0 
          \quad \text{for } 0\le r\le 1 \, .
    \end{split}
\end{equation}
Since $f_2(0)=0$ we know that 
\eqref{app:A:eq:technical_est_03} holds as soon as $f_2'(r)\ge 0$ 
for $0\le r\le 1$. Now 
\begin{equation*} 
  \begin{split}
    f_2'(r) 
      & = 
        -(p-1) \left( (1 + r)^{p-2}+ (1 - r)^{p-2} \right) 
        + 3 c 
             \left( 
               (1 + r)^{p}+(1 - r)^{p}
             \right) \\
     &\phantom{=~~}
       +cpr\left( (1 + r)^{p-1} - (1 - r)^{p-1} \right)
  \end{split}
\end{equation*} 
and using \eqref{app:A:eq:expansion} twice, the second time with $p$ replaced with $p-1$, we have 

\begin{align*}
  (1+r)^p \pm (1-r)^p 
    &= 
        (1+r)^{p-2} \pm (1-r)^{p-2} 
        + 2r\left(  
             (1+r)^{p-2} \mp (1-r)^{p-2}  
            \right) \\
    &\phantom{=~~}   
      + r^2\left(  
             (1+r)^{p-2}\pm (1-r)^{p-2}  
            \right) \, .   
\end{align*}
Hence we see that 
\begin{equation*}
  \begin{split}
    f_2'(r) 
       = 
        &\left( 3c-(p-1)\right) \left((1 - r)^{p-2} + (1 + r)^{p-2}\right) 
        \\
    &+ (6+p)cr  
             \left( 
               (1 + r)^{p-2}-(1 - r)^{p-2}
             \right) \\
     &\phantom{=~~}
       +       (3+p)c  r^2\left(  
             (1+r)^{p-2}+(1-r)^{p-2}  
            \right) \geq 0
  \end{split}
\end{equation*}
for all $0\le r\le 1$ as soon as $c\ge (p-1)/3\ge 0$. 
This proves the second inequality in \eqref{app:A:eq:technical_appendix-2} and 
finishes the proof of \eqref{app:A:eq:technical_appendix}.
\end{proof}

\end{appendix}


\providecommand{\MR}{\relax\ifhmode\unskip\space\fi MR }
\providecommand{\MRhref}[2]{%
  \href{http://www.ams.org/mathscinet-getitem?mr=#1}{#2}
}

\bibliographystyle{annotate}
\bibliography{references}

\end{document}